\documentclass[letterpaper, onecolumn, notitlepage, longbibliography, floatfix, superscriptaddress, aps, prx, unpublished, 12pt]{quantumarticle}
\pdfoutput=1
\usepackage{geometry}
\usepackage[utf8]{inputenc}
\usepackage[usenames, dvipsnames]{xcolor}
\usepackage{dsfont}
\usepackage{amsmath}
\usepackage{amsfonts}
\usepackage{amssymb}
\usepackage{graphicx}
\usepackage[pdfencoding=auto, psdextra]{hyperref} 
\hypersetup{linktocpage}
\hypersetup{colorlinks=true,citecolor=quantumpurple,linkcolor=quantumpurple, urlcolor=quantumpurple}
\usepackage{natbib}
\usepackage{environ}
\usepackage{enumitem}
\usepackage{pdfpages}
\usepackage{mathrsfs}
\usepackage{changepage}
\usepackage[caption=false]{subfig}
\usepackage{amsmath, amssymb}
\usepackage{makecell}

\NewEnviron{eqs}{%
\begin{equation}\begin{split}
    \BODY
\end{split}\end{equation}
}

\makeatletter 
\def\l@subsubsection#1#2{}
\makeatother

\hypersetup{colorlinks=true,citecolor=quantumpurple,linkcolor=quantumpurple, urlcolor=quantumpurple}

\newcommand{\ZZ}{\mathbb{Z}}

\definecolor{quantumpurple}{RGB}{82, 35, 124}

\definecolor{joe}{RGB}{127,0,127}

\definecolor{ultramarine}{RGB}{0,150,96}

\definecolor{tyler}{rgb}{1,.3,0}

\begin{document}

\title{\textcolor{quantumpurple}{Floquet codes with a twist}}

\author{Tyler~D. Ellison}
  \email{tyler.ellison@yale.edu}
\affiliation{Department of Physics, Yale University, New Haven, CT 06511, USA}
\author{Joseph Sullivan}
\email{joseph.sullivan@ubc.ca}
\affiliation{Stewart Blusson Quantum Matter Institute, University of British Columbia, Vancouver, BC, Canada V6T 1Z1}
\affiliation{Department of Physics, Yale University, New Haven, CT 06511, USA}
\author{Arpit Dua}
\affiliation{Department of Physics, California Institute of Technology, Pasadena, CA 91125, USA}
\affiliation{Institute for Quantum Information and Matter, California Institute of Technology, Pasadena, California 91125, USA}

\date{September 19, 2023}

\maketitle

\vspace{0cm}

\begin{abstract}
\noindent \footnotesize \textcolor{quantumpurple}{\textbf{\textsf{Abstract:}}}
We describe a method for creating twist defects in the honeycomb Floquet code of Hastings and Haah. 
In particular, we construct twist defects at the endpoints of condensation defects, which are built by condensing emergent fermions along one-dimensional paths. 
We argue that the twist defects can be used to store and process quantum information fault tolerantly, and demonstrate that, by preparing twist defects on a system with a boundary, we obtain a planar variant of the $\ZZ_2$ Floquet code.
Importantly, our construction of twist defects maintains the connectivity of the hexagonal lattice, requires only 2-body measurements, and preserves the three-round period of the measurement schedule. 
We furthermore generalize the twist defects to $\ZZ_N$ Floquet codes defined on $N$-dimensional qudits.~As an aside, we use the $\ZZ_N$ Floquet codes and condensation defects to define Floquet codes whose instantaneous stabilizer groups are characterized by the topological order of certain Abelian twisted quantum doubles.
\end{abstract}

\clearpage

\renewcommand{\baselinestretch}{.96}\normalsize
\tableofcontents
\renewcommand{\baselinestretch}{1.0}\normalsize

\section{Introduction} \label{sec: Introduction}

Quantum error-correcting codes are essential to ensuring that quantum computations~are reliable despite noise from the environment and faulty operations. The toric code and~its derivatives have stood out as some of the most promising candidates for this task, and substantial recent progress has been made towards their experimental implementation~{\mbox{\cite{Google2021realizing, Semeghini2021probing, Zhao2022realization, Google2022suppressing, Google2022nonabelian, xu2022digital, iqbal2023topological}}}. Nonetheless, given the importance of quantum error correction, it is imperative to continue to develop quantum error correcting codes with the goal of increasing error thresholds, reducing computational overheads, and finding codes that are better tailored to current hardware. 

Recently, Hastings and Haah introduced a new type of quantum error-correcting code, which we refer to here as the $\ZZ_2$ Floquet code~\cite{HH_dynamic_2021}. The $\ZZ_2$ Floquet code features a code space that is inherently dynamical and evolves in time through a periodic schedule of non-commuting measurements. After each round of measurements, the system is characterized by an instantaneous stabilizer group (ISG), and quantum information can be encoded in the associated instantaneous code spaces. The measurements are carefully scheduled so that the logical information is preserved throughout the dynamics and passed from one instantaneous code space to the next.
The $\ZZ_2$ Floquet code boasts high error thresholds and requires only 2-body measurements -- thus alleviating the overhead of higher-weight measurements and making it a competitive alternative to the toric code~\cite{Gidney2021faulttolerant,Gidney2022benchmarkingplanar, Paetznick2023Performance}.

Many desirable properties of the toric code are naturally transferable to the $\ZZ_2$ Floquet code, since the ISGs of the $\ZZ_2$ Floquet code are a concatenation of the hexagonal-lattice toric code with a 2-qubit repetition code. 
For example, it was shown in Ref.~\cite{Haah2022boundarieshoneycomb} that the $\ZZ_2$ Floquet code admits a planar realization encoding a single logical qubit, analogous to the surface code. It was further argued that logical Clifford gates can be implemented fault tolerantly in the planar $\ZZ_2$ Floquet code using lattice surgery. However, to adapt the $\ZZ_2$ Floquet code to a planar geometry, the approaches taken in Refs.~\cite{Haah2022boundarieshoneycomb, Gidney2022benchmarkingplanar, Paetznick2023Performance} require doubling the period of the $\ZZ_2$ Floquet code -- hence, slowing down the detection of errors near the boundary.

One valuable concept from the study of toric codes that has yet to be explored in~the context of Floquet codes is that of twist defects~\cite{Bombin2010Twist,KitaevKong2012boundaries}. Twist defects are point-like~irregularities of the code that appear at the endpoints of a defect line, which for the toric code, is created by locally modifying the stabilizer group along an open path.
In comparison~to surface codes, twist defects offer a more compact encoding of quantum information, while still admitting fault-tolerant logical operations similar to lattice surgery~\cite{hastings2014dislocations}. 
Twist defects can also be used in conjunction with surface codes to reduce the time overheads of lattice surgery~\mbox{\cite{Litinski2018latticesurgerytwist,Litinski2019gameofsurfacecodes,Campbell2022twistfree,Campbell2022twistssurgery}}. 

One drawback of twist defects, however, is that they are typically created by adding a dislocation to the lattice -- thereby changing the connectivity of the qubits and complicating the measurement of the stabilizer syndrome. 
For the $\ZZ_2$ Floquet code, there is an additional obstacle to inserting twist defects, given the inherent dynamics of the system. In particular, the modifications needed to create twist defects for one of the ISGs, might fail to create twist defects for the subsequent ISGs.

Here, we overcome these challenges by devising a method for inserting twist defects in the $\ZZ_2$ Floquet code, which preserves the connectivity of the qubits and the period of the measurement schedule. Our construction requires only a mild change to the $\ZZ_2$ Floquet code, consisting of measuring a different set of 2-body operators along the defect line. We show that our twist defects are indeed able to encode quantum information throughout the evolution of the code, by identifying a set of logical operators that are invariant under the dynamics. 
Further, we describe how topological charge measurements, i.e., measurements of anyon string operators, can be used to implement the full Clifford group on the logical qubits. 

The key to our construction of twist defects is the concept of one-dimensional condensation defects introduced in Ref.~\cite{Roumpedakis2022Higher}. As described from a field-theoretic perspective in Ref.~\cite{Roumpedakis2022Higher}, twist defects in 2D topological orders are hosted at the endpoints of condensation defects, constructed by condensing anyons along open paths. For example, the twist defects in the topological order of the toric code appear at the endpoints of a condensation defect created by condensing emergent fermions along an open path. In Appendix~\ref{app: TC condensation defects}, we demonstrate that this construction holds at the lattice level by explicitly condensing emergent fermions of the toric code along a path (see also Ref.~\cite{Wootton2015Matching}). This approach to creating twist defects is especially well suited for the $\ZZ_2$ Floquet code of Ref.~\cite{HH_dynamic_2021}, since the string operators of the emergent fermions are invariant under the dynamics~\cite{Kitaev_2006}. This implies that, in each of the ISGs, the same string operator can be used to create emergent fermions. Therefore, condensing the emergent fermions along a path produces twist defects in each of the~ISGs. 

In this work, we also show that the $\ZZ_2$ Floquet code can be generalized to $\ZZ_N$ Floquet codes defined on $N$-dimensional qudits (see also Ref.~\cite{sullivan2023floquet}). These Floquet codes have the property that their ISGs have the same topological order as that of $\ZZ_N$ toric codes. Our construction of twist defects carries over immediately to the case of $\ZZ_N$ Floquet codes. We furthermore leverage the $\ZZ_N$ Floquet codes and our method for inserting defect lines to construct Floquet codes with ISGs that exhibit even more exotic topological orders, namely certain Abelian twisted quantum doubles (TQDs)~\cite{Ellison2022stabilizer}.

The paper is organized as follows. 
In Section~\ref{sec: review}, we recall the details of the $\ZZ_2$ Floquet code of Ref.~\cite{HH_dynamic_2021}. 
In Section~\ref{sec: Z2 defects}, we describe how to modify the $\ZZ_2$ Floquet code to insert twist defects. 
In Section~\ref{sec: ZN defects}, we then generalize the $\ZZ_2$ Floquet code to $\ZZ_N$ Floquet codes built on systems of $N$-dimensional qudits and describe how twist defects can be inserted in this case.
In Appendix~\ref{app: TC condensation defects}, we motivate our approach to constructing twist defects in Floquet codes by constructing twist defects in the $\ZZ_2$ toric code via the condensation of emergent fermions. In Appendix~\ref{app: inferring stabilizers}, we fill in the technical details related to inferring the measurement outcomes of the instantaneous stabilizers in the vicinity of defect lines. 

\section{Review of the $\ZZ_2$ Floquet code} \label{sec: review}

We begin by reviewing the Floquet code introduced in Ref.~\cite{HH_dynamic_2021}, which we refer to here as the $\ZZ_2$ Floquet code.\footnote{This nomenclature is motivated by the fact that each instantaneous code space is the same as the ground state subspace of a deconfined $\ZZ_2$ gauge theory, i.e., the toric code, as demonstrated in Section~\ref{sec: review ISGs}.} We encourage readers that are familiar with the $\ZZ_2$ Floquet code to proceed to Section~\ref{sec: Z2 defects} for the construction of twist defects. 

\begin{figure*}[t]
\centering
\subfloat[\label{fig: xyz edges}]{\includegraphics[width=.45\textwidth]{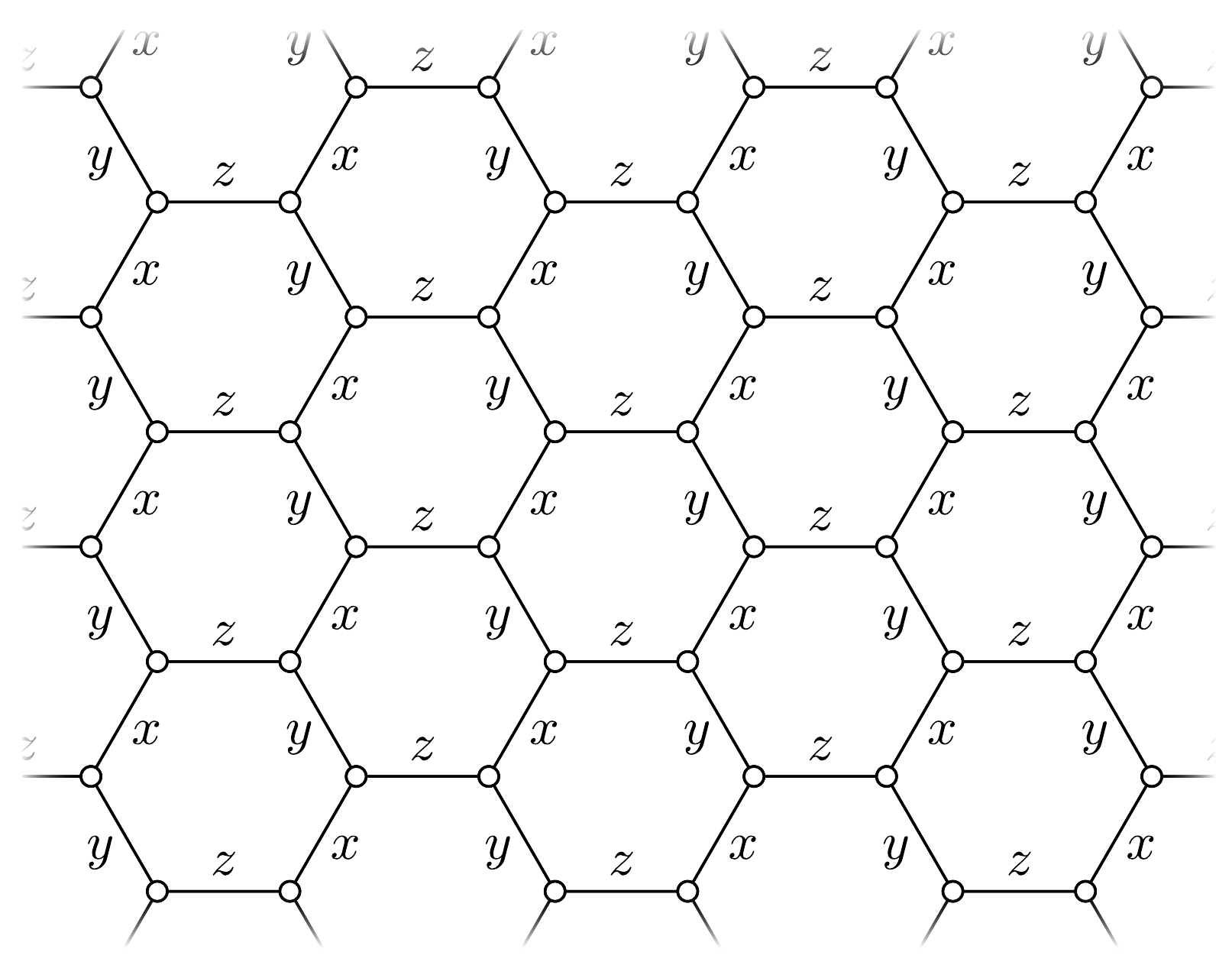}} \quad
\subfloat[\label{fig: 012 edges}]{\includegraphics[width=.45\textwidth]{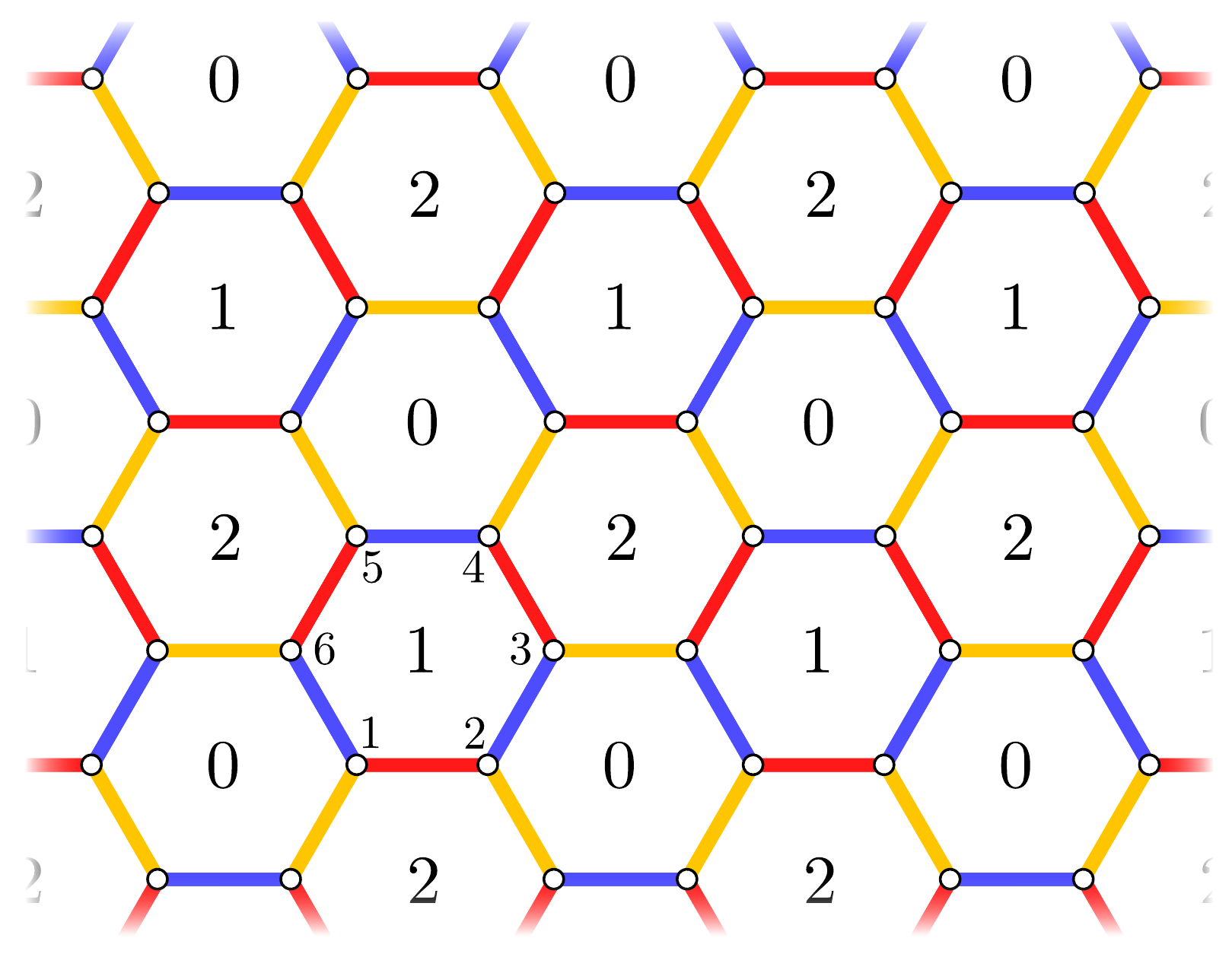}}
\caption{(a) To define the operators $K_{ij}$ in Eq.~\eqref{eq: Z2 checks}, each edge of the hexagonal lattice is labeled as an $x$-, $y$-, or $z$-edge. (b) To specify the measurement schedule, we label the plaquettes as $0$-, $1$-, and $2$-plaquettes, such that neighboring pairs of plaquettes have distinct labels.~The $0$-edges~(red), $1$-edges (yellow), and $2$-edges (blue) connect the $0$-, $1$-, and $2$-plaquettes, respectively. The vertices of the plaquettes are labeled 1-6 to define the plaquette stabilizer in Eq.~\eqref{eq: plaquette stabilizer vertices}.}
\label{fig: xyz,012 lattice}
\end{figure*}

The $\ZZ_2$ Floquet code is defined on a hexagonal lattice with periodic boundary conditions and a qubit at each vertex, as shown in Fig.~\ref{fig: xyz edges}. We write the Pauli $X$, $Y$, and $Z$ operators at a site $i$ as $X_i$, $Y_i$, and $Z_i$, respectively. The $\ZZ_2$ Floquet code is then specified by two pieces of data: (i) a set of 2-body operators known as the check operators, and (ii) a measurement schedule dictating the order in which the check operators are measured. In Section~\ref{sec: review measurement schedule} we explicitly define the check operators and the measurement schedule of the $\ZZ_2$ Floquet code. In Sections~\ref{sec: review ISGs}~and~\ref{sec: review logical operators}, we then argue that there are logical qubits that emerge from the dynamics of the measurement schedule. 

\subsection{What to measure and when} \label{sec: review measurement schedule}

To define the check operators of the $\ZZ_2$ Floquet code, we label each edge of the hexagonal lattice as an $x$-, $y$-, or $z$-edge according to Fig.~\ref{fig: xyz edges}. We then define the operator $K_{ij}$ for each edge $\langle ij \rangle$ connecting neighboring vertices $i$ and $j$:
    \begin{align} \label{eq: Z2 checks}
        K_{ij} \equiv 
        \begin{cases}
            X_iX_j & \text{ if } \langle ij \rangle \in x\text{-edges}, \\
            Y_iY_j & \text{ if } \langle ij \rangle \in y\text{-edges}, \\
            Z_iZ_j & \text{ if } \langle ij \rangle \in z\text{-edges}.
        \end{cases}
    \end{align}
The operators $K_{ij}$ can be graphically represented as:
\begin{align}
    \vcenter{\hbox{\includegraphics[scale=.6]{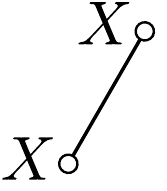}}}, \qquad 
    \vcenter{\hbox{\includegraphics[scale=.6]{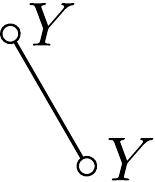}}}, \qquad
    \vcenter{\hbox{\includegraphics[scale=.6]{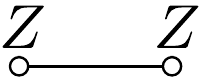}}}.
\end{align}
These operators are precisely the check operators of the $\ZZ_2$ Floquet code. We note that, if the check operators are viewed as the generators of a gauge group of a subsystem code, then the subsystem code encodes no logical qubits~\cite{HH_dynamic_2021}. In this sense, the logical qubits of the $\ZZ_2$ Floquet code, described in Section~\ref{sec: review logical operators}, are dynamically generated. 
 
Next, to specify the measurement schedule, 
we label each plaquette of the hexagonal lattice with an element from the set $\{0,1,2\}$, such that any two plaquettes sharing an edge are labeled by a different value (see Fig.~\ref{fig: 012 edges}). We refer to the plaquettes labeled by $r$ in $\{0,1,2\}$ as the $r$-plaquettes. We then define the $r$-edges to be the edges connecting the $r$-plaquettes and define the $r$-checks to be the set of check operators associated to the $r$-edges, as shown in Fig.~\ref{fig: 012 edges}. Explicitly, the $r$-checks are the following set of operators:
\begin{align}
    r\text{-checks} \equiv \{ K_{ij} : \langle ij \rangle \in r\text{-edges} \}.
\end{align}

The measurement schedule consists of measuring 0-, 1-, and 2-checks, starting with a series of measurements needed to initialize the code. The $\ZZ_2$ Floquet code is initialized by four rounds of measurements, in which we measure the 2-, 0-, and 1-checks, followed by a second round of 2-checks. After the initialization, the schedule is periodic and proceeds by measuring 0-, 1-, and 2-checks consecutively. We write the full schedule as:
    \begin{align} \label{eq: Z2 measurement schedule}
        [2,0,1,2](0,1,2)(0,1,2)(0,1,2)\ldots,
    \end{align}
where the period of initialization is in square brackets and each subsequent period is in parentheses.

\subsection{Instantaneous stabilizer groups} \label{sec: review ISGs}

The information that can be inferred from our measurements at a given instance of time is captured by the ISG. Before making any measurements, we take the ISG to be the identity. This represents the fact that we have not obtained any information, and thus there are no constraints on the system. As we make measurements, we update the stabilizer group according to the methods of the Gottesman-Knill theorem. That is, after measuring the $r$-checks, we append the $r$-checks (multiplied by the measurement outcome) to the set of generators of the stabilizer group and remove all of the stabilizers that fail to commute with the $r$-checks. This then defines our new ISG after the $r$-checks. 

After the four rounds of measurements needed to initialize the Floquet code, the ISGs become periodic, up to the values of the measurement outcomes -- which we ignore moving forwards. We label the ISG after measuring the $r$-checks as $\mathcal{S}_r$ and refer to it as the $r$-ISG. To make the $r$-ISG explicit, we define a stabilizer $S_p$ for each plaquette $p$ as:
\begin{align} \label{eq: plaquette stabilizer vertices}
     S_p \equiv X_1Y_2Z_3X_4Y_5Z_6,
\end{align}
where the vertices are ordered around the plaquette $p$ as in Fig.~\ref{fig: 012 edges}. The $r$-ISG is then generated by both the $r$-checks and the set of all plaquette stabilizers. Graphically, we represent the r-ISG as:
\begin{align} \label{eq: r-ISG}
    \mathcal{S}_r \equiv \left \langle 
    \vcenter{\hbox{\includegraphics[scale=.5]{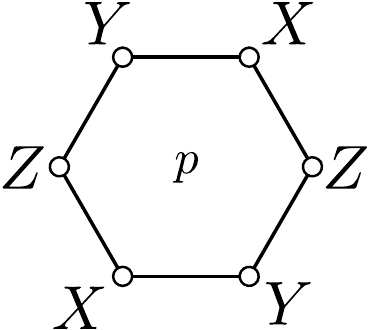}}}, \quad 
    \underbrace{\vcenter{\hbox{\includegraphics[scale=.6]{Figures/X_Kij.pdf}}}, \quad
    \vcenter{\hbox{\includegraphics[scale=.6]{Figures/Y_Kij.pdf}}}, \quad
    \vcenter{\hbox{\includegraphics[scale=.6]{Figures/Z_Kij.pdf}}}}_{\text{$r$-checks}}
    \right \rangle.
\end{align}
We note that the measurement outcome of the plaquette stabilizer at an $r$-plaquette can be inferred from the previous measurements of the $(r-2)$- and $(r-1)$-checks, with $r-2$ and $r-1$ computed modulo 3. 

\begin{figure*}[t]
\centering
\includegraphics[width=.3\textwidth]{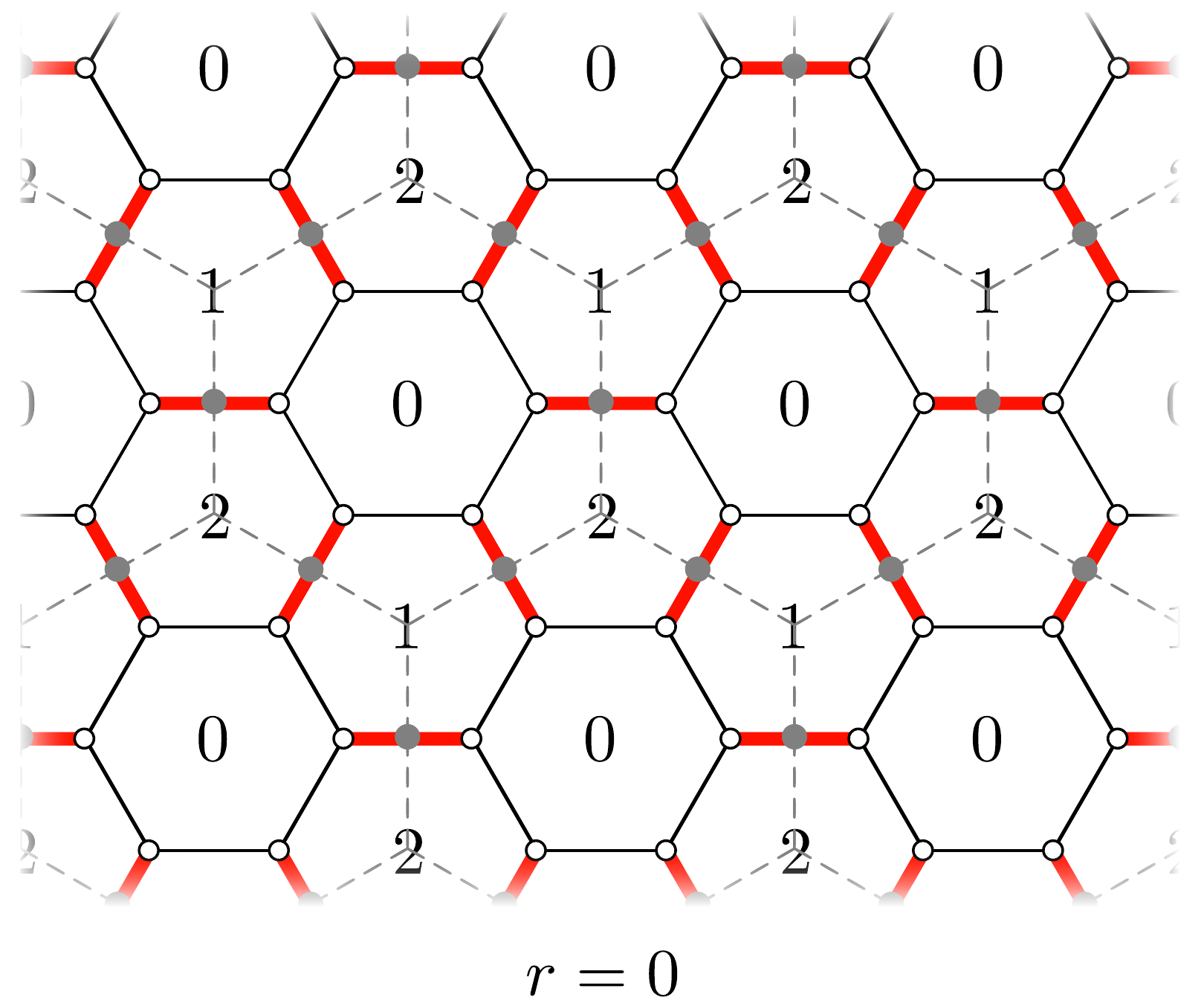}
\raisebox{.42cm}{\rule{.5pt}{3.5cm}}
\includegraphics[width=.3\textwidth]{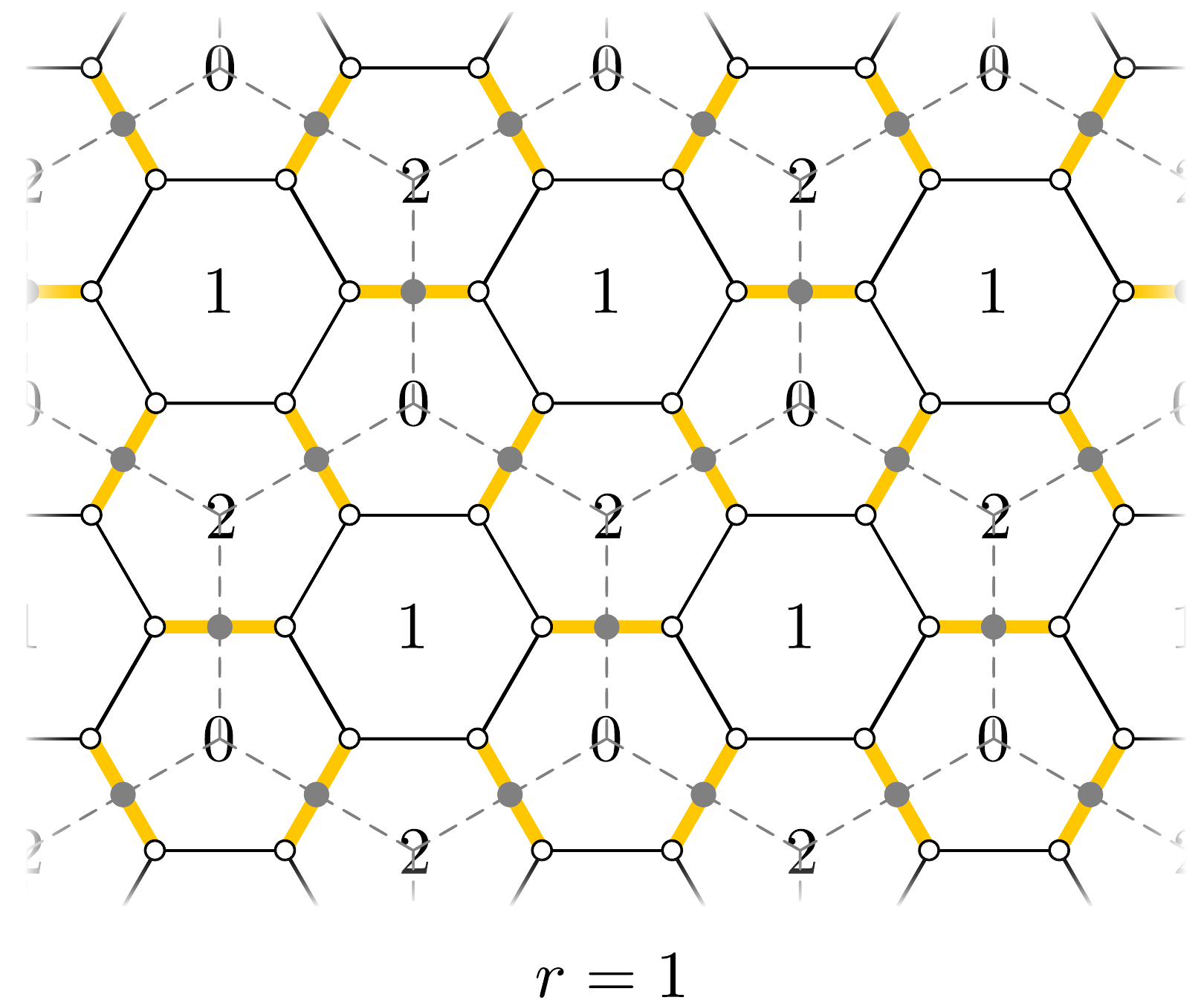}
\raisebox{.42cm}{\rule{.5pt}{3.5cm}}
\includegraphics[width=.3\textwidth]{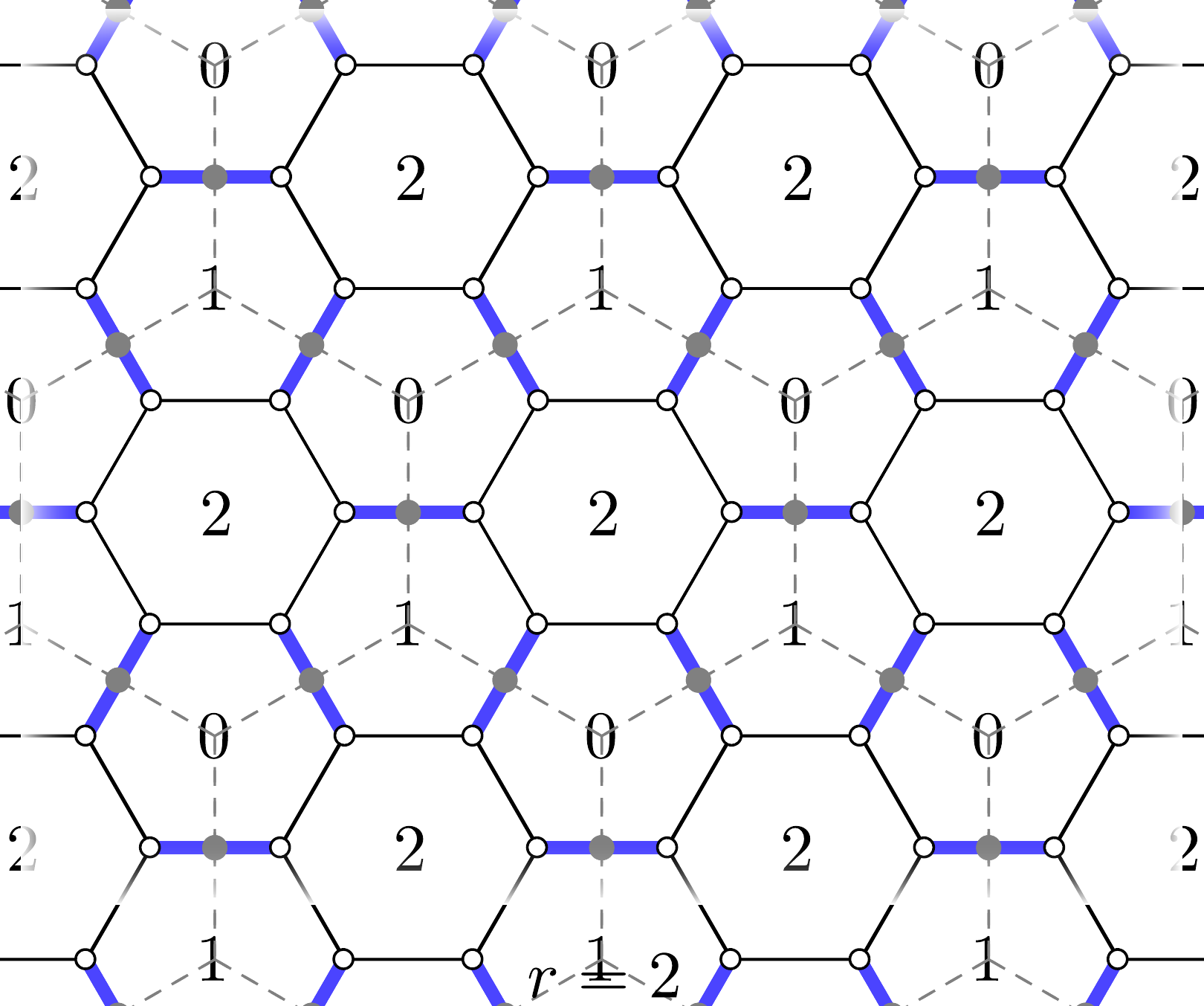}
\caption{After each round of measurements, we interpret the $+1$ eigenspace of the 2-body checks as an effective qubit on the edges (gray dots). The effective qubits live on the edges of a hexagonal super-lattice (dashed gray).}
\label{fig: effective lattices}
\end{figure*}

\begin{table}[tb]
    \centering
    \begin{tabular}{c|c|c|c}
         \hspace*{0mm}$\langle ij \rangle$ edge type \hspace*{0mm} & \hspace*{0mm}Subspace\hspace*{0mm} & \hspace*{3mm} $X_e$\hspace*{3mm} & \hspace*{3mm} $Z_e$ \hspace*{3mm}     
\\
         \hline \hline
        $x\text{-edge}$ & $X_i X_j = 1$  & $Y_iZ_j = Z_iY_j$ & $X_i I_j = I_i X_j$\\
        $y\text{-edge}$ & $Y_i Y_j = 1$ & $Z_i X_j = X_i Z_j$ & $Y_iI_j = I_i Y_j$\\
        $z\text{-edge}$ & $Z_i Z_j = 1$ & $X_i Y_j = Y_i X_j$ & $Z_i I_j = I_i Z_j$\\
    \end{tabular}
    \caption{After measuring the $r$-checks, we define an effective qubit at each $r$-edge by restricting to the two-dimensional subspace where $X_iX_j$, $Y_iY_j$, or $Z_iZ_j$ is the identity, depending on whether the $r$-edge $\langle ij \rangle$ is an $x$-, $y$-, or $z$-edge, respectively. In this subspace, we define an effective Pauli $X$ operator $X_e$ and an effective Pauli $Z$ operator $Z_e$.}
    \label{tab: effective qubits}
\end{table}

As argued in Ref.~\cite{HH_dynamic_2021}, the $r$-ISG has the same underlying topological order as the toric code. This can be made explicit by viewing the (two-dimensional) $+1$ eigenspace of the $r$-check as an effective qubit at the $r$-edge.
We think of the effective qubits as living on the edges of a hexagonal super-lattice, whose plaquettes are centered on the $r$-plaquettes, as pictured in Fig.~\ref{fig: effective lattices}. We then define effective Pauli operators $X_e$ and $Z_e$ that act on the effective qubits. The effective Pauli operators at an $r$-edge $\langle ij \rangle$ can be expressed in terms of the Pauli operators at vertices $i$ and $j$ according to Table~\ref{tab: effective qubits}. Using the definitions of $X_e$ and $Z_e$ in Table~\ref{tab: effective qubits}, the plaquette stabilizers at the $r$-plaquettes can be written as a product of $Z_e$ operators around a plaquette of the super-lattice, while the plaquette stabilizers at the $(r-2)$- and $(r-1)$-plaquettes can be written as products of $X_e$ operators around the vertices of the super-lattice:
\begin{align}
    \vcenter{\hbox{\includegraphics[scale=.5]{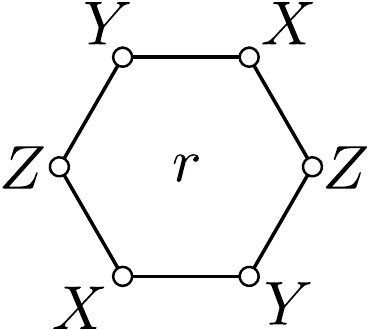}}} \, \longleftrightarrow \vcenter{\hbox{\includegraphics[scale=.5]{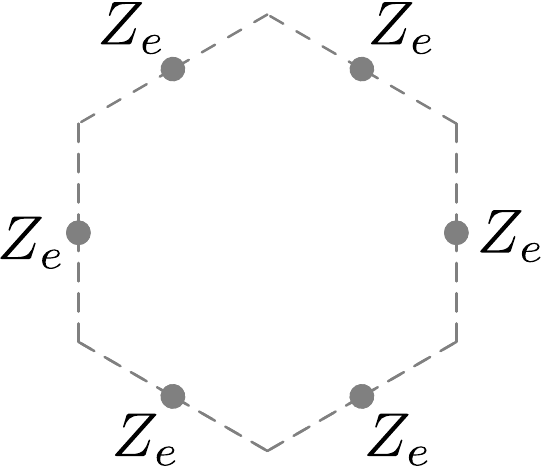}}},  \qquad
    \vcenter{\hbox{\includegraphics[scale=.5]{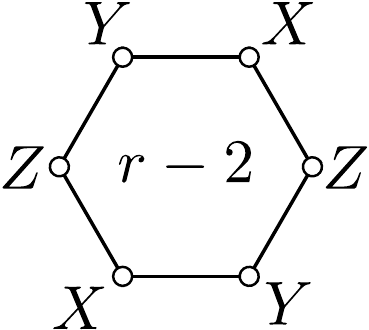}}}, \,\, \vcenter{\hbox{\includegraphics[scale=.5]{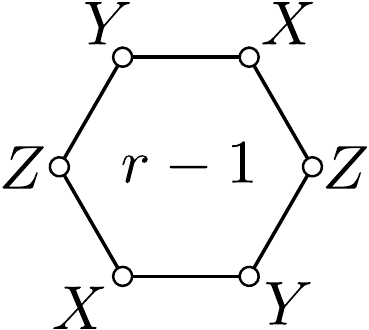}}} \, \longleftrightarrow \!\!\! \vcenter{\hbox{\includegraphics[scale=.5]{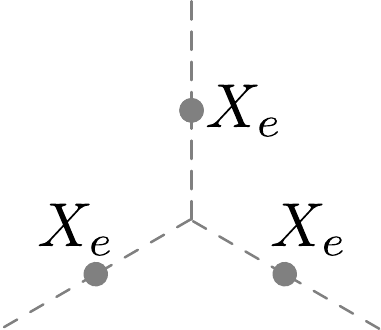}}}.
\end{align}
Thus, in the $+1$ eigenspace of the $r$-checks, the $r$-ISG is nothing more than a toric code on a hexagonal super-lattice. 

\begin{figure*}[t]
\centering
\subfloat[\label{fig: e string example}]{\includegraphics[width=.38\textwidth]{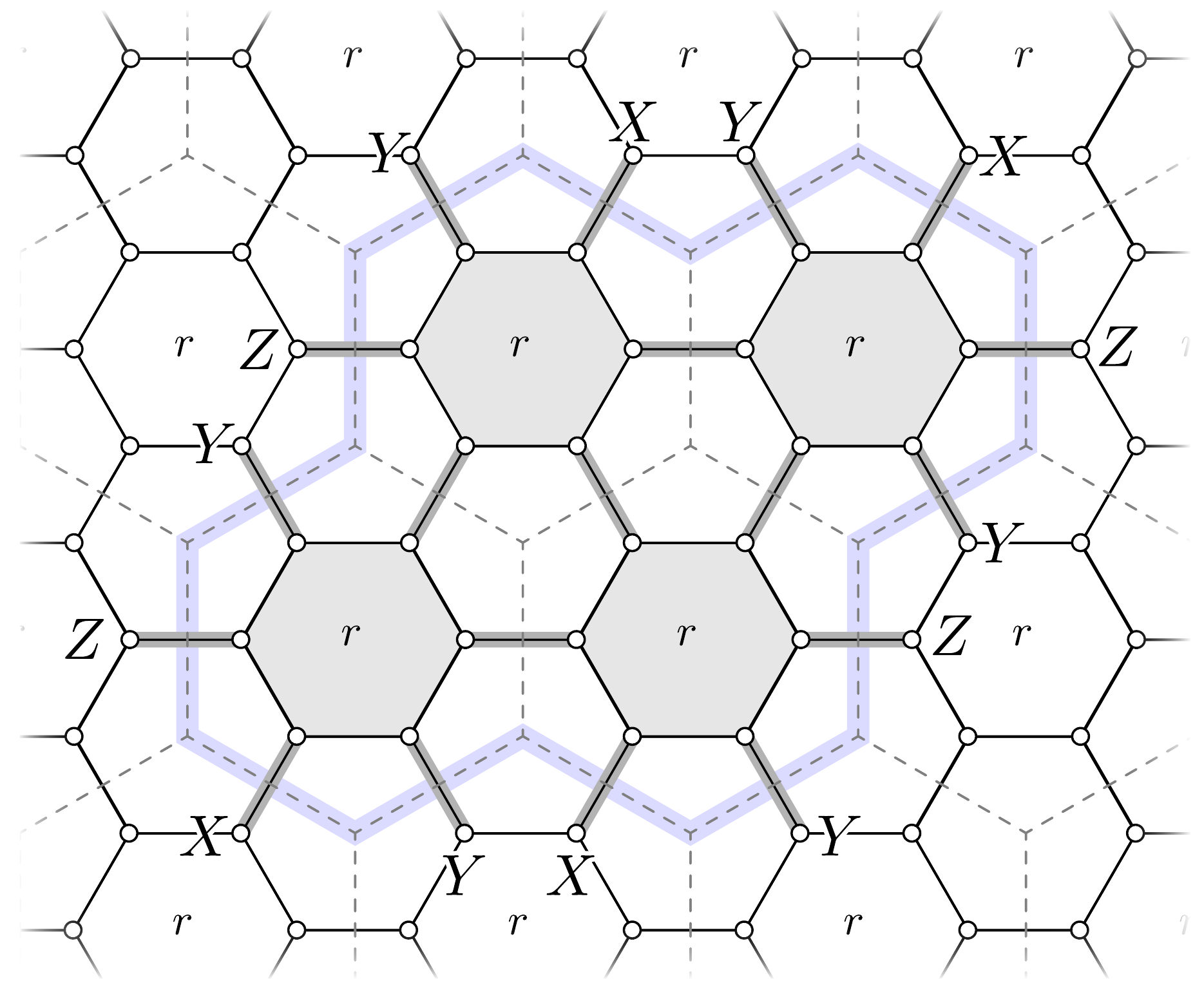}} \quad
\subfloat[\label{fig: m string example}]{\includegraphics[width=.38\textwidth]{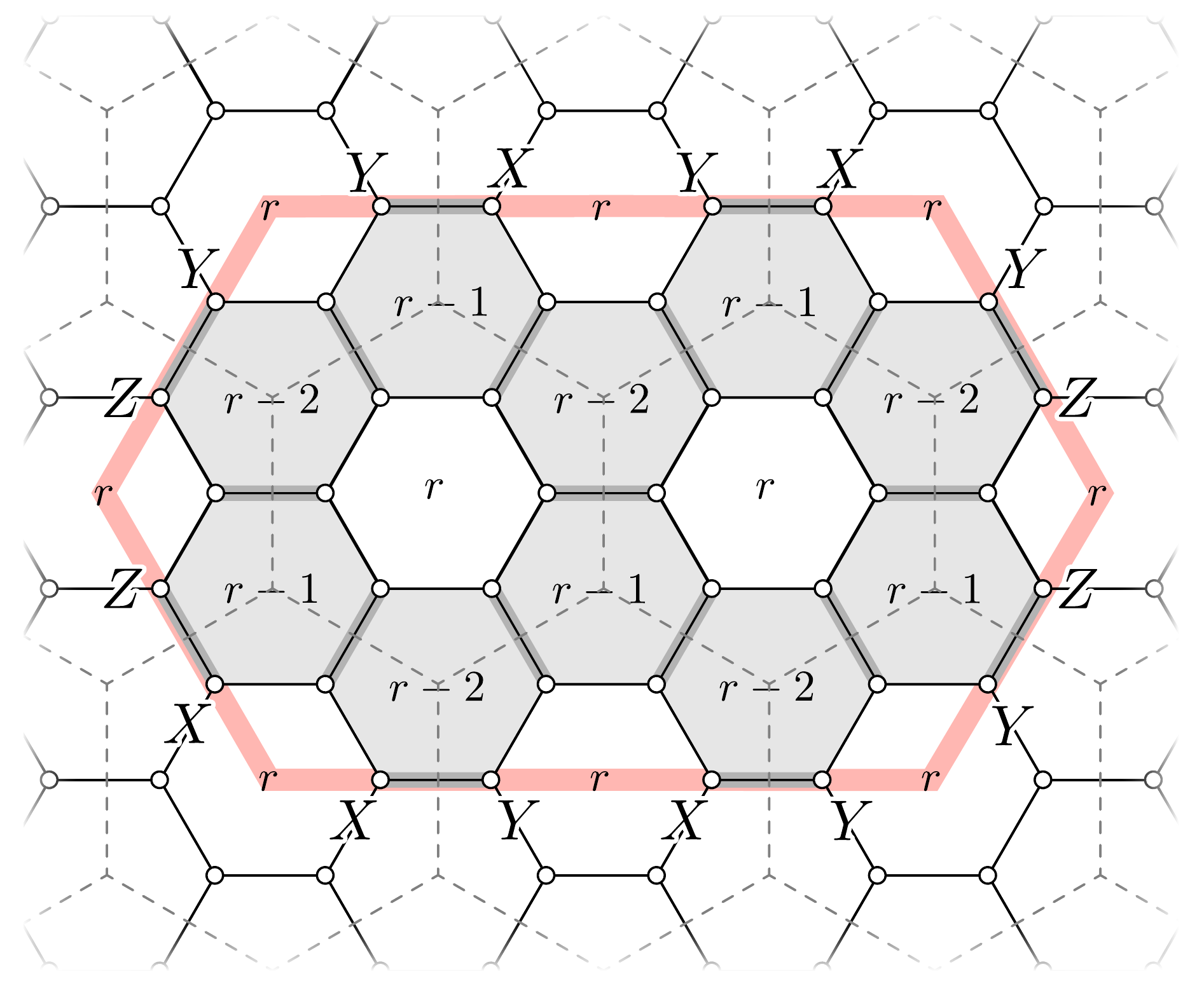}}
\caption{(a) After measuring the $r$-checks, the $e$ string operator (light blue) along a contractible path is a product of the $r$-plaquettes (gray) and $r$-checks (gray) enclosed by the path. (b) The $m$ string operator (light red) along a contractible path is a product of the $(r-2)$-plaquettes, $(r-1)$-plaquettes (both gray), and the $r$-checks (gray) enclosed by the path.}
\label{fig: e and m string example}
\end{figure*}

This implies that the $r$-ISG is characterized by the same anyon types as the toric code. We label the (nontrivial) anyon types as $e$, $m$, and $\psi$. We use the convention that a violation of an $r$-plaquette stabilizer is an $m$ anyon and a violation of an $(r-2)$- or $(r-1)$-plaquette stabilizer is an $e$ anyon. These correspond to a violation of a plaquette stabilizer or a vertex stabilizer of the effective toric code, respectively. Given a contractible path $\gamma$, the string operator of an $e$ anyon along $\gamma$ is a product of the $r$-plaquette stabilizers and $r$-checks enclosed by $\gamma$, while the string operator of an $m$ anyon along $\gamma$ is a product of the $(r-2)$- and $(r-1)$-plaquette stabilizers and the $r$-checks enclosed by $\gamma$. The string operators for $e$ and $m$ anyons can be seen in Fig.~\ref{fig: e and m string example}.

The $\psi$ anyon is a composite of an $e$ anyon and $m$ anyon and thus corresponds to a violation of both an $r$-plaquette stabilizer and an $(r-2)$- or $(r-1)$-plaquette stabilizer. 
The string operator for a $\psi$ anyon is a product of the string operators for an $e$ anyon and an $m$ anyon. This can be written more simply in terms of $K_{ij}$ operators, as made explicit in Section~\ref{sec: Z2 defects}. We note here that $\psi$ has fermionic exchange statistics. Since Pauli operators satisfy bosonic commutation relations and there are no bona fide physical fermions in the system, we sometimes refer to $\psi$ as being an emergent fermion. 

\subsection{Logical operators} \label{sec: review logical operators}

\begin{figure}[t]
\centering
\includegraphics[width=.23\textwidth]{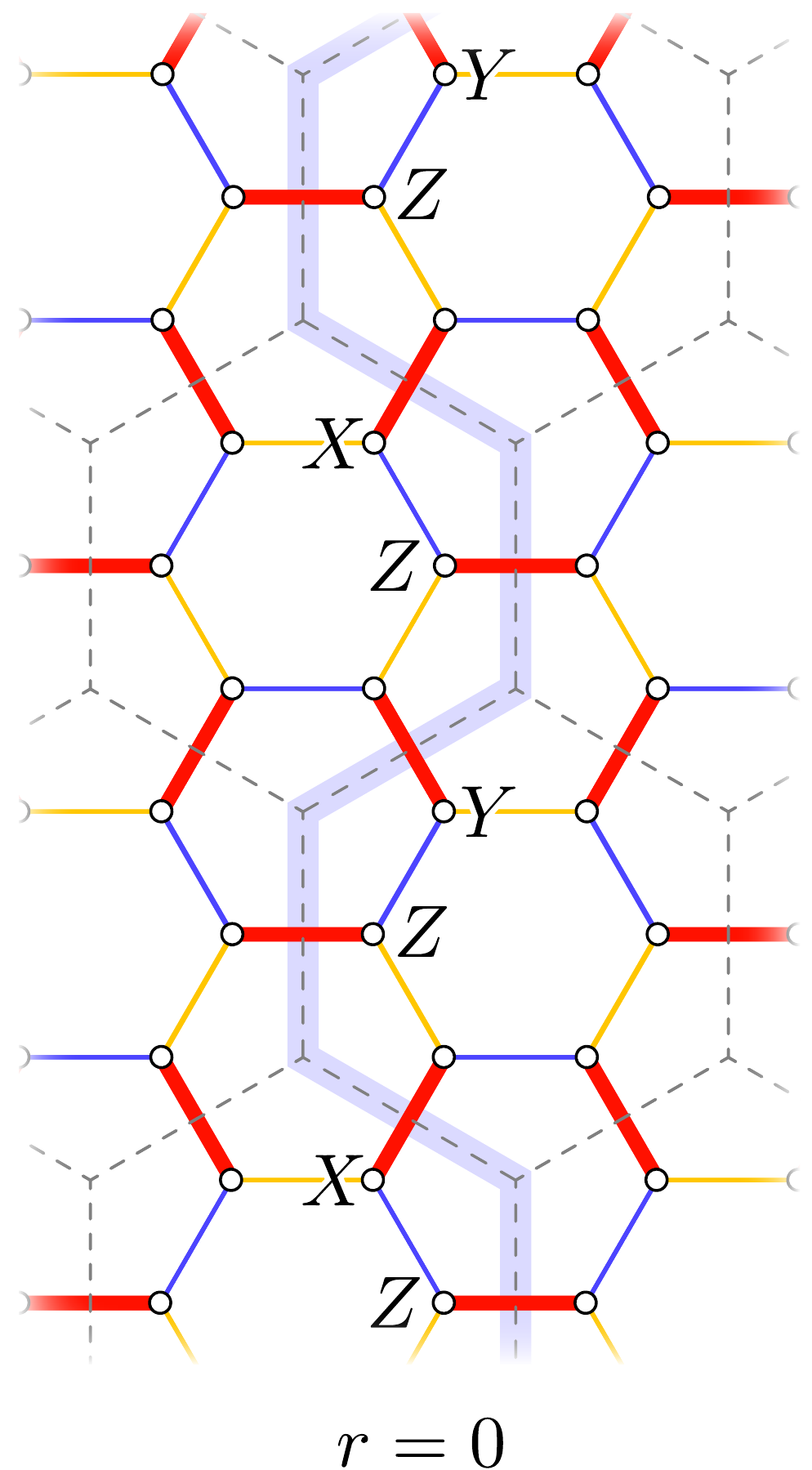}\,
\raisebox{.53cm}{\rule{.5pt}{6.1cm}}
\includegraphics[width=.23\textwidth]{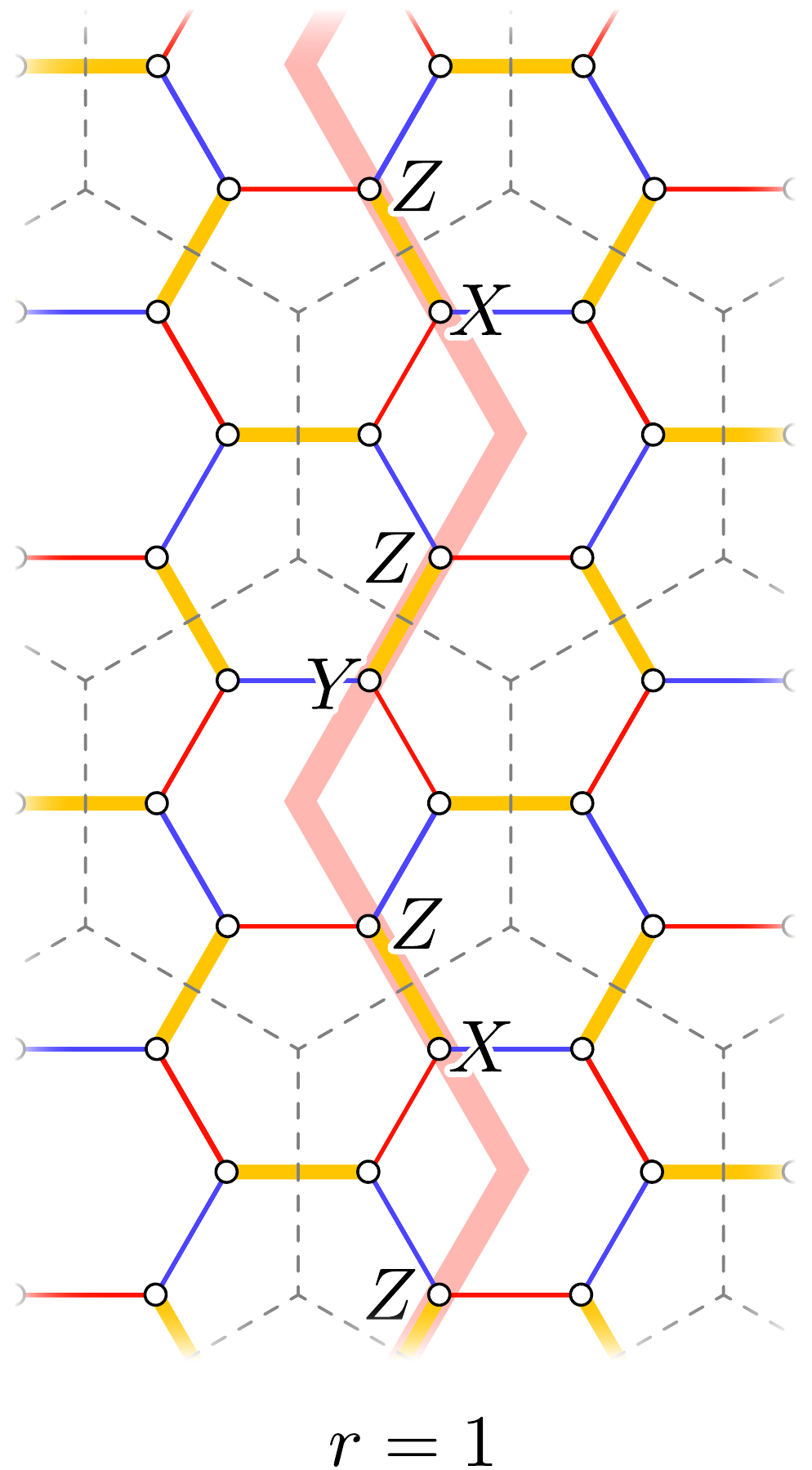}\,
\raisebox{.53cm}{\rule{.5pt}{6.1cm}}
\includegraphics[width=.23\textwidth]{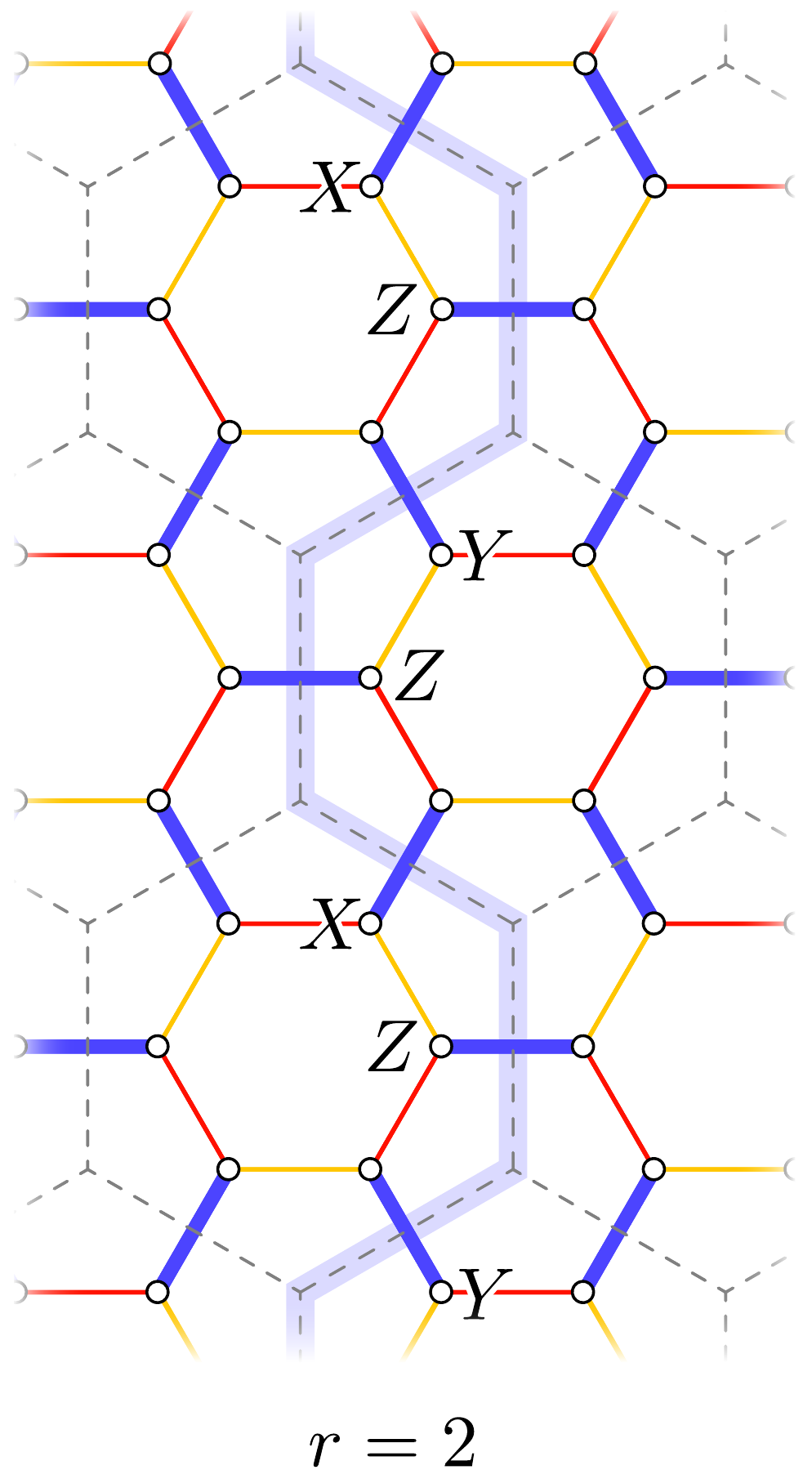}\,
\raisebox{.53cm}{\rule{.5pt}{6.1cm}}
\includegraphics[width=.23\textwidth]{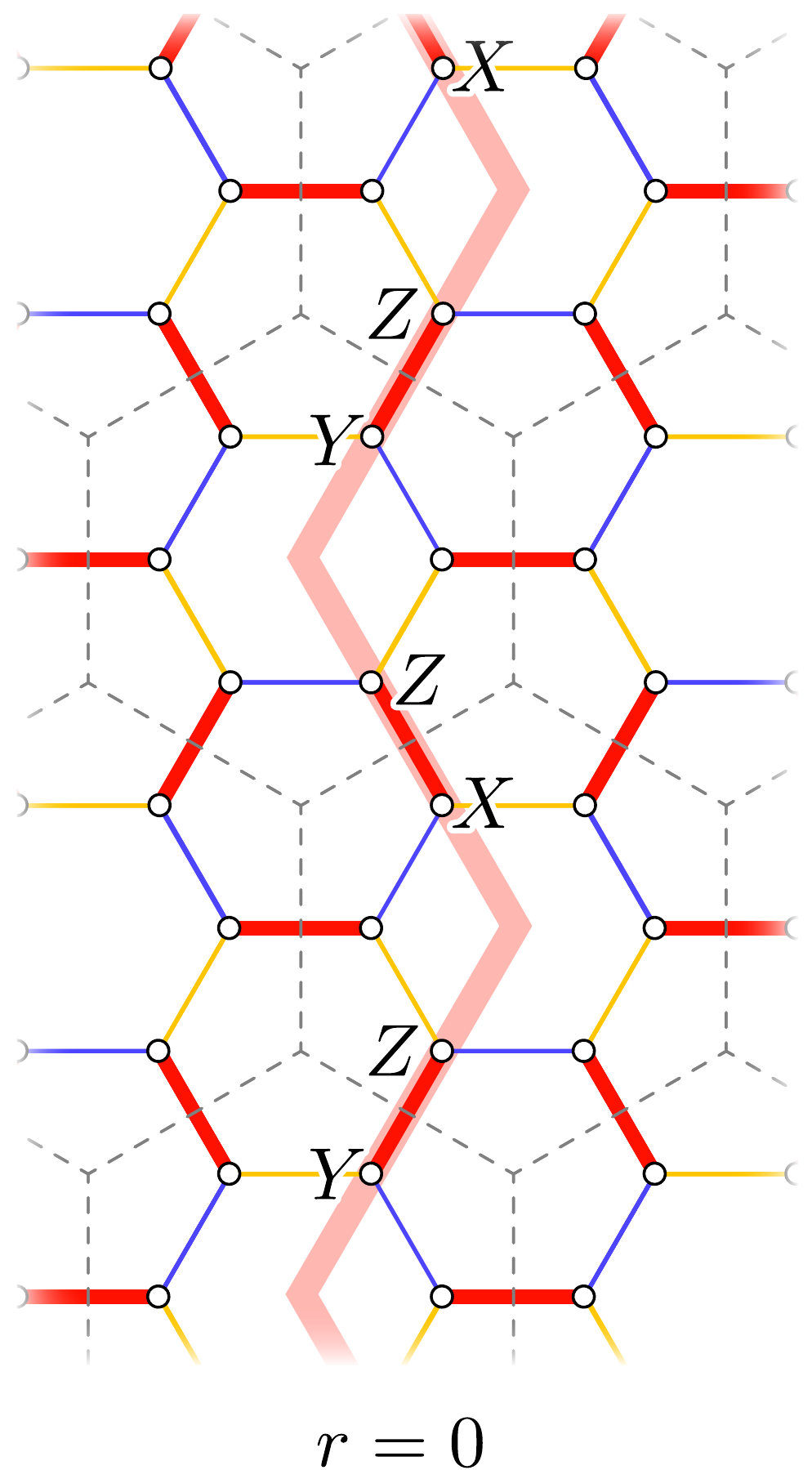}
\caption{Starting with an $e$ string operator (light blue) in the $r=0$ round, we multiply by $0$-checks (red) to find a representation of the logical operator that commutes with the $1$-checks (yellow). The logical operator now corresponds to an $m$ string operator (light red) in the $r=1$ round. In general, we multiply by $r$-checks to find a representation of the logical operator that commutes with the $(r+1)$-checks. The logical operator corresponding to an $e$ string operator transforms into an $m$ string operator after a single period.}
\label{fig: logical transformation}
\end{figure}

Quantum information is stored in the instantaneous code spaces defined by the ISGs. The logical operators after measuring the $r$-checks are simply the logical operators associated with the $r$-ISG. These can be determined from the logical operators of the toric code on the hexagonal super-lattice by inverting the mapping to effective qubits in Table~\ref{tab: effective qubits}. The logical operators of the toric code correspond to string operators that move $e$, $m$, and $\psi$ anyons around non-contractible paths. Consequently, the instantaneous logical operators can be represented by non-contractible anyon string operators. Explicit logical operators of the $\ZZ_2$ Floquet code can be seen in Fig.~\ref{fig: logical transformation}.

To verify that the quantum information is passed from one instantaneous code space to the next, we need to check that, after measuring the $r$-checks, each logical operator can be multiplied by elements of the $r$-ISG to find a representation that commutes with the $(r+1)$-checks. The evolution of a logical operator throughout the measurement schedule is shown in Fig.~\ref{fig: logical transformation}. After a full period, the logical operator that corresponds to an $e$ string operator is mapped to an $m$ string operator and vice-versa. In this sense, the dynamics implement a nontrivial automorphism of the anyons~\cite{HH_dynamic_2021}. We note that, in particular, the logical operator defined by the string operator of the $\psi$ anyon type is invariant under the dynamics, since it commutes with all of the check operators. This observation is key to our construction of twist defects, as described in the next section.

\section{$\ZZ_2$ Floquet code with twist defects} \label{sec: Z2 defects}

We now describe how the $\ZZ_2$ Floquet code can be modified to accommodate twist defects. Our construction is inspired by the notion of condensation defects, introduced in~Ref.~\cite{Roumpedakis2022Higher}. In particular, we create a pair of twist defects by condensing emergent fermions along an open path $\gamma$. 
To be more precise, here, we say that a fermion $\psi$ has been condensed along $\gamma$, if there exists fermion string operators $W^\psi_{ij}$ between the vertices $i,j \in \gamma$, such that the expectation value of $W^\psi_{ij}$ in a code state (or a ground state in more general contexts) goes to a constant $C$ for large separations between $i$ and $j$:
\begin{align} \label{eq: fermion condensation definition}
    \langle W_{ij}^\psi \rangle \to C \text{ in the limit of large }|i-j|.
\end{align}
This is to say that, for a sufficiently large length scale, there is a finite expectation value for creating (or annihilating) a pair of fermions along the line. In contrast, for a code state of the toric code (without any defect lines), this expectation value is exactly zero along any open path.\footnote{More generally, if we perturb from the fixed point toric code Hamiltonian, then we expect that the expectation value decays exponentially in the separation of $i$ and $j$.}
The expression in Eq.~\eqref{eq: fermion condensation definition} has the intuition of being a so-called Fredenhagen-Marcu order parameter along the path $\gamma$.\footnote{We thank Meng Cheng for the suggestion to define fermion condensation along a path in terms of a Fredenhagen-Marcu order parameter.} Such order parameters are typically used to probe the condensation of bosonic anyons over an entire 2D system~\cite{Fredenhagen1983charged,Fredenhagen1988dual,Gregor2011diagnosing,Verresen2021prediction}, but we have adapted it here to capture emergent fermions condensing along a line. 

At an intuitive level, the condensation of fermions along a path produces a 1D topological superconductor, i.e., a Majorana wire, built out of emergent fermions~\cite{nonabelian_twist_2015}. 
This can be understood from the fact that the string operators $W^\psi_{ij}$ in Eq.~\eqref{eq: fermion condensation definition} map to string order parameters of the Majorana wire under the generalized Jordan-Wigner transformation in Ref.~\cite{Yuan2018bosonization}.
As described below, the emergent 1D topological superconductors can then be used to store and process quantum information by manipulating the endpoints, which are nothing other than the twist defects. 

Operationally, the condensation of the emergent fermions is accomplished by modifying the check operators so that the emergent fermion string operators $W^\psi_{ij}$ along $\gamma$ belong to each of the ISGs. This implies that the condition for fermion condensation in Eq.~\eqref{eq: fermion condensation definition} is satisfied, since the string operators $W_{ij}^\psi$ would have expectation value $+1$ throughout the dynamics. To demonstrate this construction in a more familiar setting, we show in Appendix~\ref{app: TC condensation defects} that this approach can be used to construct twist defects in the usual toric code. 

Our construction of twist defects thus relies on the following special property of the $\ZZ_2$ Floquet code: \textit{for each of the ISGs, the exact same string operator can be used to create a pair of emergent fermions.}
More explicitly, for any path $\gamma$ on the hexagonal lattice, we define the string operator $W^\psi_\gamma$ to be a product of $K_{ij}$ operators [see Eq.~\eqref{eq: Z2 checks}] along $\gamma$:\footnote{Here, the string operator $W^\psi_\gamma$ is only defined up to a sign, due to the ambiguity in the ordering of the operators. However, any choice of ordering suffices for our purposes.}
\begin{align} \label{eq: fermion string operator def}
    W^\psi_\gamma \equiv \prod_{\langle ij \rangle \in \gamma} K_{ij}.
\end{align}
The string operator $W^\psi_\gamma$ commutes with all of the stabilizers of the ISGs along the path $\gamma$ and only fails to commute with stabilizers near the endpoints. By computing the exchange statistics using the methods of Refs.~\cite{Levin2003Fermions,Kawagoe2020microscopic,Ellison2022subsystem}, it can be checked that the violations created at the endpoints of $\gamma$ are indeed emergent fermions. 
We note that the contractible string operators are products of the plaquette stabilizers $S_p$. Hence, $S_p$ can be interpreted as a loop of emergent fermion string operator around the plaquette $p$.\footnote{The check operators have an anomalous $\ZZ_2$ $1$-form symmetry (see Ref.~\cite{Qi2021higherform} or Appendix~B of Ref.~\cite{Ellison2022subsystem}).} In other words, the closed paths of the fermion string operator belong to the stabilizer group of the underlying subsystem code. 

\subsection{What to measure and when} \label{sec: Z2 defect measurement schedule}

\begin{figure}[t]
\centering
\includegraphics[width=.23\textwidth]{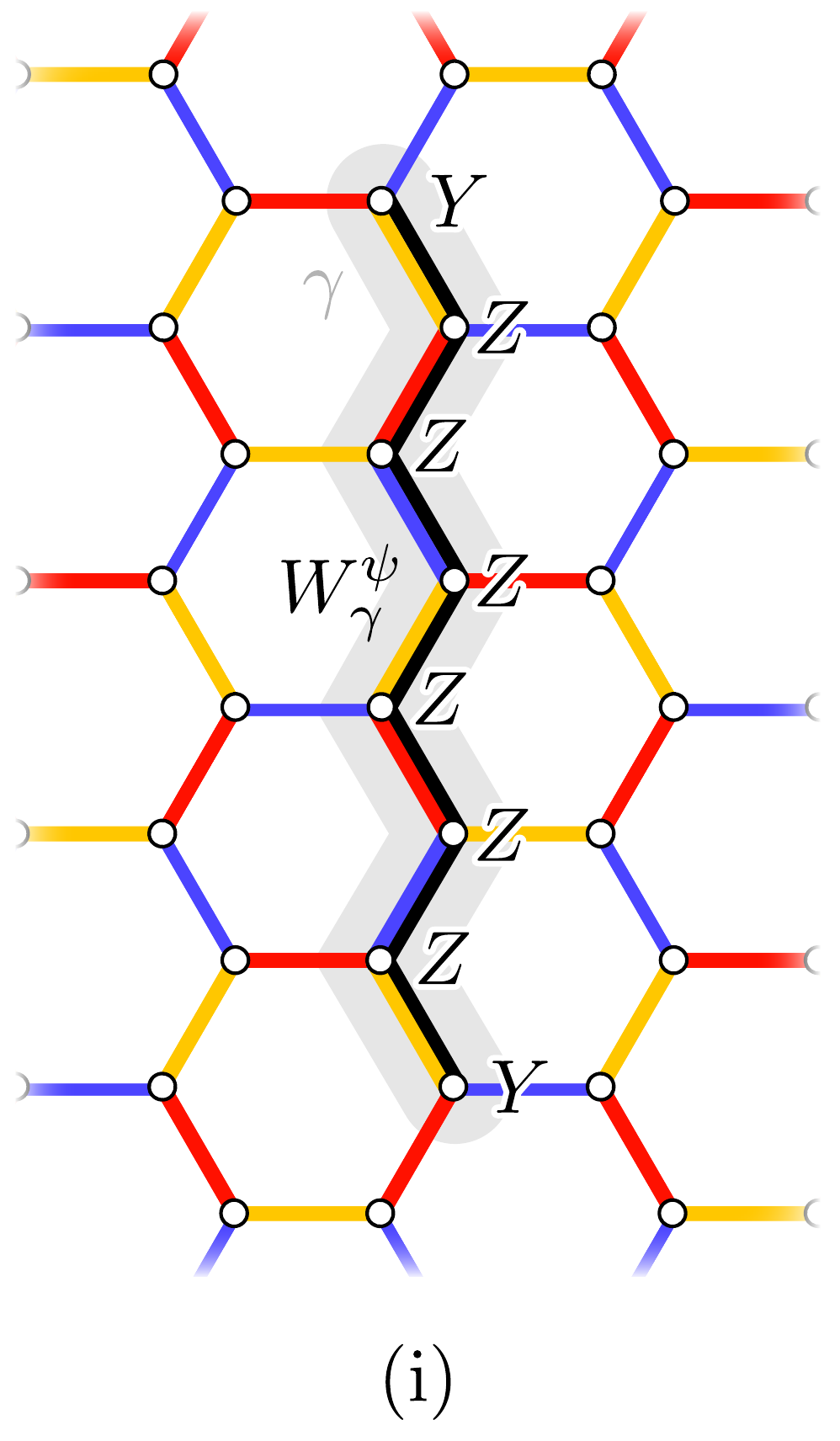}
\raisebox{.683cm}{\rule{.5pt}{5.545cm}}
\includegraphics[width=.23\textwidth]{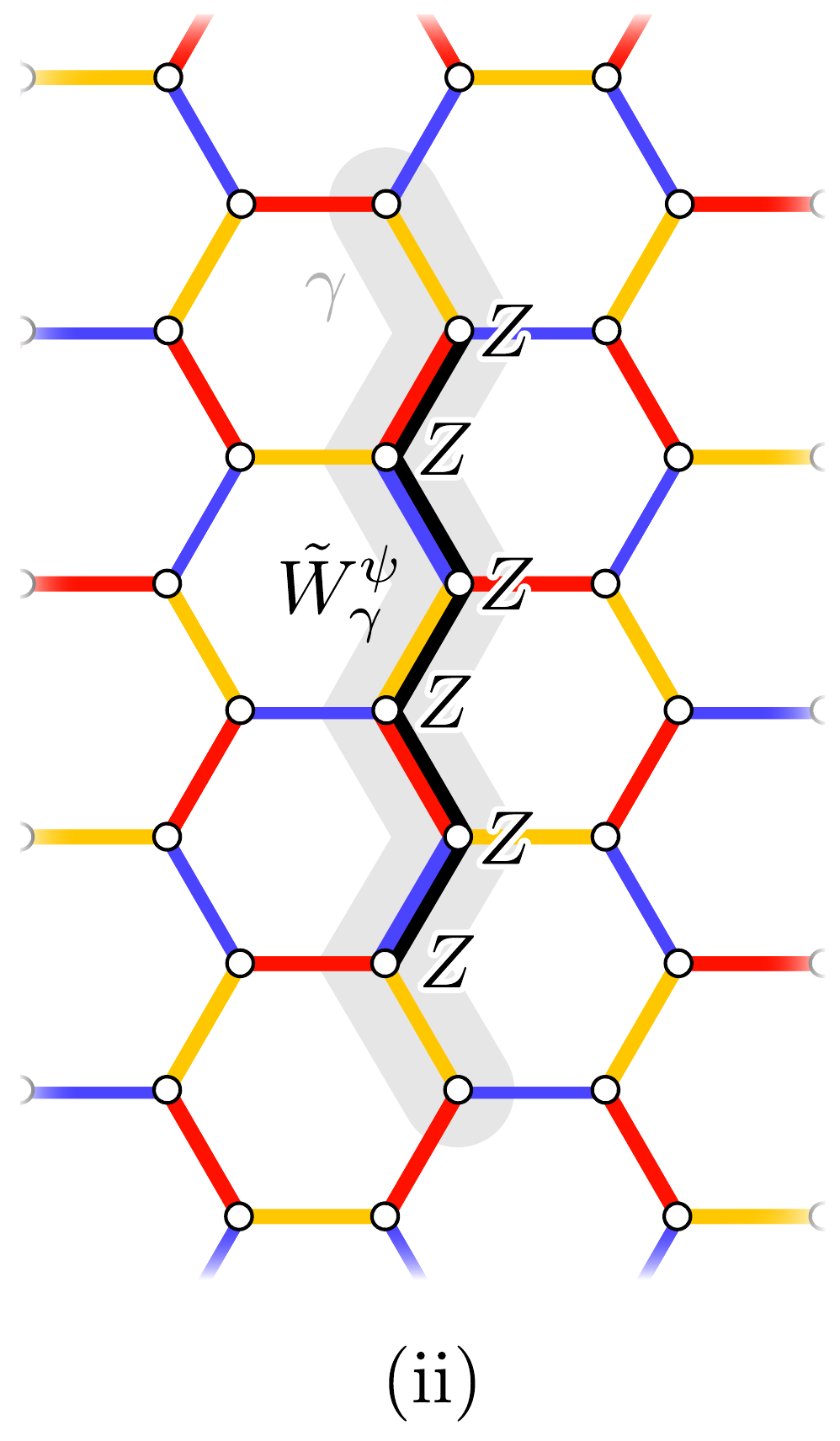}
\raisebox{.683cm}{\rule{.5pt}{5.545cm}}
\includegraphics[width=.23\textwidth]{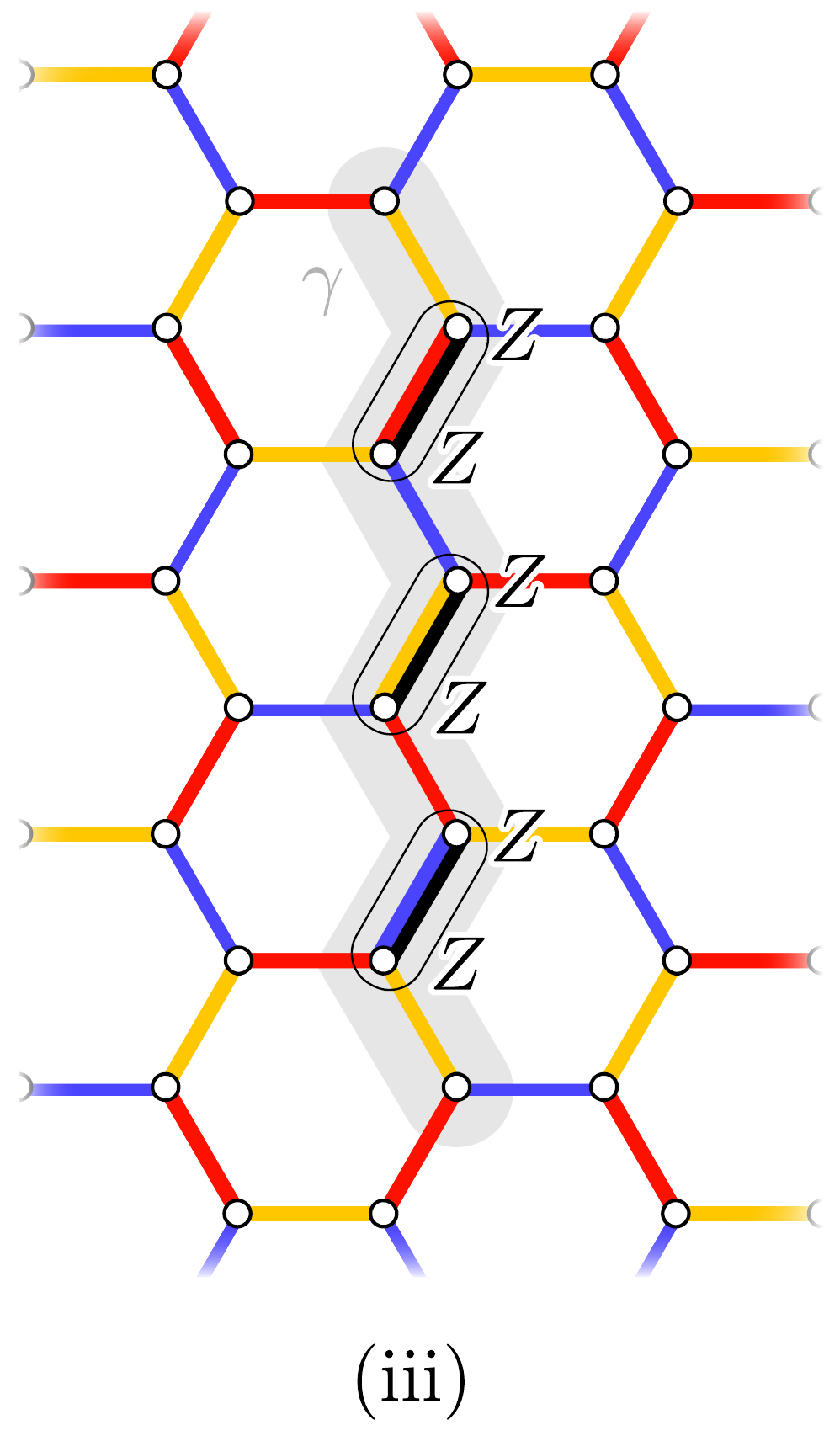}
\raisebox{.683cm}{\rule{.5pt}{5.545cm}}
\includegraphics[width=.23\textwidth]{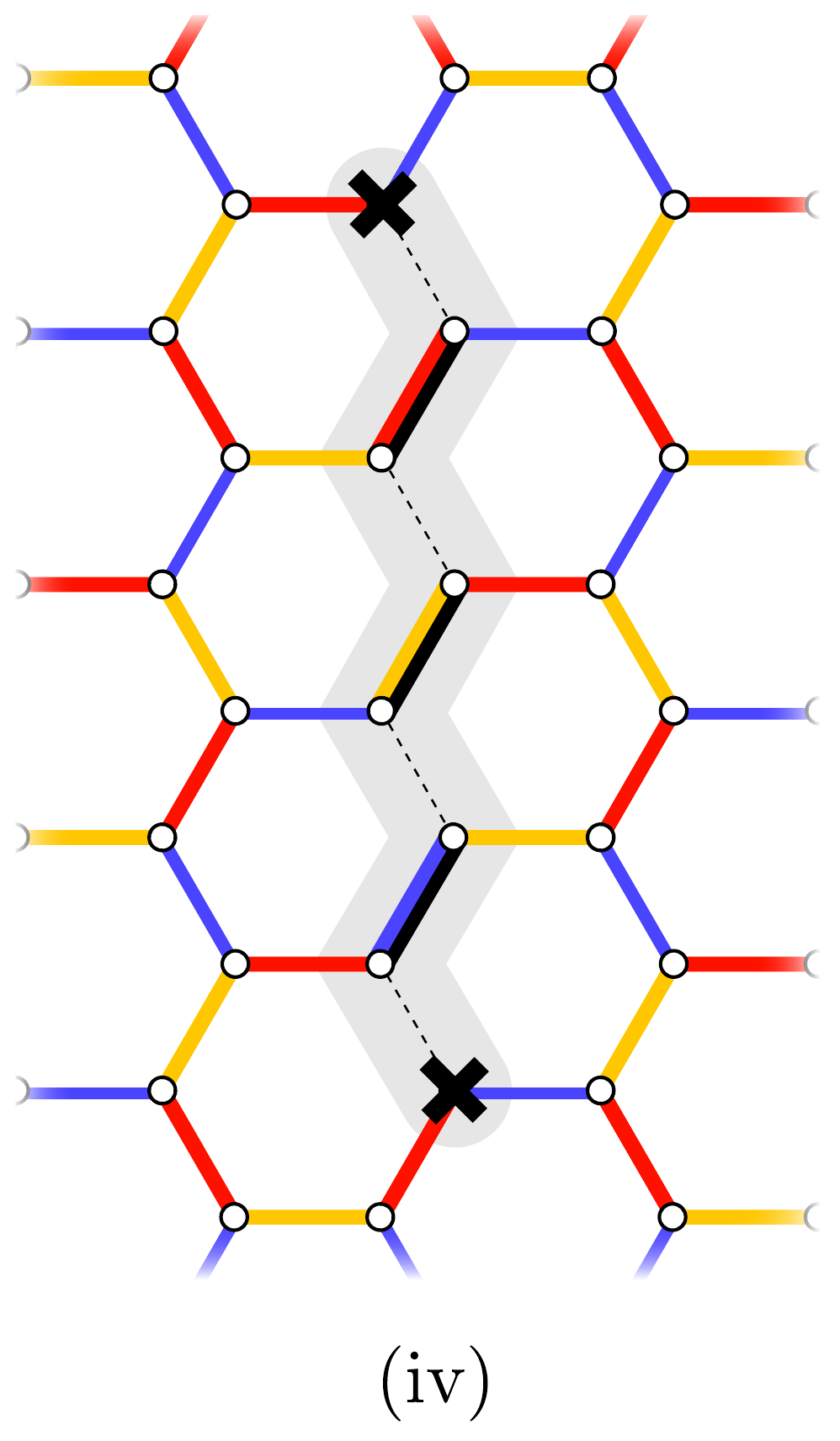}
\caption{The procedure for inserting a defect line along an open path $\gamma$ (light gray) involves four steps. (i) We define a fermion string operator $W^\psi_\gamma$ along $\gamma$. (ii) We remove the two Pauli operators at the endpoints to define the string operator $\tilde{W}^\psi_\gamma$. (iii) We divide $\tilde{W}^\psi_\gamma$ into 2-body short string operators. (iv) We define the 2-body short string operator as defect checks (bold black) and remove the checks that fail to commute with the defect checks (dashed lines). The twist defects (bold black crosses) are hosted at the endpoints of $\gamma$.}
\label{fig: defect construction}
\end{figure}

To insert twist defects at the endpoints of an open (connected, not self-intersecting) path $\gamma$, we modify the check operators of the $\ZZ_2$ Floquet code along $\gamma$. For our construction,
we assume that the path $\gamma$ has an odd length -- i.e., it is comprised of an odd number of edges.\footnote{For an even-length path $\gamma$, the na\"{i}ve generalization of our construction requires a weight three check operator.} We then use the following algorithm to modify the check operators along $\gamma$, as illustrated in Fig.~\ref{fig: defect construction}:
\begin{enumerate}[label={(\roman*)}]
\item We define the string operator $W^\psi_\gamma$ along $\gamma$, as in Eq.~\eqref{eq: fermion string operator def}.
\item We truncate $W^\psi_\gamma$ to a string operator $\tilde{W}^\psi_\gamma$ by removing a Pauli operator from each endpoint.\footnote{This step is only used to simplify the statement of step (iv).}
\item We decompose $\tilde{W}^\psi_\gamma$ into 2-body operators. This is always possible, given that $\gamma$~has odd length, and thus, $\tilde{W}^\psi_\gamma$ has an even weight. These 2-body operators define new check operators along $\gamma$. Intuitively, they are short string operators that create pairs of emergent fermions. 
\item We remove all of the check operators of the $\ZZ_2$ Floquet code that fail to commute with the check operators introduced in step (iii). 
\end{enumerate}
In summary, to construct a defect line along an open path $\gamma$ of length $2l+1$, we modify the checks along $\gamma$ by introducing $l$ new 2-body check operators, as defined in steps (i)-(iii), and remove $l+1$ of the original check operators, as described in step (iv). Away from the defect line, the check operators are the same as those of the $\ZZ_2$ Floquet code. We refer to the new check operators along $\gamma$ as the defect checks, and refer to the set of $r$-checks that no longer include the checks removed in step (iv) as the $r^*$-checks. 

\begin{figure*}[t]
\centering
\includegraphics[width=.23\textwidth]{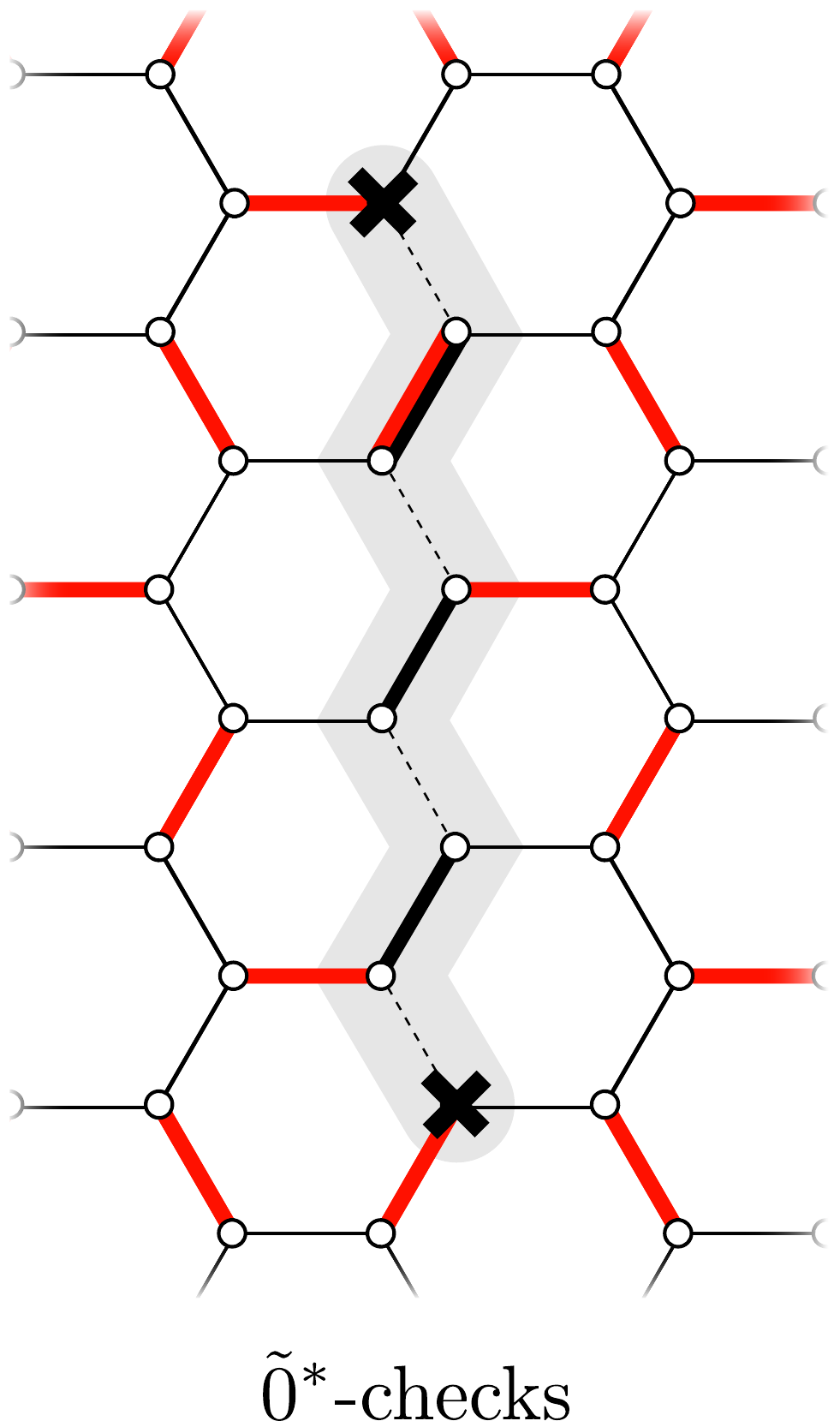}
\raisebox{.42cm}{\rule{.5pt}{5.8cm}}
\includegraphics[width=.23\textwidth]{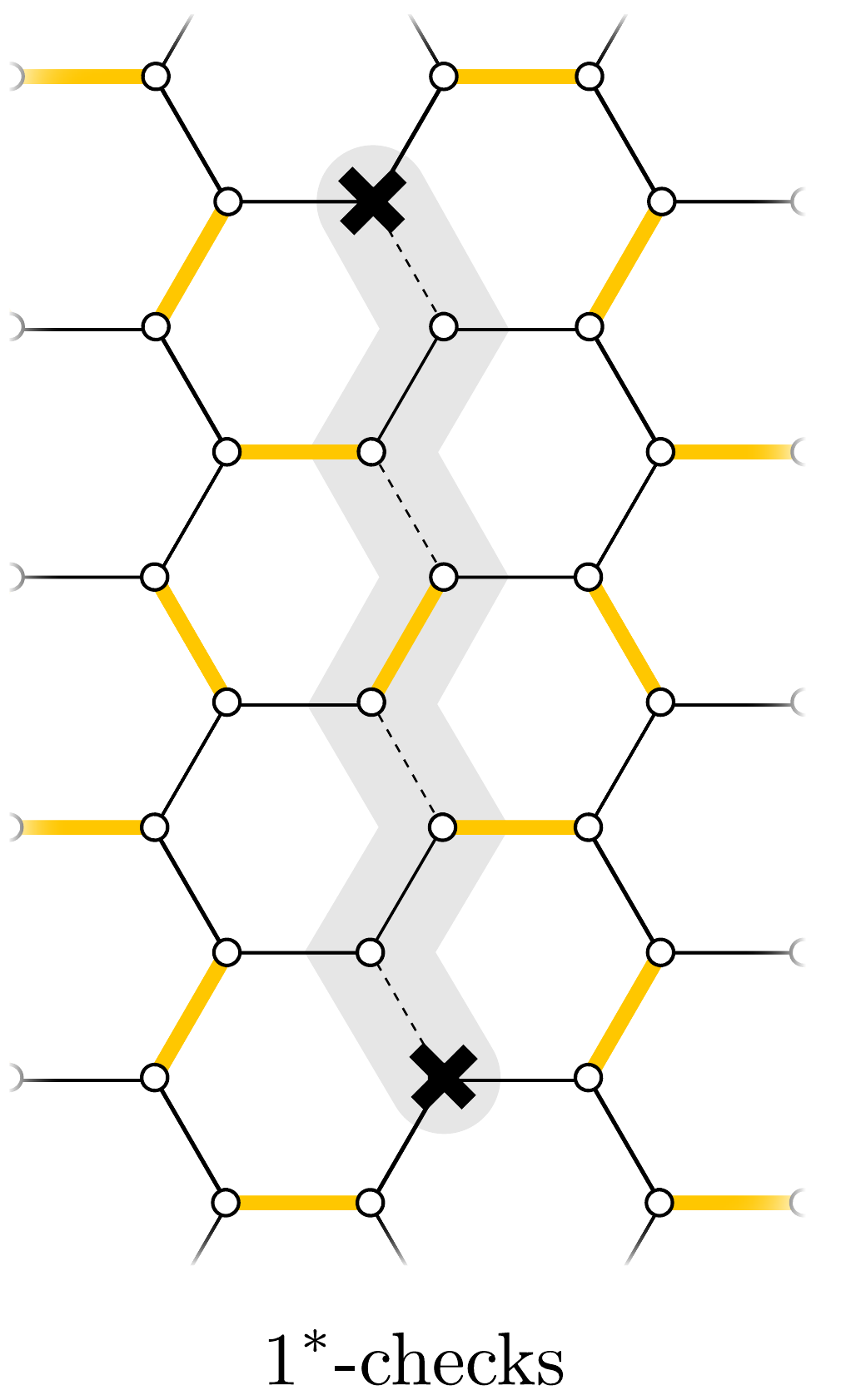}
\raisebox{.42cm}{\rule{.5pt}{5.8cm}}
\includegraphics[width=.23\textwidth]{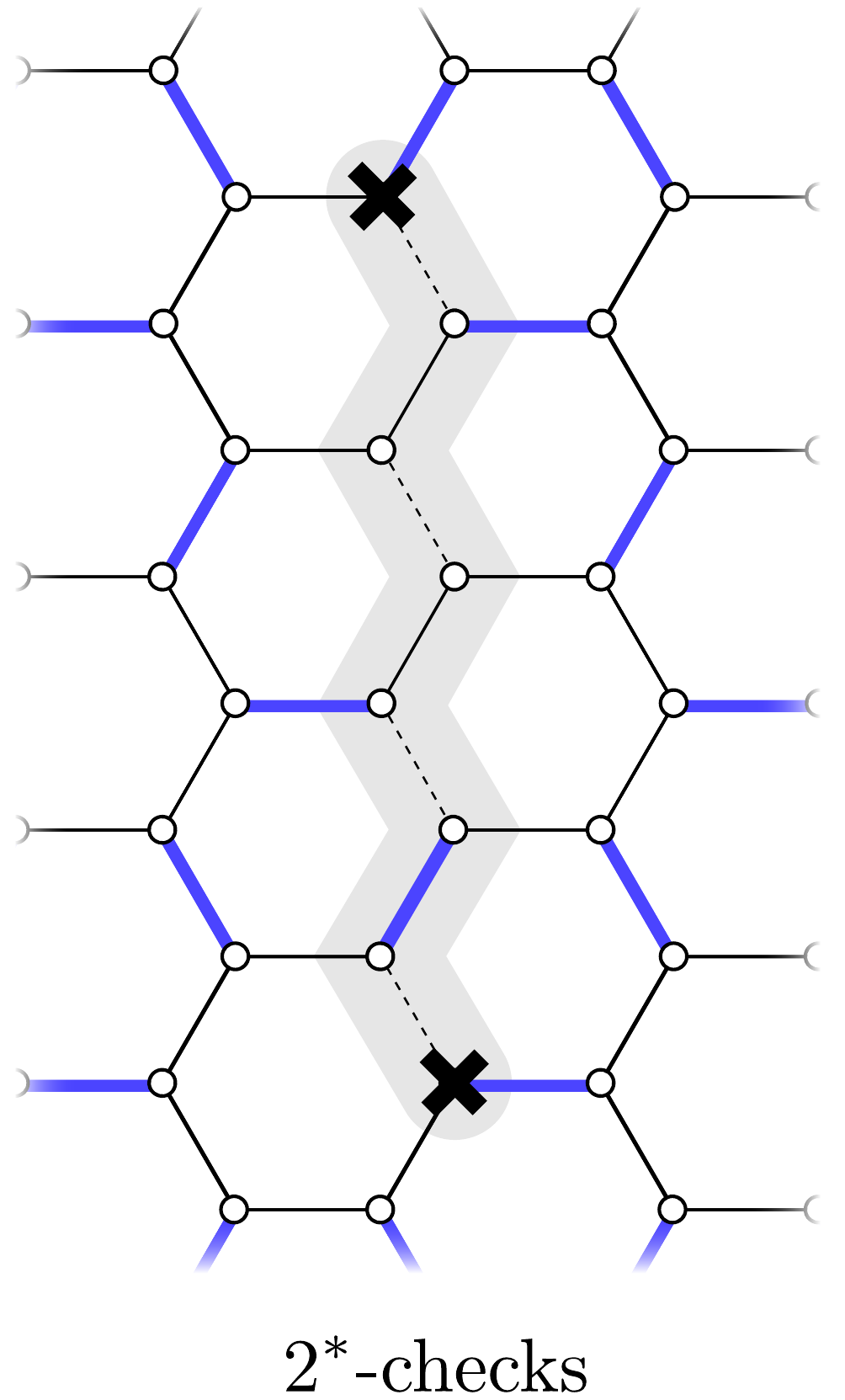}
\caption{The periodic part of the schedule for the $\ZZ_2$ Floquet code with twist defects consists of measuring $\tilde{0}^*$-checks (red and bold black), $1^*$-checks (yellow), and $2^*$-checks (blue). The $\tilde{0}^*$-checks are the $0^*$-checks (red) and the defect checks (bold black). Since the defect checks commute with all of the checks, they can be measured in any round.}
\label{fig: r*checks}
\end{figure*}

\begin{figure}[t]
\centering
\includegraphics[width=.6\textwidth]{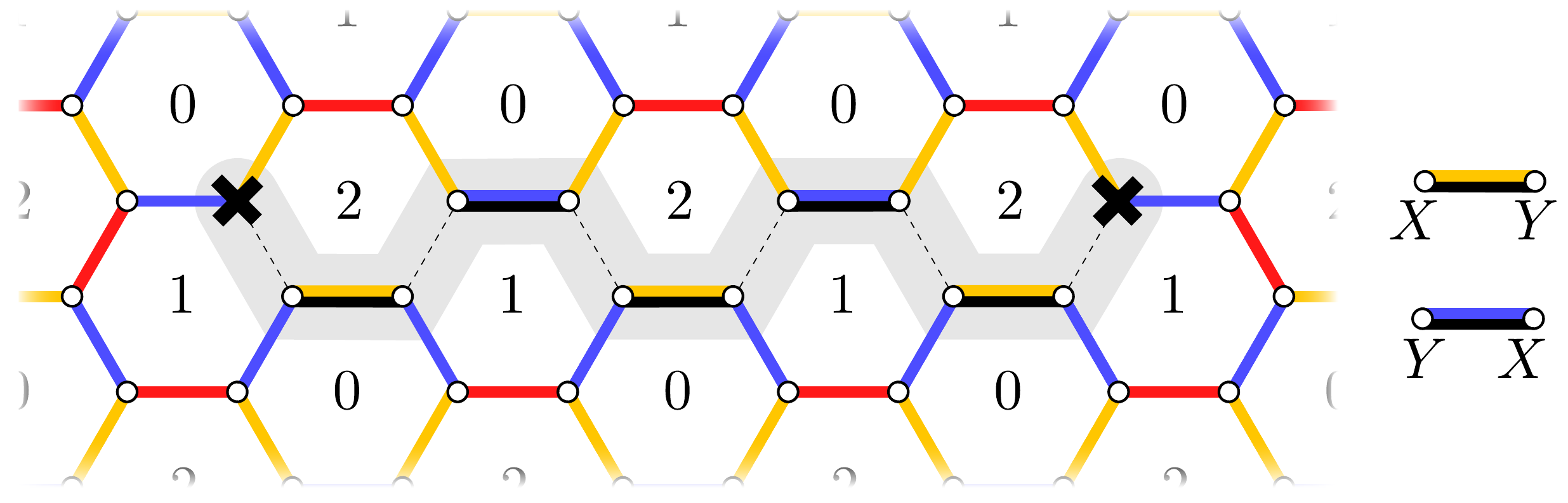}
\caption{In some cases, the defect checks (bold black lines) have non-overlapping support with the $0^*$-checks (red). For such a defect line, there are nonlocal stabilizers given by a product of all of the plaquette stabilizers sharing edges with removed checks (dashed lines). Here, the defect checks are $XY$ if they share an edge with a $1^*$-check (yellow) and $YX$ if they share an edge with a $2^*$-check (blue).}
\label{fig: 0edge defect}
\end{figure}

The measurement schedule of the $\ZZ_2$ Floquet code is then modified to include the defect checks.  
To initialize, however, we first initialize the $\ZZ_2$ Floquet code without twist defects and perform $d-1$ periods of the measurement schedule in Eq.~\eqref{eq: Z2 measurement schedule}, where $d$ is the code distance. These extra periods are necessary to fault-tolerantly initialize the twist defects in a definite logical state, as described further in Sections~\ref{sec: Z2 defects ISGs} and \ref{sec: Z2 defects logical operators}. The twist defects are then inserted by measuring the defect checks and measuring the $r^*$-checks in sequence (see Fig.~\ref{fig: r*checks}). Explicitly, the schedule is given by:
\begin{align} \label{eq: Z2 defect measurement schedule}
    [2,0,1,2]\underbrace{(0,1,2)\ldots(0,1,2)}_{d-1}(\tilde{0}^*,1^*,2^*)(\tilde{0}^*,1^*,2^*)\ldots
\end{align}
Here, $\tilde{0}^*$ represents simultaneously measuring the defect checks and the $0^*$-checks, while $1^*$ and $2^*$ represent measuring the $1^*$-checks and the $2^*$-checks, respectively. Note that, by construction, the defect checks commute with all of the other $r^*$-checks. Therefore, they can be measured during any round of the measurement schedule. We have included the defect checks with the $0^*$-checks for concreteness. We point out that, although the defect checks generally have overlapping supports with the $0^*$-checks, if the removed checks along the defect line are all $0$-checks, as shown in Fig.~\ref{fig: 0edge defect}, then the supports of the defect checks are disjoint from the supports of the $0^*$-checks.

\subsection{Instantaneous stabilizer groups} \label{sec: Z2 defects ISGs}

We now describe how the ISGs are affected by the changes made to the check operators along the path $\gamma$. We argue that each of the subsequent ISGs hosts a defect line along $\gamma$ and twist defects at the endpoints. We do so by showing that the $e$ and $m$ anyons of the instantaneous toric codes are permuted across $\gamma$ and that the emergent fermions can be condensed at its endpoints. To simplify the discussion, we focus on the example shown in Fig.~\ref{fig: defect construction}. Although we do not give a proof here, we expect that the discussion carries over to a construction of defect lines along more general paths. 

Prior to measuring the defect checks, the ISGs are the same as those in Eq.~\eqref{eq: r-ISG}. After measuring the defect checks along the path $\gamma$, the $r$-ISG becomes:
\begin{align} \label{eq: r*-ISG}
    \mathcal{S}_{r^*} \equiv \left \langle 
        \vcenter{\hbox{\includegraphics[scale=.5]{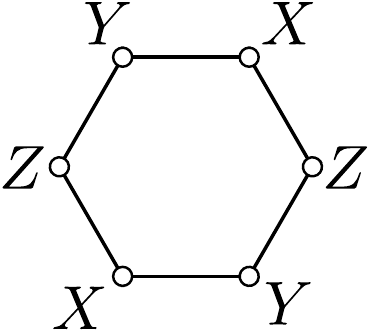}}}, \,\,\,
        \underbrace{\vphantom{ \left(\frac{{{{{{{{{{a^5}}^5}^5}^5}^5}^5}^5}^5}^{.5}}{5}\right) }\vcenter{\hbox{\includegraphics[scale=.6]{Figures/X_Kij.pdf}}}, \,\,\,
        \vcenter{\hbox{\includegraphics[scale=.6]{Figures/Y_Kij.pdf}}}, \,\,\,
        \vcenter{\hbox{\includegraphics[scale=.6]{Figures/Z_Kij.pdf}}}}_{\text{$r^*$-checks}}, \,\,\,
        \underbrace{\vcenter{\hbox{\includegraphics[scale=.6]{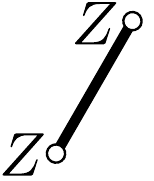}}}, \,\,\, 
        \vcenter{\hbox{\includegraphics[scale=.47]{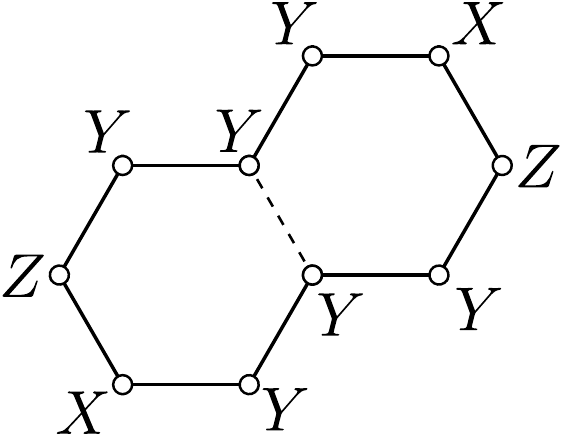}}}, \,\,\,
        \vcenter{\hbox{\includegraphics[scale=.5]{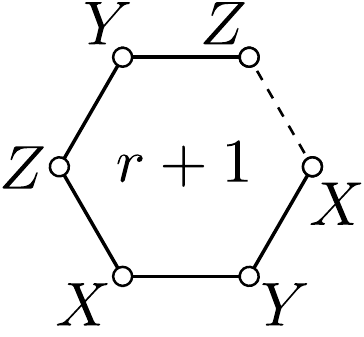}}}}_{\text{Along defect line}}
    \right \rangle
\end{align}
where $\mathcal{S}_{{r}^*}$ denotes the $r^*$-ISG resulting from the corresponding $\tilde{0}^*$, $1^*$, or $2^*$ round of measurements in Eq.~\eqref{eq: Z2 defect measurement schedule}. The first four generators of the $r^*$-ISG are the same as those in the absence of twist defects. These correspond to:
\begin{enumerate}
\item plaquette operators $S_p$ for every plaquette $p$ not sharing an edge with a removed check operator,
\item $r^*$-checks.
\end{enumerate}
The remaining generators are supported along $\gamma$ and can be summarized as: 
\begin{enumerate}
\setcounter{enumi}{2}
\item defect checks,
\item products of $S_p$ operators sharing edges with a removed check operator,
\item products of checks that partially form the operators in 4. 
\end{enumerate}
We note that the stabilizers of type {1}, {3}, and {4}~are common to all of the ISGs, after the period of initialization. For completeness, we argue that these stabilizers can be inferred from the measurement schedule, in Appendix~\ref{app: inferring stabilizers}.

\begin{figure*}[t]
\centering
\subfloat[\label{fig: em permute a}]{\includegraphics[width=.245\textwidth]{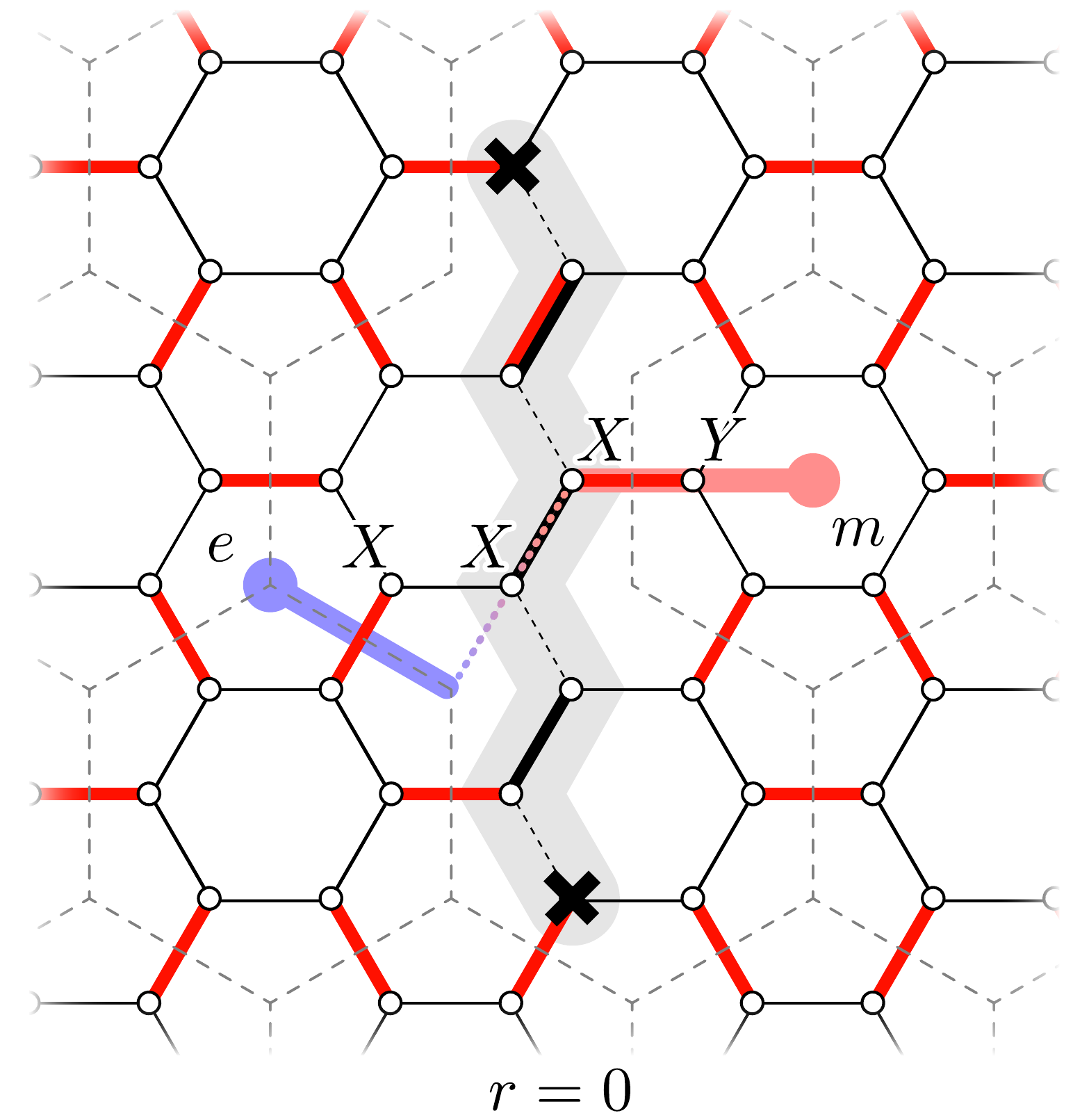}}\,
\raisebox{.25cm}{\rule{.5pt}{3.76cm}}
\subfloat[\label{fig: em permute b}]{\includegraphics[width=.245\textwidth]{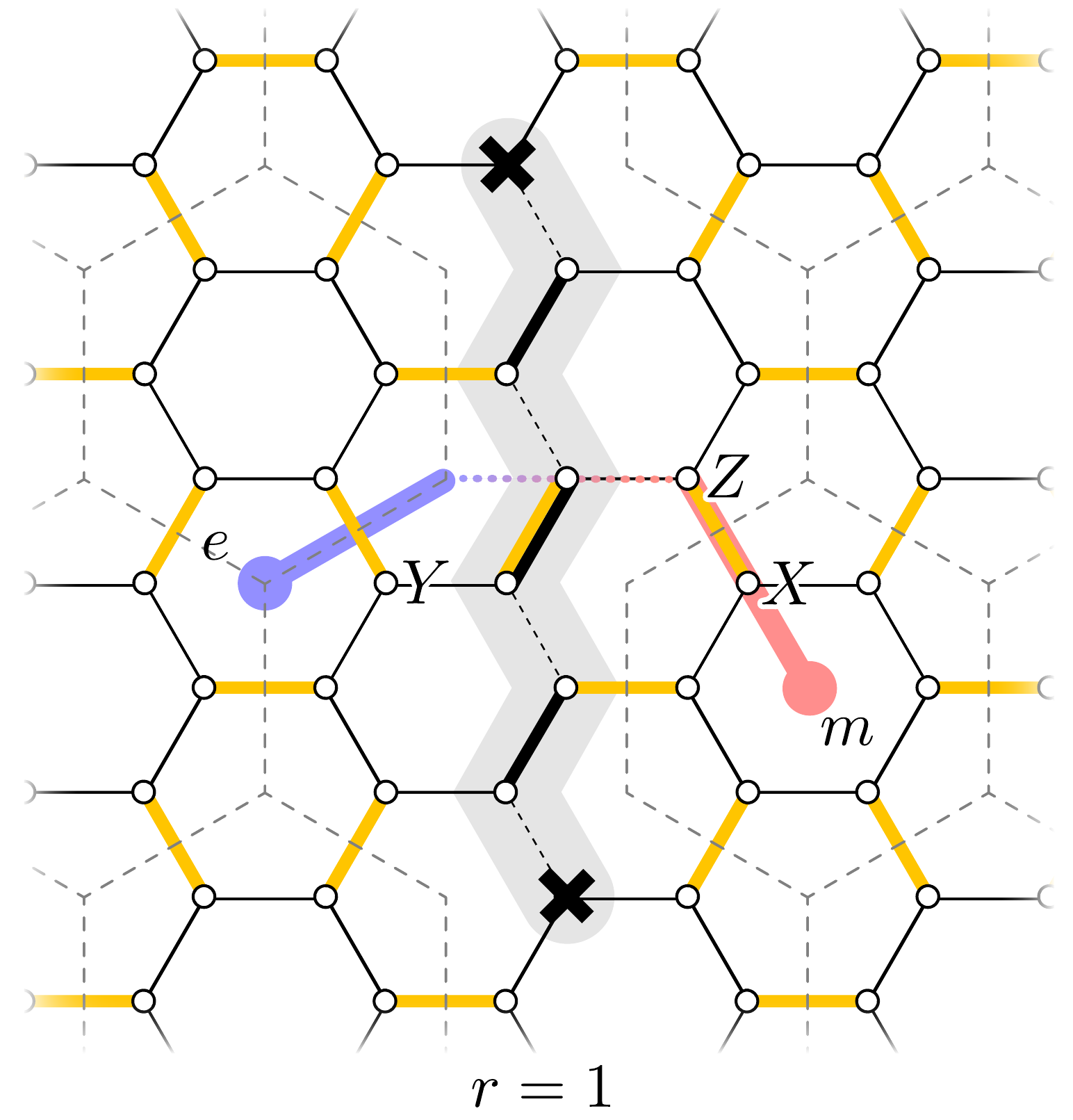}}\,
\raisebox{.25cm}{\rule{.5pt}{3.76cm}}
\subfloat[\label{fig: em permute c}]{\includegraphics[width=.245\textwidth]{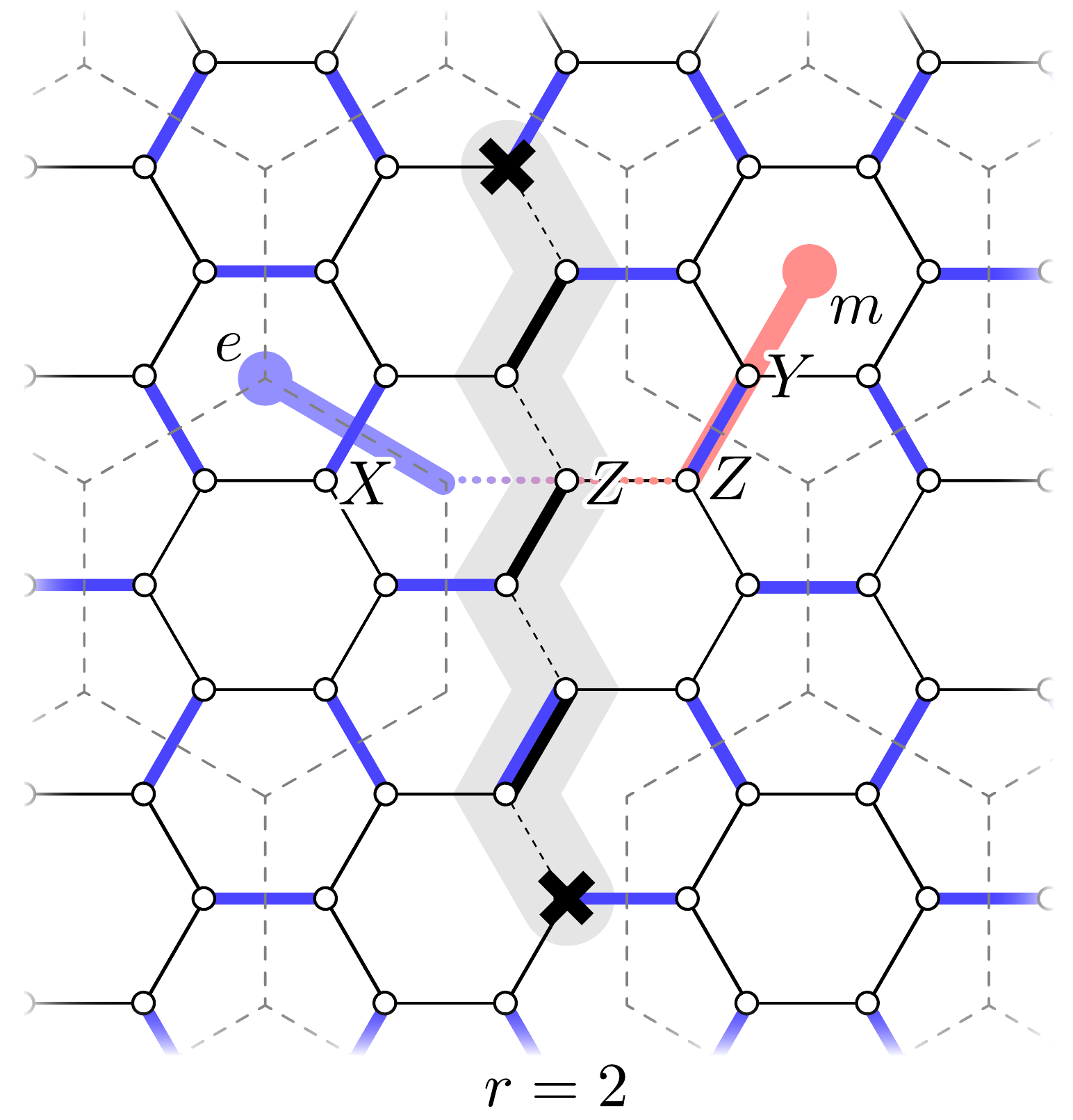}}\,\,
\subfloat[\label{fig: fermion on end}]{\raisebox{.18cm}{\includegraphics[width=.2\textwidth]{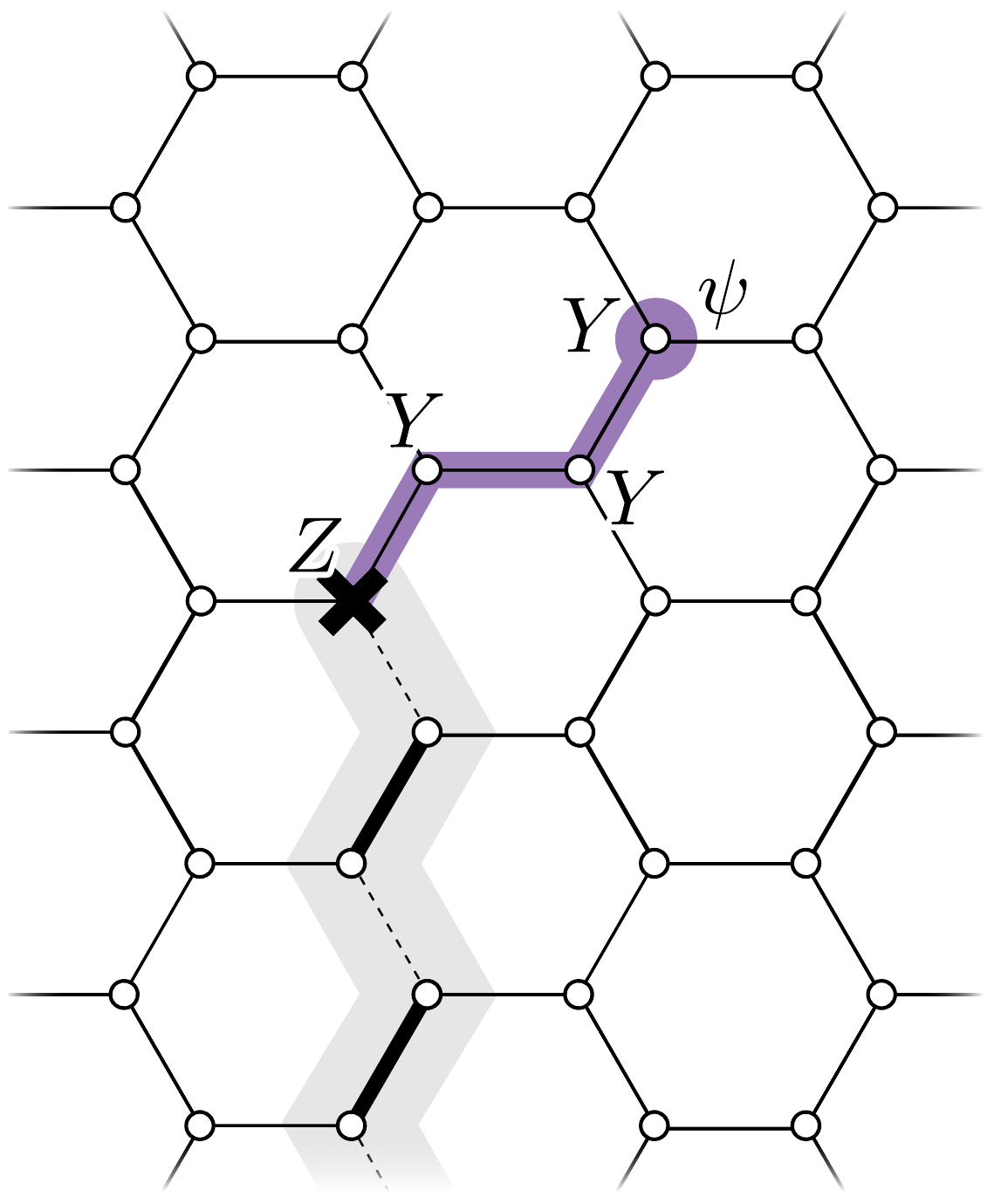}}}
\caption{The construction produces a defect line in each of the ISGs. This is evidenced by the fact that the $e$ and $m$ anyons are permuted across the defect line and the emergent fermion $\psi$ can be condensed at the twist defect. (a-c) We identify string operators (light blue and light red) that create an $e$ anyon to the left of the defect line and an $m$ anyon to the right of the defect line, respectively. There are no instantaneous stabilizers violated along the length of the string operator. (b) We identify a string operator (purple) that creates a single $\psi$ anyon and commutes with all of the checks in the vicinity of the twist defect. Note that the string operator differs from a product of $K_{ij}$ operators only by Pauli operators at the endpoints. These Pauli operators at the endpoints do not change the anyon type. }
\label{fig: defect line properties}
\end{figure*}

To see that the ISGs indeed host a defect line along $\gamma$, we consider the string operators shown in Figs.~\ref{fig: em permute a}-\ref{fig: em permute c}. The string operators intersect the path $\gamma$ and only fail to commute with instantaneous stabilizers at the endpoints. Using the mapping to an effective toric code on a super-lattice, as described in Section~\ref{sec: review ISGs} and Table~\ref{tab: effective qubits}, we see that the string operators create an $e$ anyon to the left of $\gamma$ and an $m$ anyon to the right. 
Similarly, by multiplying with an emergent fermion string operator, we can construct string operators that create an $m$ anyon on the left and an $e$ anyon on the right. Therefore, the $e$ and $m$ anyons are permuted across $\gamma$, as expected for a defect line in the toric code. Furthermore, the emergent fermions can be condensed at the endpoints of the defect line. This is illustrated in Fig.~\ref{fig: fermion on end}, where a fermion string operator is terminated at the endpoint of the defect line in such a way that it commutes with all of the check operators in the vicinity of the twist defect. This establishes that we have created a pair of twist defects at the endpoints of $\gamma$.

To store quantum information, we need to insert multiple defect lines. Following the prescription given above, we consider inserting $2k$ twist defects at the endpoints of the paths $\gamma_1, \ldots, \gamma_k$, pictured in Fig.~\ref{fig: multiple logicals}. The resulting ISGs, after the initialization in Eq.~\eqref{eq: Z2 defect measurement schedule}, are:
\vspace{-.75cm}
\begin{eqs} \label{eq: r*-ISG multiple defects}
        \mathcal{S}^{\text{ini}}_{{r}^*} \equiv \Bigg \langle
        \vcenter{\hbox{\includegraphics[scale=.5]{Figures/Sp_no_p.pdf}}}, \,\,
        \underbrace{\vphantom{ \left(\frac{{{{{{{{{{{{{{{{{a^5}^5}^5}^5}^5}^5}^5}^5}^5}^5}^5}^5}^5}^5}^5}^5}^5}{5}\right) }\vcenter{\hbox{\includegraphics[scale=.55]{Figures/X_Kij.pdf}}}, \,\,
        \vcenter{\hbox{\includegraphics[scale=.55]{Figures/Y_Kij.pdf}}}, \,\,
        \vcenter{\hbox{\includegraphics[scale=.55]{Figures/Z_Kij.pdf}}}}_{\text{$r^*$-checks}}, \,\,
        {\underbrace{\vcenter{\hbox{\includegraphics[scale=.55]{Figures/defect_check.pdf}}}, \,\, 
        \vcenter{\hbox{\includegraphics[scale=.45]{Figures/prodSp.pdf}}}, \,\,
        \vcenter{\hbox{\includegraphics[scale=.45]{Figures/horseshoe.pdf}}}, \,\,
        {\raisebox{-1.47cm}{\hbox{\includegraphics[scale=.4]{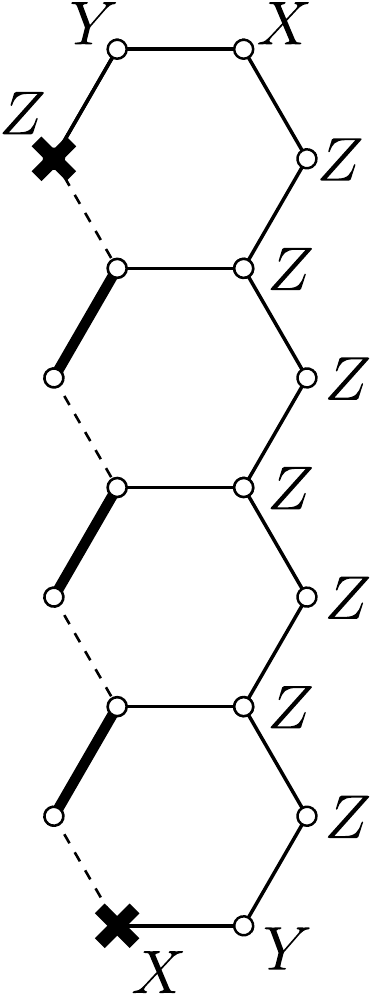}}}}\,\,}_{\text{For each defect line}}
        \Bigg\rangle.}
\end{eqs}
Here, the last generator of the ISG $\mathcal{S}^{\text{ini}}_{{r}^*}$ is a fermion string operator that connects between the endpoints of one of the paths $\gamma_1, \ldots, \gamma_k$. The ISG implicitly includes $k$ such generators. We refer to the string operator connecting the endpoints of $\gamma_i$ as $W^\psi_{\tilde{\gamma}_i}$, where the string operator is supported along the path $\tilde{\gamma}_i$. In general, the product of all $W^\psi_{\tilde{\gamma}_i}$ operators can be generated by $r$-plaquette stabilizers, $r^*$-checks, and the other generators near the defect lines. As a consequence, one of the $W^\psi_{\tilde{\gamma}_i}$ string operators can be removed from the set of generators since it is redundant. For this reason, the string operator $W^\psi_{\tilde{\gamma}_i}$ does not need to be included as a generator for a single defect line as in Eq.~\eqref{eq: r*-ISG}.

The measurement outcomes of all of the instantaneous stabilizers can be inferred from the measurement schedule in Eq.~\eqref{eq: Z2 defect measurement schedule}, except for the $W_{\tilde{\gamma}_i}$ operators.\footnote{Note that the product of the $W_{\tilde{\gamma}_i}$ operators can be inferred from the measurements.} 
As described in the next section, we interpret these operators as logical operators of the $\ZZ_2$ Floquet code with twist defects. The fact that the string operators $W^\psi_{\tilde{\gamma}_i}$ appear in the ISG $\mathcal{S}^{\text{ini}}_{{r}^*}$ is a reflection of the fact that we have initialized the Floquet code in a particular logical state. Ignoring the logical operators from initialization, the $r^*$-ISG for multiple defect lines is:
\begin{eqs} \label{eq: r*-ISG multiple defects 2}
        \mathcal{S}_{{r}^*} \equiv \Bigg \langle
        \vcenter{\hbox{\includegraphics[scale=.5]{Figures/Sp_no_p.pdf}}}, \,\,
        \underbrace{\vphantom{ \left(\frac{{{{{{{{{{a^4}}^4}^4}^4}^4}^4}^4}^4}^{.5}}{4}\right) }\vcenter{\hbox{\includegraphics[scale=.55]{Figures/X_Kij.pdf}}}, \,\,
        \vcenter{\hbox{\includegraphics[scale=.55]{Figures/Y_Kij.pdf}}}, \,\,
        \vcenter{\hbox{\includegraphics[scale=.55]{Figures/Z_Kij.pdf}}}}_{\text{$r^*$-checks}}, \,\,
        {\underbrace{\vcenter{\hbox{\includegraphics[scale=.55]{Figures/defect_check.pdf}}}, \,\, 
        \vcenter{\hbox{\includegraphics[scale=.45]{Figures/prodSp.pdf}}}, \,\,
        \vcenter{\hbox{\includegraphics[scale=.45]{Figures/horseshoe.pdf}}}, \,\,}_{\text{For each defect line}}
        \Bigg\rangle.}
\end{eqs}

\subsection{Logical operators} \label{sec: Z2 defects logical operators}

We next discuss how quantum information can be encoded using the twist defects of the $\ZZ_2$ Floquet code. To get started, we give an intuitive description of the logical Pauli operators based on Ref.~\cite{Landahl2023Majorana}. In particular, when viewed as 1D topological superconductors constructed from emergent fermions, each defect line encodes the Hilbert space of a single physical fermion, split between the two ``emergent Majorana modes'' at the endpoints. This implies that $2k$ twist defects encode $k$ physical fermions -- giving us an instantaneous code space of dimension $2^k$. However, the total number of emergent fermions must have an even parity. Therefore, the instantaneous code space has dimension $2^{k-1}$, implying that $2k$ twist defects encode $k-1$ logical qubits. The logical Pauli operators can then be generated by operators that measure the fermion parity of a defect line or transfer fermion parity between defect lines. These are precisely the fermion string operators that connect the twist defects. 

To see that $k$ defect lines encode $k-1$ logical qubits more concretely, we consider the number of constraints imposed by the $r^*$-ISG. After measuring the defect checks, we add them as generators of the ISG and remove one non-commuting generator for each defect check. As a consequence, the total number of constraints imposed by the instantaneous stabilizers does not change. Hence, the ISG $\mathcal{S}^{\text{ini}}_{{r}^*}$ in Eq.~\eqref{eq: r*-ISG multiple defects}, immediately following initialization, encodes only two logical qubits -- corresponding to the two logical qubits in the absence of twist defects. 

\begin{figure*}[t]
\centering
\subfloat[\label{fig: logical Zs}]{\includegraphics[width=.4\textwidth]{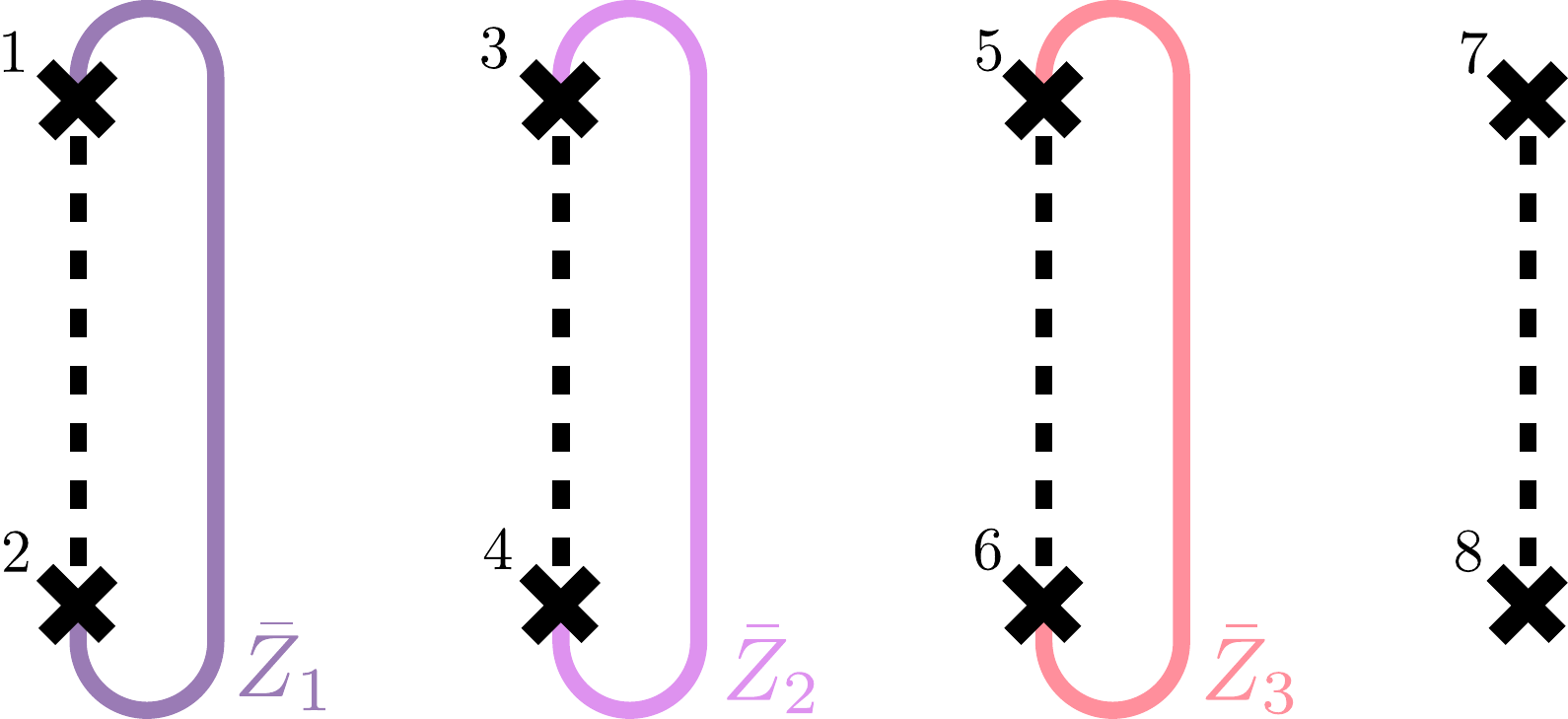}} 
\qquad \,\, \rule{.5pt}{2.9cm} \qquad
\subfloat[\label{fig: logical Xs}]{\raisebox{.21cm}{\includegraphics[width=.4\textwidth]{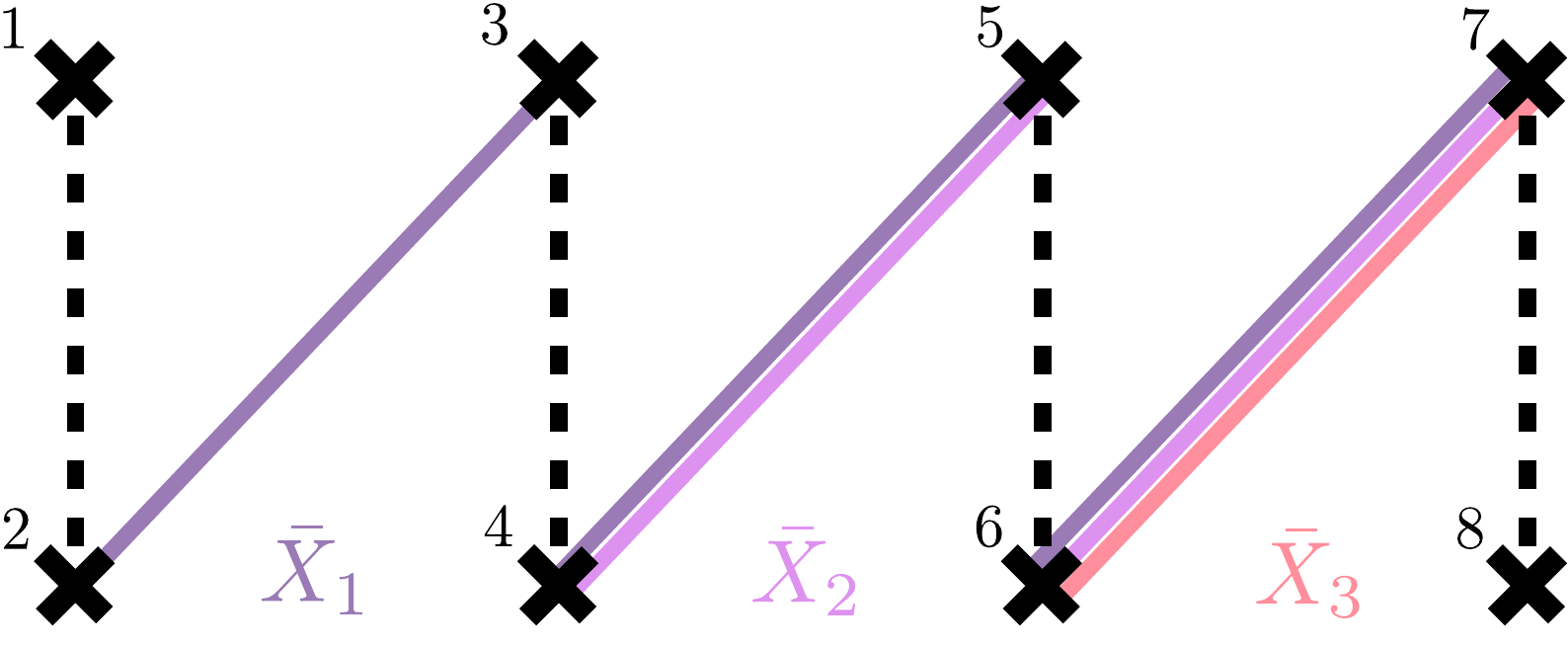}}}
\caption{A set of $k=4$ defect lines (dashed lines) encodes three logical qubits. (a) The logical $Z$ operators denoted $\bar{Z}_1$, $\bar{Z}_2$, and $\bar{Z}_3$, can be represented by fermion string operators (purple and pink) that connect the endpoints of the defect lines. (b) The logical $X$ operators, denoted $\bar{X}_1$, $\bar{X}_2$, and $\bar{X}_3$, can also be represented by fermion string operators. Note that, if the eight twist defects are identified with eight Majorana operators $\eta_1, \ldots, \eta_8$, then the encoding can be read off from the Jordan-Wigner transformation, e.g., $\bar{Z}_i \to \eta_i\eta_{i+1}$, and $\bar{X}_i \to \prod_{j=2i}^7 \eta_j$.}
\label{fig: multiple logicals}
\end{figure*}

Nevertheless, we can indeed encode quantum information using the twist defects, since not all of the stabilizer generators of the ISG $\mathcal{S}^{\text{ini}}_{{r}^*}$ are inferred from the subsequent rounds of measurements. More specifically, the fermion string operators $W^\psi_{\tilde{\gamma}_i}$ cannot be inferred from the measurement schedule. This follows from the fact that for each string operator $W^\psi_{\tilde{\gamma}_i}$, we can find a fermion string operator $W^\psi_{\gamma'}$ between  two different twist defects such that $W^\psi_{\gamma'}$ commutes with all of the checks but fails to commute with $W^\psi_{\tilde{\gamma}_i}$, as shown in Fig.~\ref{fig: multiple logicals}. This implies that the measurement outcome of $W^\psi_{\gamma'}$ is unchanged by the measurements of the check operators. This is only possible if the measurement outcome of $W^\psi_{\tilde{\gamma}_i}$ cannot be inferred from the measurement schedule.

Consequently, we can interpret the $W^\psi_{\tilde{\gamma}_i}$ string operators as logical operators. However, given that the product of all of the $W^\psi_{\tilde{\gamma}_i}$ is equivalent to a product of the remaining $r^*$-ISG stabilizer generators, there are only $k-1$ independent $W^\psi_{\tilde{\gamma}_i}$ string operators. Therefore, the $k$ twist defects give $k-1$ logical qubits (besides the two logical qubits from the periodic boundary conditions). In Fig.~\ref{fig: multiple logicals}, we represent the logical Pauli $Z$ operators by the first $k-1$ $W^\psi_{\tilde{\gamma}_i}$ string operators. The ISG $\mathcal{S}^{\text{ini}}_{{r}^*}$ then tells us that the Floquet code is initialized in a logical $Z$ eigenstate, for all of the logical qubits.  

Since the fermion string operators commute with all of the checks, the logical information is trivially passed between the instantaneous code spaces, and thus, the information is preserved throughout the dynamics. Despite this, we can consider other representations of the logical operators, such as a loop of $e$ or $m$ string operator that encircles a defect line. Given that these representations generically fail to commute with the subsequent round of measurements, they need to be multiplied by instantaneous stabilizers to find a representation that commutes with the upcoming measurements. Following the same logic as in the absence of twist defects, one possibility is that a loop of $e$ ($m$) string operator has mapped to a loop of $m$ ($e$) string operator after a single period. However, this mapping is ambiguous in the presence of twist defects, since the loop of $e$ or $m$ operator encircling the defect line could alternatively be mapped (by multiplying with instantaneous stabilizers) to a fermion string operator that connects the endpoints of the defect line. As for the logical operators that wrap around a non-contractible path, the dynamics still implements a nontrivial automorphism, as in the case without twist defects. 

\subsection{Construction of boundaries} \label{sec: Z2 defects boundaries}

Thus far, we have considered systems with periodic boundary conditions. These boundary conditions present challenges for quantum computing platforms such as superconducting processors, where the qubits are most naturally arranged in a planar geometry and interact with their nearest neighbors. It is therefore desirable to construct Floquet codes that can be embedded in a plane in such a way that both the checks and the noise model are local. This necessitates studying Floquet codes in systems with boundaries.

As recognized in Ref.~\cite{HH_dynamic_2021}, the construction of a $\ZZ_2$ Floquet code on a planar geometry is subtle. Before getting into the subtlety, we note that, since the ISGs of the $\ZZ_2$ Floquet code have the same topological order as the toric code, there are two types of boundaries: rough boundaries ($e$ condensing) and smooth boundaries ($m$ condensing). Similar to the surface code, spatially alternating rough and smooth boundaries can, in principle, be used to encode a single logical qubit. However, the nontrivial automorphism exhibited by the $\ZZ_2$ Floquet code is at odds with the fact that the boundary conditions do not change after a single period. An $e$ string operator connecting two rough boundaries should transform into an $m$ string operator after a single period, but $m$ anyons can only condense at smooth boundaries.  

One solution, presented in Refs.~\cite{Haah2022boundarieshoneycomb, Gidney2022benchmarkingplanar, Paetznick2023Performance}, is to extend the measurement schedule to six rounds of measurements per period. This ensures that there is no nontrivial automorphism after a full period but comes at the cost of slowing down the extraction of the stabilizer syndrome near the boundary.~Another solution, described in Ref.~\cite{vuillot2021planar}, is to allow the logical operators to rotate under the dynamics. In this way, a logical $e$ string operator transforms into a rotated logical $m$ string operator connecting two smooth boundaries. However, due to the nonlocal evolution of the logical operators, this construction of a planar $\ZZ_2$ Floquet code is, unfortunately, not fault-tolerant. 

The results of Ref.~\cite{Aasen2023measurement} suggest that a compromise is inevitable. In particular, it was proven in Ref.~\cite{Aasen2023measurement} that, for any planar construction of the $\ZZ_2$ Floquet code satisfying a technical condition called ``local reversibility'', there is either period doubling or constant-weight logical operators that propagate along the boundary. 
We refer to Ref.~\cite{Aasen2023measurement} for the definition of local reversibility and simply note here that local reversibility guarantees that the logical operators have local update rules -- i.e., for any logical operator, we can multiply by stabilizers near its support to construct a logical operator for the subsequent ISG. 
The boundary construction of Ref.~\cite{vuillot2021planar} breaks the local reversibility condition, leading to nonlocal transformations of the logical operators after a period and resulting in constant-weight undetectable errors.

\begin{figure*}[t]
\centering
\subfloat[\label{fig: 0 boundaries}]{\includegraphics[width=.53\textwidth]{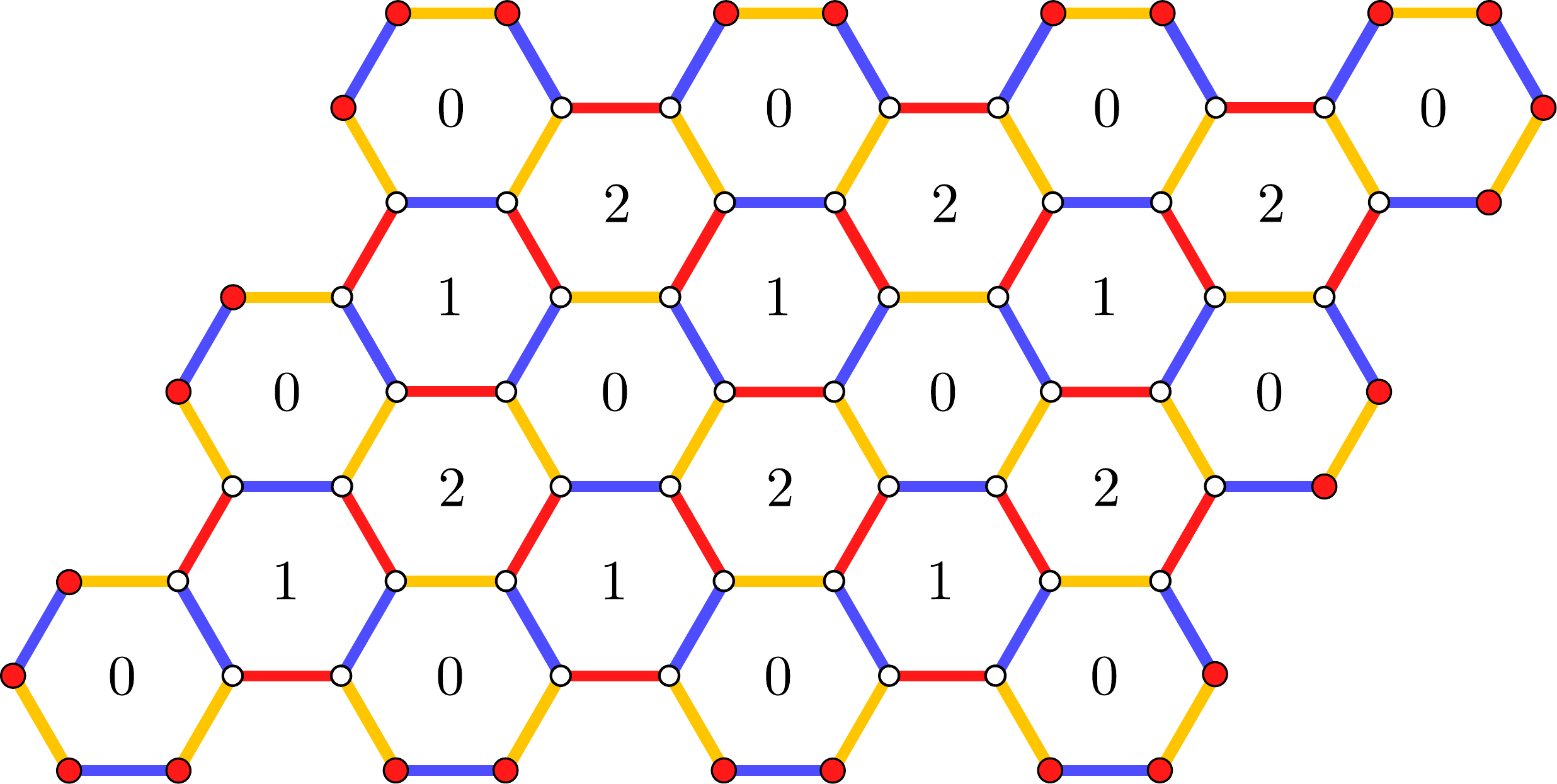}} \!\!\!\!\!\!\!\!\!\!\!\!\!\!\!\!
\subfloat[\label{fig: 0 boundaries defects}]{\includegraphics[width=.53\textwidth]{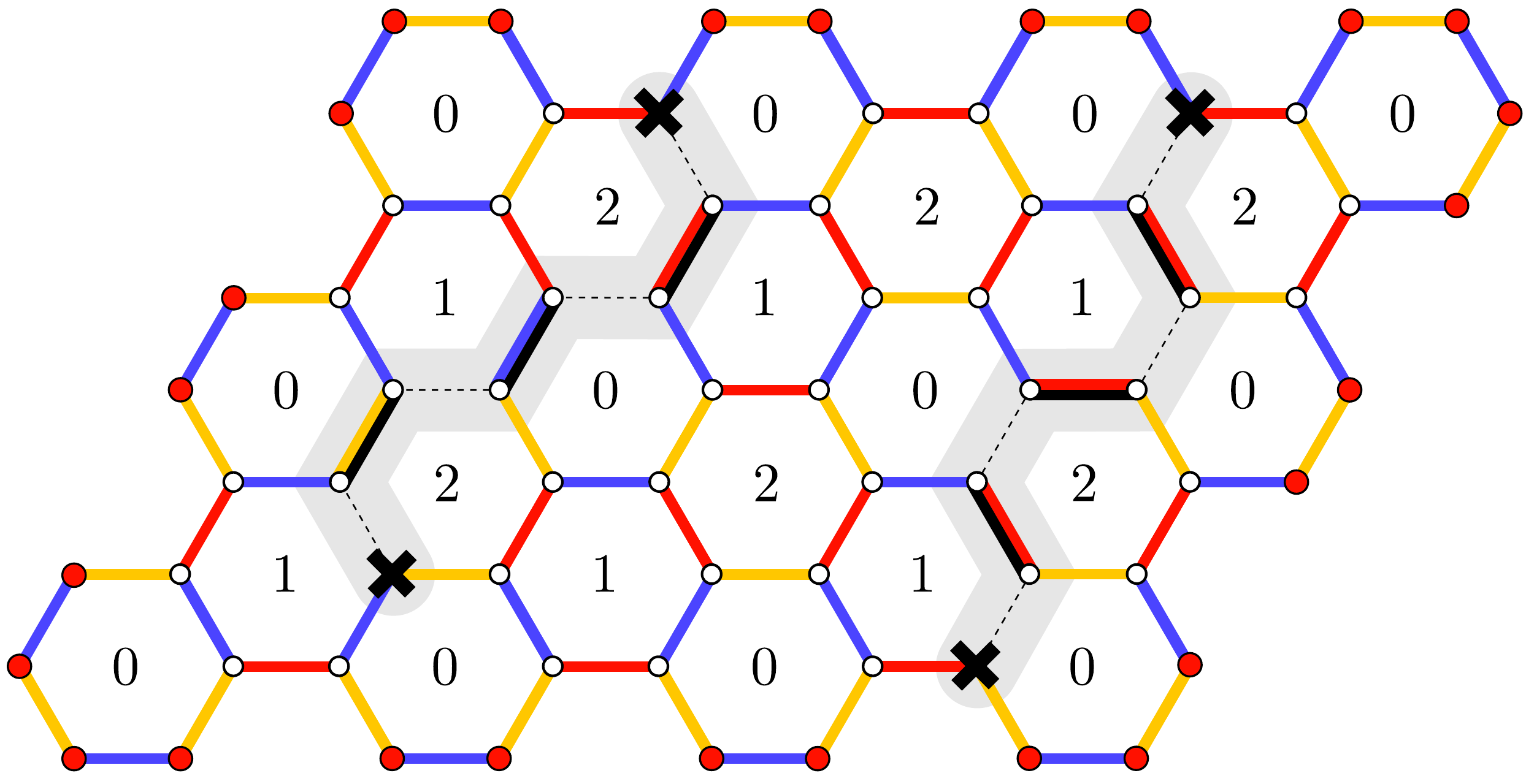}}
\caption{(a) We define the $\ZZ_2$ Floquet code on a system with $0$-plaquettes on the boundary by introducing single-site $0$-checks (red dots) for all of the bivalent vertices.
(b) We then add twist defects following the prescription in Section~\ref{sec: Z2 defect measurement schedule}. 
}
\label{fig: 0 boundaries w and wo defects}
\end{figure*}

Here, we employ twist defects to encode quantum information in a $\ZZ_2$ Floquet code with a planar geometry. Notably, our construction with twist defects has a three-round period and no transport of quantum information along the boundary. In agreement with Ref.~\cite{Aasen2023measurement}, our construction violates the local reversibility condition. However, in contrast to Ref.~\cite{vuillot2021planar}, we identify representations of the logical operators that are invariant under the dynamics -- trivially implying that they have local transformations after three rounds of measurements.  
In the subsequent section, we argue that our $\ZZ_2$ Floquet code with twist defects is fault tolerant, as long as the twist defects are sufficiently far from the boundary. Therefore, we trade the time overhead of six rounds of measurements for spatial overhead to ensure that the twist defects are far from the boundary. In the next section, we describe another measurement schedule with a six-round period, which is fault tolerant, even if the twist defects are placed on or close to the boundary.

\begin{figure}[t]
\centering
\includegraphics[width=\textwidth]{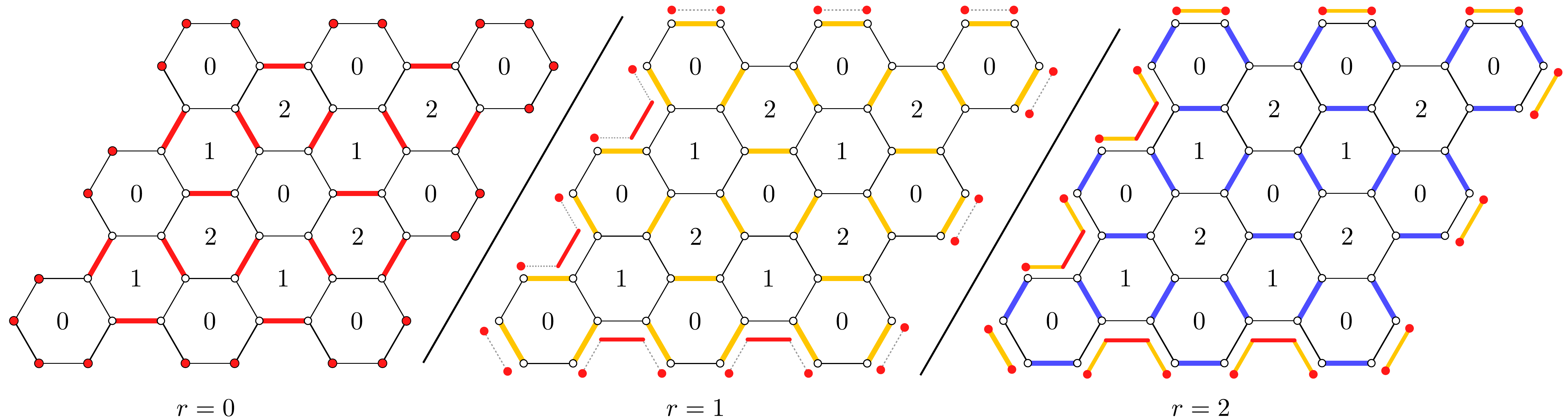}
\caption{The $0$-ISG is generated by the $0$-checks (red) and the plaquette stabilizers. The $1$-ISG is generated by the $1$-checks (yellow), the plaquette stabilizers, and local products of $0$-checks at the boundary (red connected by gray dotted lines). The $2$-ISG is generated by the $2$-checks (blue), the plaquette stabilizers, and local products of $0$-checks and $1$-checks at the boundary (red connected by yellow). In general, the $r$-ISG for the system with a boundary is a truncation of the $r$-ISG on a system without boundaries. }
\label{fig: 0 boundaries ISGs}
\end{figure}

We now describe our construction of a $\ZZ_2$ Floquet code on a planar geometry with a three-round measurement schedule.
We start by considering the planar $\ZZ_2$ Floquet code depicted in Fig.~\ref{fig: 0 boundaries}. In the bulk, the check operators are the same as those in the case of periodic boundary conditions. On the boundary, we introduce single-site 0-checks for all of the bivalent vertices. The single-site checks can be interpreted as truncations of 2-body 0-checks to a single site. The measurement schedule proceeds as before, starting with four rounds to initialize the code and followed by measuring the 0-, 1-, and 2-checks in sequence:
\begin{align}
    [2012](012)(012)(012)\ldots
\end{align}
The 0-, 1-, and 2-ISGs are represented in Fig.~\ref{fig: 0 boundaries ISGs}. 

This Floquet code does not encode any logical qubits. This is because, after measuring the 0-checks or the 2-checks, the entire boundary is a rough boundary, and after measuring the 1-checks, the entire boundary is a smooth boundary. More explicitly, the product of the single site checks at the boundary of the 0-ISG can be interpreted as a loop of $e$ string operator. Indeed, the string operator is a product of $0$-plaquette stabilizers and the~bulk $0$-checks, agreeing with the characterization of the $e$ string operator in Section~\ref{sec: review ISGs}. The $0$-checks at the boundary can be thought of as $e$ short string operators, indicating that the $e$ anyons have been condensed. Similarly, the product of the $1$-checks and the $0$-checks on the boundary of the $1$-ISG is a loop of $m$ string operator, and the product of $2$-checks and the remaining product of $0$- and $1$-checks on the boundary of the $2$-ISG is a loop of $e$ string operator.
We note that this $\ZZ_2$ Floquet code with boundaries is not locally reversible. Nonetheless, since there are no logical operators, there are no update rules to discuss.

We next introduce twist defects to encode information, following the same steps as in Section~\ref{sec: Z2 defect measurement schedule}. We position the defect lines as in Fig.~\ref{fig: 0 boundaries defects} so that the removed check operators do not overlap with any of the boundary edges.
The measurement schedule is then the same as the schedule without boundaries [Eq.~\eqref{eq: Z2 defect measurement schedule}]:
\begin{align} \label{eq: Z2 defect measurement schedule boundaries}
    [2,0,1,2]\underbrace{(0,1,2)\ldots(0,1,2)}_{d-1}(\tilde{0}^*,1^*,2^*)(\tilde{0}^*,1^*,2^*)\ldots,
\end{align}
where again, we fault-tolerantly initialize the code without twist defects, before inserting them in the $\tilde{0}^*$-round. The instantaneous stabilizers in the vicinity of the defect lines take the same form as those described below Eq.~\eqref{eq: r*-ISG}.

The four twist defects shown in Fig.~\ref{fig: 0 boundaries defects} encode a single logical qubit. Of course, further logical qubits can be encoded by inserting additional twist defects, such as in Fig.~\ref{fig: computation basis}. Similar to Section~\ref{sec: Z2 defects logical operators}, $2k$ twist defects encode $k-1$ logical qubits. The logical Pauli $X$ and Pauli $Z$ operators can be represented by fermion string operators that connect the twist defects, analogous to Fig.~\ref{fig: multiple logicals}. These string operators commute with all of the checks and thus are invariant under the dynamics. 

\subsection{Fault-tolerance} \label{sec: fault tolerance}

Here, we provide an argument that the $\ZZ_2$ Floquet code with twist defects is fault-tolerant as a quantum memory. We argue that (i) the code distance grows with the separation of the twist defects, and (ii) the decoding problem can be mapped to a statistical-mechanical model with an order-to-disorder phase transition. Together, these two points are sufficient to show that the Floquet code has a nonzero error threshold. We emphasize here that, for Floquet codes, (i) requires showing that there are no constant-weight undetectable errors in both space and time. Indeed, the Floquet code of Ref.~\cite{vuillot2021planar} fails to be fault tolerant due to constant-weight 
undetectable errors in spacetime, despite each ISG having an extensive code distance.

To address whether the $\ZZ_2$ Floquet code with twist defects has undetectable errors with constant weight, it is convenient to adopt a spacetime perspective, as in Refs.~{\mbox{\cite{Williamson2022spacetime, bombin2023unifying, bauer2023topological}}}. From this perspective, the instantaneous stabilizer codes correspond to discrete time slices of spacetime. As the Floquet code evolves in time, defect lines and twist defects trace out worldsheets and worldlines, respectively. 
We refer to the worldsheets as defect membranes and call the worldlines at their boundaries the twist defect lines. It is also meaningful to consider the worldlines of anyons. The worldline of an anyon corresponds to a combination of Pauli errors, which move the anyon through space, and measurement errors, which allow the anyon to go undetected between ISGs.

Given that the ISGs have the same topological order as that of the toric code, Pauli errors and measurement errors can be interpreted as worldlines of anyons in spacetime. The undetectable errors, in particular, can be decomposed as worldlines of fermions that connect between different twist defect lines or worldlines of $e$ and $m$ anyons that encircle a set of twist defect lines.
This implies that, in the absence of boundaries, the undetectable errors have a weight that grows with the separation of the twist defects. Therefore,
there are no constant-weight undetectable errors with periodic boundary conditions. 

\begin{figure}[t]
\centering
\includegraphics[width=.8\textwidth]{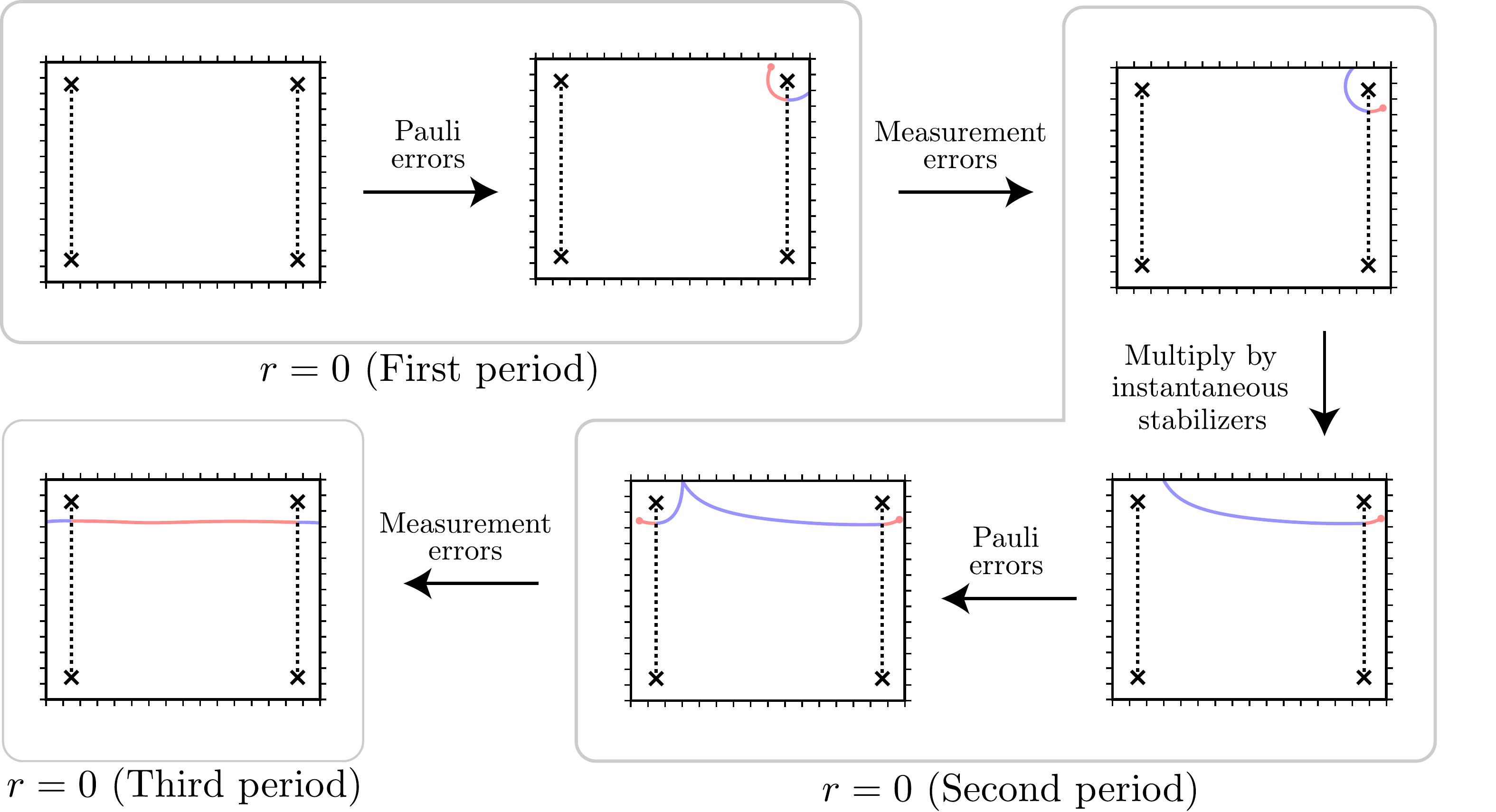}
\caption{
If the twist defects are a constant distance from the spatial boundary, then there is a constant-weight undetectable error. One such example of a constant-weight logical operator is produced as follows. In an $r=0$ round, when the boundaries are rough (hashed boundary), a constant weight Pauli error can create an $e$ anyon and move it across a defect line (light blue) to transform it into an $m$ anyon (light red). The $m$ anyon is unable to be reabsorbed into the boundary, but after three rounds of measurements (the subsequent $r=0$ round), we can multiply the string operator by local stabilizers to turn it into an $e$ string operator, at which point it can condense into the boundary. Given that the $e$ anyons are condensed at the boundary, we can extend the string operator across the system to another twist defect. Then a second constant-weight Pauli error can move the $e$ anyon from the boundary and pass it through a second defect line. At the subsequent $r=0$ round, the $e$ anyon can be absorbed into the boundary. Throughout, this process, between $r=0$ rounds, we assume that there are a constant number of measurement errors that make the endpoints of the string operator undetectable.}
\label{fig: constant weight error}
\end{figure}

On a system with boundary, more care is needed, since the boundary type can fluctuate over time between rough and smooth -- incommensurate with the automorphism of anyon types in the bulk after a period.
In fact, if the twist defects are moved too close to the boundary, then there are constant-weight undetectable errors, as illustrated in Fig.~\ref{fig: constant weight error}. To see this, suppose that there are twist defect lines near the spatial boundary. In the case when the boundaries are rough, a Pauli error can create an $e$ anyon from the boundary and move it through the defect membrane. The resulting $m$ anyon can then go undetected with a constant number of measurement errors. At some later time, the anyon can be reabsorbed into the boundary and re-emerge on the other side of the system nearby another twist defect line, where a second constant-weight Pauli error can move the anyon around that twist defect line. Finally, at a later time step, the anyon can be reabsorbed into the boundary, having encircled two different twist defect lines. This spacetime process produces a constant-weight logical operator. Therefore, if we want to maintain a code distance that grows with the separation of the twist defects, the twist defects need to be positioned at a distance $O(d)$ from the boundary. This, however, implies that there is an additional qubit overhead at the boundary, as compared to the boundary constructions in Refs.~\cite{Haah2022boundarieshoneycomb, Gidney2022benchmarkingplanar, Paetznick2023Performance}. 

To avoid the qubit overhead at the boundary and maintain fault tolerance, we can alternatively extend the schedule to six rounds per period. Specifically, after an appropriate initialization, we measure the $r^*$-checks in the sequence:
\begin{align} \label{eq: six round schedule}
    \cdots (\tilde{0}^*, 1^*, 2^*, 1^*, \tilde{0}^*, 2^*)(\tilde{0}^*, 1^*, 2^*, 1^*, \tilde{0}^*, 2^*)(\tilde{0}^*, 1^*, 2^*, 1^*, \tilde{0}^*, 2^*)\cdots.
\end{align}
This schedule has the benefit that, using the checks in Fig.~\ref{fig: 0 boundaries w and wo defects}, the boundary type alternates between rough (on all sides) and smooth (on all sides) after each measurement round. 
The error depicted in Fig.~\ref{fig: constant weight error} is no longer an issue since the anyon cannot be reabsorbed into the boundary after passing through the defect membrane. This suggests that the twist defects can be positioned at a constant distance from the boundary using this schedule. One caveat of this schedule is that, in order to infer all of the stabilizers along the defect line, the defect line needs to be constructed in such a way that only $0$- and $1$-checks are removed. This is a consequence of the asymmetry of the measurement schedule and is made more explicit in Appendix~\ref{app: inferring stabilizers}.

To finish the fault-tolerance argument, we describe how the decoding problem can be mapped to a statistical-mechanical model with a phase transition. 
Following the logic in Ref.~\cite{HH_dynamic_2021}, we consider a simplified error model, in which a Pauli error of type $X$, $Y$, or $Z$ can only occur immediately after measuring an $XX$, $YY$, or $ZZ$ check, respectively. As noted in Ref.~\cite{HH_dynamic_2021}, this noise model is sufficient for proving that the $\ZZ_2$ Floquet code has a threshold, even in the presence of measurement errors.\footnote{Here, measurement errors can be reproduced from correlated Pauli errors.} However, note that to construct defect lines, we remove the check operators that fail to commute with the defect checks. If we remove a $YY$ check, for example, then this error model suggests that Pauli $Y$ errors do not occur on the neighboring vertices. To compensate for this, we assume that the measurements of defect checks may be faulty. A Pauli $Y$ error is then equivalent to a combination of an $X$ error, a $Z$ error, and depending on the time at which the $Y$ error occurs, a faulty measurement of a defect check. 

As described in~Ref.~\cite{HH_dynamic_2021}, the $\ZZ_2$ Floquet code without twist defects can be decoded entirely based on the measurement outcomes of the plaquette stabilizers, inferred from the 2-body measurements. These stabilizers belong to each of the ISGs, so their measurement outcomes are consistent between periods, at least, in the absence of errors. Similarly, the $\ZZ_2$ Floquet code with twist defects can be decoded based on the measurement outcomes of the defect checks and the products of plaquette stabilizers that commute with them. We refer to these stabilizers as the static stabilizers.

To make the decoding problem concrete, we consider the defect line shown in Fig.~\ref{fig: defect construction}. We assume that the fault-tolerance argument can be extended straightforwardly to more general configurations of defect lines. The generators of the static stabilizers in the case of the defect line in Fig.~\ref{fig: defect construction} can be written as:
\begin{align} \label{eq: static stabilizer generators}
        \Bigg \{
        \vcenter{\hbox{\includegraphics[scale=.5]{Figures/Sp_no_p.pdf}}}, \,\,
        {\underbrace{\vcenter{\hbox{\includegraphics[scale=.55]{Figures/defect_check.pdf}}}, \,\, 
        \vcenter{\hbox{\includegraphics[scale=.45]{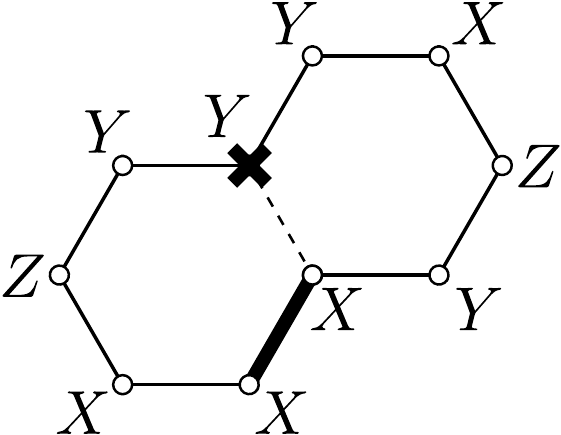}}}, \,\,
        \vcenter{\hbox{\includegraphics[scale=.45]{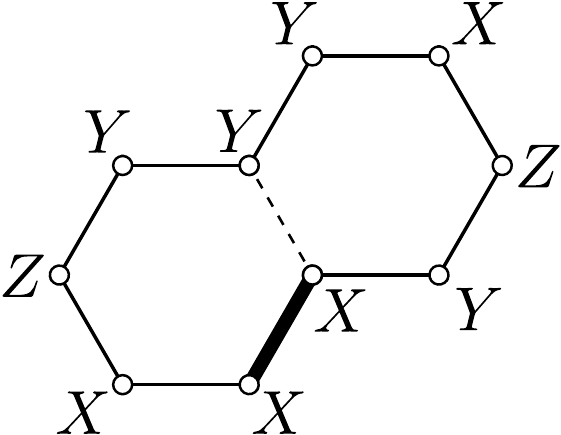}}}, \,\,
        \vcenter{\hbox{\includegraphics[scale=.45]{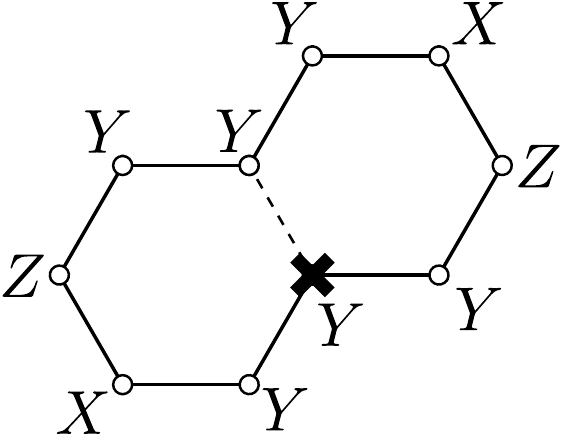}}} \,\,}_{\text{Along defect line}}
        \Bigg\}.}
\end{align}
Note that, compared to Eq.~\eqref{eq: r*-ISG}, we have multiplied the products of plaquettes along the defect line by the defect checks. This choice of generators ensures that the (uncorrelated) errors of our simplified noise model violate pairs of static stabilizers instead of triplets. 

With this choice of generators for the static stabilizers, we define the syndrome graph associated with the simplified noise model. 
We begin by recalling the syndrome graph of the case without twist defects, as described in Refs.~\cite{HH_dynamic_2021, Haah2022boundarieshoneycomb}. The vertices of the syndrome graph mark the spacetime coordinates where the plaquette stabilizers are inferred. To be more specific, we let $t$ denote a discrete time coordinate, such that, at time $t$, we measure the $r$-checks with $r = t \text{ mod } 3$. After measuring the $r$-checks, we can infer the measurement outcomes of all of the plaquette stabilizers corresponding to the $(r+1)$-plaquettes. The vertices of the syndrome graph at time $t$ are thus positioned at the $(r+1)$-plaquettes. For a system with boundary, we further define a ``terminal'' vertex at each time $t$ \cite{Haah2022boundarieshoneycomb}. This vertex can be thought of as the vertex corresponding to the product of all of the plaquette stabilizers.

The edges of the syndrome graph are determined by the error model. We connect two vertices by an edge if there is an (uncorrelated) error that flips the measurement outcomes of the corresponding plaquette stabilizers. The effects of the (uncorrelated) errors on the measurement outcomes are given in the table below, reproduced from Ref.~\cite{Haah2022boundarieshoneycomb}:
\begin{center}
\begin{tabular}{| c || c | c | c | c | c | c | c |} 
 \hline
 Time step & $t=0$ & $t=1$ & $t=2$ & $t=3$ & $t=4$ & $t=5$ & $\cdots$ \\ 
 \hline
 Error after $t=0$ & 1 & \textcolor{red}{2} & 0 & \textcolor{red}{1} & 2 & 0 & $\cdots$ \\ 
 \hline
 Error after $t=1$ & 1 & 2 & \textcolor{red}{0} & 1 & \textcolor{red}{2} & 0 & $\cdots$ \\ 
 \hline
 Error after $t=2$ & 1 & 2 & 0 & \textcolor{red}{1} & 2 & \textcolor{red}{0} & $\cdots$ \\ 
 \hline
\end{tabular}
\end{center}
Here, the entries are the $0$-, $1$-, and $2$-plaquette labels for the plaquette stabilizers inferred at that time step. The plaquette labels in red imply that the eigenvalue of the plaquette stabilizer has been flipped for the first time. As an example, in the row labeled ``Error after $t=0$'', we assume that a Pauli error ($X$, $Y$, or $Z$ determined by the edge type) occurs after the $t=0$ round. After measuring the 1-checks at $t=1$, the measurement outcome of a plaquette stabilizer on one of the neighboring 2-plaquettes has been flipped. After measuring the 0-checks again at $t=3$, we find that the plaquette stabilizer on the neighboring 1-plaquette has had its eigenvalue flipped. This tells us to define an edge between the vertex at the 2-plaquette ($t=1$) and the vertex at the 1-plaquette ($t=3$). On a system with a boundary, if there is an error that flips only one plaquette stabilizer in spacetime, we connect the corresponding vertex to the terminal vertex at that time $t$. This captures the fact that a single anyon can be created at the boundary if the boundary conditions allow for it. 

For a $\ZZ_2$ Floquet code with twist defects, we can use the same prescription as above to define a syndrome graph. That is, we define the vertices of the syndrome graph to be spacetime coordinates at which the static stabilizers are inferred. We then add an edge between two vertices if an uncorrelated error flips the eigenvalue of the two stabilizers associated to the endpoints. Importantly, the generators of the static stabilizers in Eq.~\eqref{eq: static stabilizer generators} are such that each uncorrelated error always flips a pair of generators. We further add an edge between pairs of vertices corresponding to defect checks at neighboring time steps. These edges capture the potential for faulty measurements of the defect checks. 

Now that we have defined the syndrome graph, a standard matching algorithm can be used to decode the $\ZZ_2$ Floquet code with twist defects. To see that this has a positive threshold, we map the decoding problem onto a statistical-mechanical model. The statistical-mechanical model is defined on the syndrome graph by placing a classical spin on each vertex. We define an interaction between each pair of neighboring spins, with an interaction strength determined by the probability of the error associated with that edge (see Ref.~\cite{Dennis2002quantummemory}). The energetic penalty for inserting a domain wall reflects the likelihood of the corresponding error along that domain wall. Since there are no constant-weight undetectable errors, the undetectable errors are associated with domain walls that are extensive in the separation of the twist defects.~It follows from a comparison of the energy cost and entropy gain of inserting a domain wall (i.e., a Peierls argument~\cite{Peierls1936Ising}), that this statistical-mechanical model has a phase transition from an ordered phase to a disordered phase. This phase transition is induced by increasing the error rate, which in turn increases the interaction strengths.~In the disordered phase, the domain walls fluctuate wildly, and there is a nonzero expectation value for inserting a domain wall that is extensive in the separation of the twist defects. The critical point of this statistical-mechanical model determines a (positive) lower bound on the error threshold of the Floquet code~\cite{Dennis2002quantummemory}. Thus, the $\ZZ_2$ Floquet code with twist defects is fault tolerant. 

\subsection{Computation with twist defects} \label{sec: Z2 defect gate set}

Quantum information that is encoded in twist defects can be manipulated by braiding~the twist defects and making topological charge measurements. Topological charge measurements, in this context, amount to measuring the fermion parity of an even number of twist defects. This can be done, for example, by measuring the string operator that moves an $e$ or $m$ anyon around an even number of twist defects. An outcome of $+1$ indicates that the twist defects have an even fermion parity, while a $-1$ outcome indicates that they have an odd fermion parity. Topological charge measurements alone are, in fact, able to reproduce all of the effects of braiding~\cite{Bonderson2008Measurementonly, Zheng2016Measurementonly}. For twist defects in the $\ZZ_2$ toric code, this means that topological charge measurements with ancillary twist defects are sufficient to implement the entire Clifford group. By further supplementing with magic state injection, we can realize a universal gate set. 

In this section, we leverage the construction of boundaries in Section~\ref{sec: Z2 defects boundaries} to perform certain topological charge measurements. These allow us to measure an arbitrary tensor product of logical Pauli $X$ and $Z$ operators. We then describe a protocol for implementing a logical $S$ gate, defined in the computational basis as:
\begin{align}
    S \equiv \begin{pmatrix}
1 & 0 \\
0 & i
\end{pmatrix}.
\end{align}
Together, these operations are sufficient to measure an arbitrary tensor product of Pauli operators (including Pauli $Y$ operators) and thereby implement any logical Clifford gate with the help of ancillary twist defects. 

\begin{figure*}[t]
\centering
\subfloat[\label{fig: computation basis}]{\includegraphics[width=.4\textwidth]{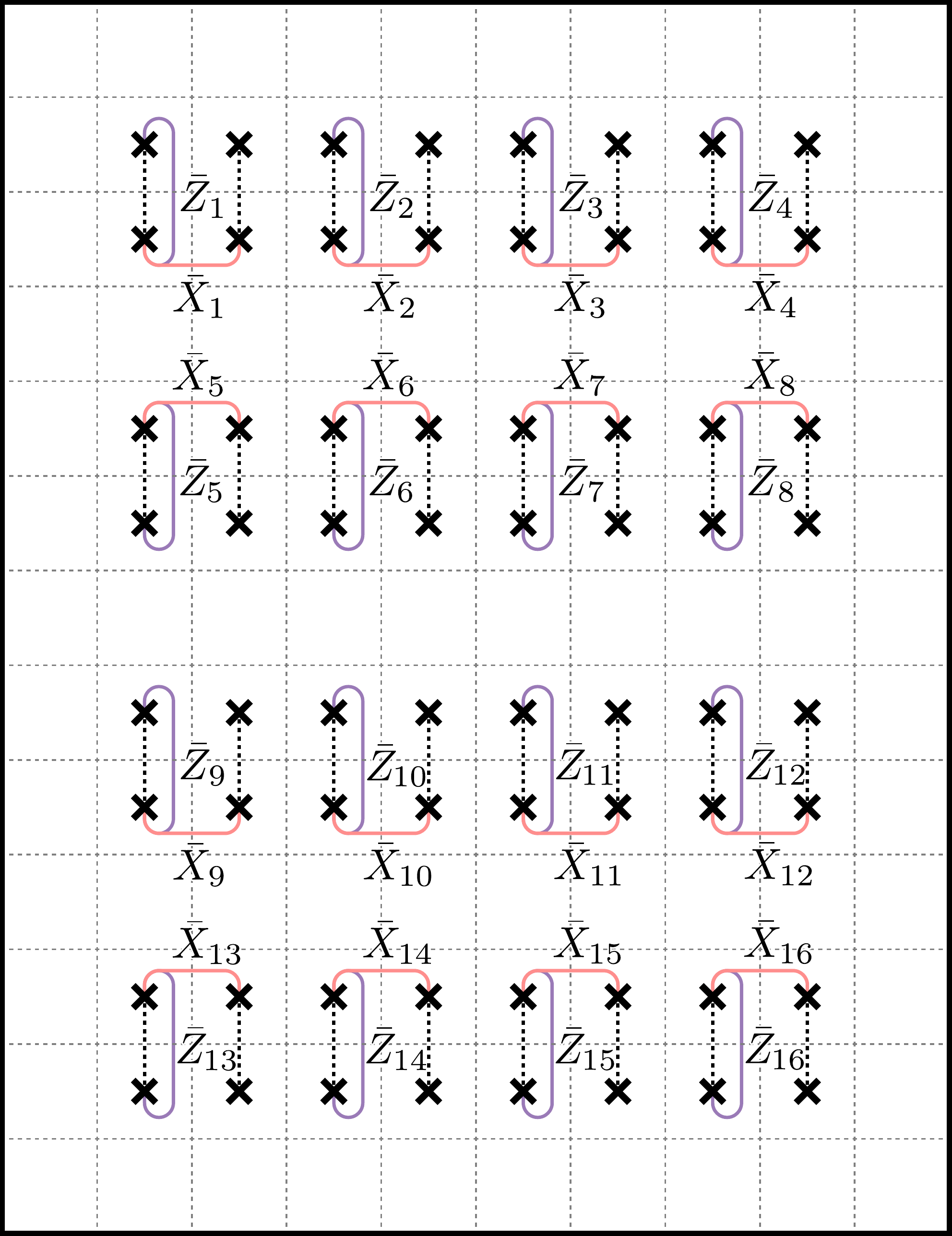}} \qquad
\subfloat[\label{fig: logical measurement}]{\includegraphics[width=.4\textwidth]{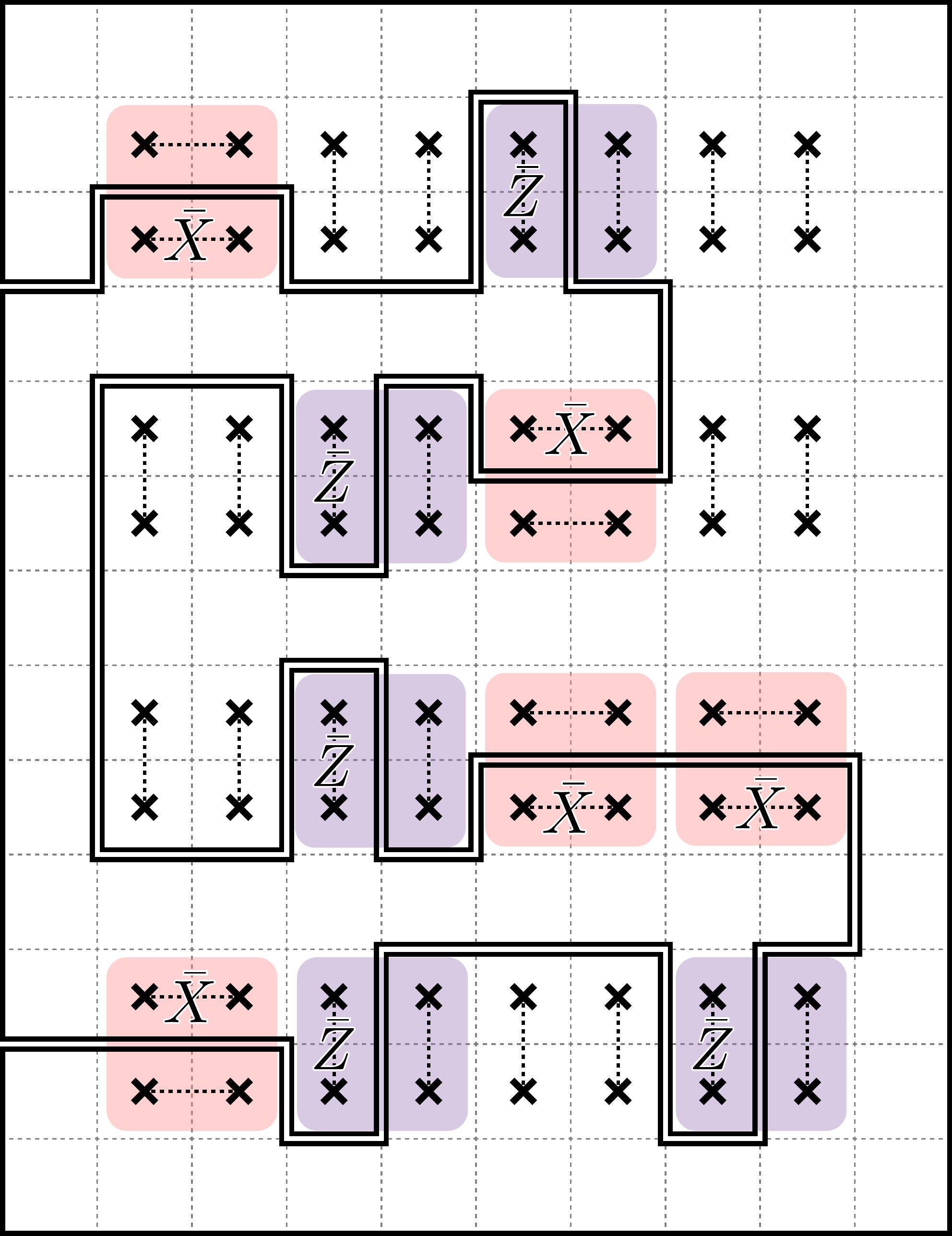}}
\caption{We consider a system with boundary (black border) divided into squares of width $O(d)$. We place a twist defect (bold black cross) in a subset of the squares. Empty squares are used as routing space. (a) The twist defects are connected by vertical defect lines (black dashed lines). A set of four twist defects defines a logical qubit. The logical $X$ and $Z$ operators are represented by fermion string operators (pink and purple, respectively). (b) To measure a tensor product $\bar{P}$ of $\bar{X}$ operators (pink shaded) and $\bar{Z}$ operators (purple shaded), we connect the twist defects involved in the $\bar{X}$ measurements by horizontal defect lines. We then insert boundaries (black lines) to isolate the pairs of twist defects corresponding to the tensor factors of $\bar{P}$. In effect, we measure an $e$ or $m$ string (depending on the measurement round) encircling the twist defects, which is a representation of $\bar{P}$.}
\label{fig: logical gates}
\end{figure*}

To simplify the discussion, we consider the layout of twist defects shown in Fig.~\ref{fig: computation basis}. In particular, a system with a boundary is partitioned into squares of width $O(d)$, and twist defects are inserted in the centers of a subset of the squares. The squares without twist defects are needed as ``routing space'', similar to that used in lattice surgery~\cite{Campbell2022twistfree, Campbell2022twistssurgery}. As a matter of convention, we connect the twist defects by vertical defect lines. 
We then use four twist defects to encode each logical qubit and represent the logical $X$ and $Z$ operators as fermion string operators connecting between the four twist defects as in Fig.~\ref{fig: computation basis}. The logical operators at a given instantaneous time can alternatively be represented by loops of $e$ or $m$ string operators encircling the twist defects. We denote the logical Pauli operators of the $i$th logical qubit as $\bar{X}_i$, $\bar{Y}_i$, and $\bar{Z}_i$.
We remark that, even though $4k$ twist defects are able to encode $2k-1$ logical qubits, we only use $k$ of them in the computational scheme described below. The remaining $k-1$ qubits can be interpreted as gauge qubits, in the sense of subsystem codes~\cite{Knill2000theory,Kribs2005unified,Poulin2005stabilizer,Kribs2006operator,Bacon2006operator}.

With this setup, we can measure an arbitrary tensor product $\bar{P}$ of $\bar{X}$ and $\bar{Z}$ operators by introducing boundaries, as described in Section~\ref{sec: Z2 defects boundaries}. The first step is to redefine the defect lines for each factor of $\bar{X}$ in $\bar{P}$. More specifically, if $\bar{P}$ includes an $\bar{X}$ on the $i$th logical qubit, then we reconnect the four twist defects of the $i$th logical qubit by horizontal defect lines (see Fig.~\ref{fig: logical measurement}). This makes it so that the representation of $\bar{X}_i$ is parallel with the defect lines. More importantly, the $\bar{X}_i$ operator can now be represented by an $e$ or $m$ string operator that does not pass through any defect lines. This implies that, after reconnecting the twist defects, $\bar{P}$ can also be represented by an $e$ or $m$ string operator that encircles twist defects without passing through any defect lines. 
We note that, since defect lines have an odd length by construction, in order to reconnect the twist defects, we require that they are separated by paths of odd length in both the vertical direction and the horizontal direction. For example, this is the case for the twist defects shown in Fig.~\ref{fig: 0 boundaries defects}.

The next step is to measure the $e$ or $m$ string operator that represents $\bar{P}$. This can be accomplished by constructing boundaries to isolate the pairs of twist defects corresponding to the tensor factors of $\bar{P}$, as illustrated in Fig.~\ref{fig: logical measurement}. As noted in Section~\ref{sec: Z2 defects boundaries}, the check operators at the boundary can be interpreted as short string operators for an $e$ or $m$ string operator, depending on the measurement round. Thus, after introducing the boundaries, 
the representation of $\bar{P}$ as an $e$ or $m$ string operator encircling the twist defects becomes a product of local instantaneous stabilizers. This implies that the measurement outcome of $\bar{P}$ can be inferred from the measurement schedule in Eq.~\eqref{eq: Z2 defect measurement schedule boundaries}. However, we point out that, due to the nontrivial automorphism of the $\ZZ_2$ Floquet code, it takes six rounds of measurements for each subsequent measurement of the $e$ or $m$ string operator. We also point out that the routing space is needed to both maintain the code distance and ensure that the tensor factors of $\bar{P}$ are not measured independently. To reduce the space and time overhead, it may be advantageous to remove the routing space and make 2-body logical measurements of neighboring logical qubits.

\begin{figure*}[t]
\centering
\subfloat[\label{fig: S gate 1}]{\includegraphics[width=.13\textwidth]{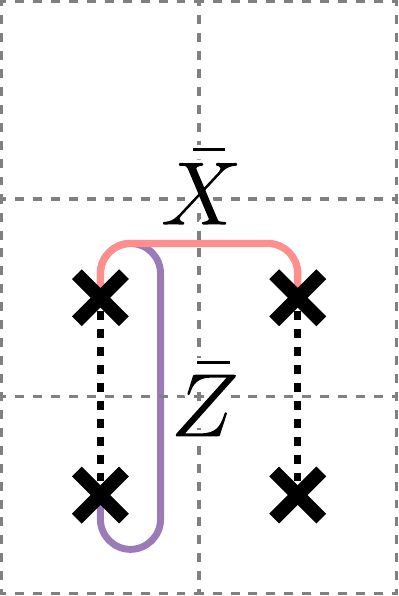}} \qquad
\subfloat[\label{fig: S gate 2}]{\includegraphics[width=.13\textwidth]{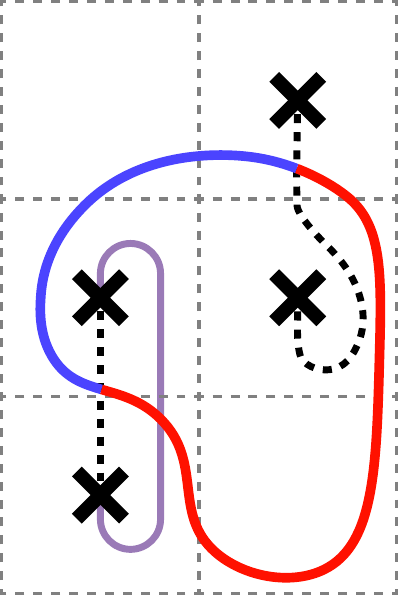}} \qquad
\subfloat[\label{fig: S gate 3}]{\includegraphics[width=.13\textwidth]{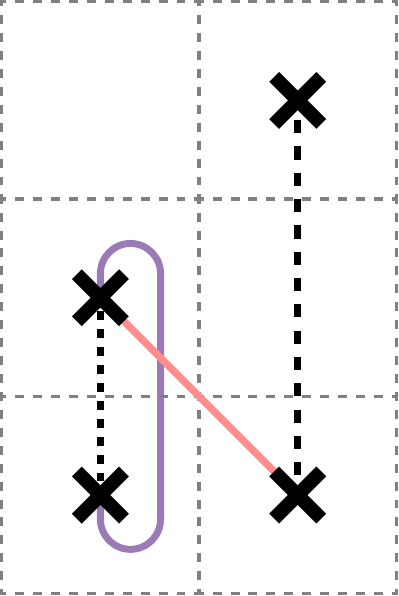}} \qquad
\subfloat[\label{fig: S gate 4}]{\includegraphics[width=.13\textwidth]{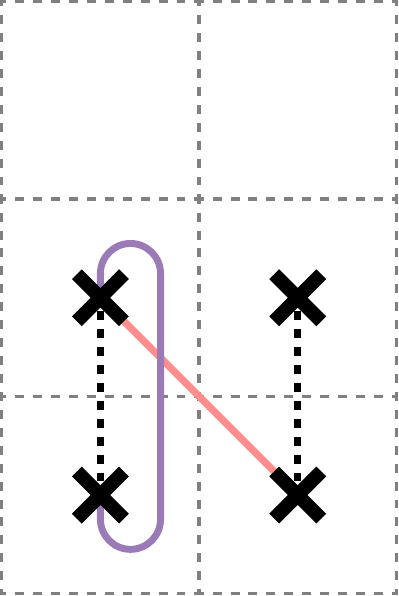}}
\caption{An $S$ gate can be implemented in three steps. (a) Prior to implementing the $S$ gate, we represent the logical $X$ and $Z$ operators for a set of four twist defects by fermion string operators that connect the twist defects (pink and purple, respectively). (b) The first step of performing an $S$ gate is to move the twist defect from the bottom right corner into the routing space above. The logical operators can still be represented by the fermion string operators in (a). Alternatively, the logical $X$ operator can be represented by an $e$ string operator (blue) that turns into an $m$ string operator (red) when it crosses the defect line. (c) Next, the twist defect on the bottom of the rightmost pair can be moved downwards. The logical operators can be represented by the same operators depicted in (b); thus, the logical information is preserved. Alternatively, the logical operators can be represented by fermion string operators as shown. (d) Finally, we shift the twist defect on the top of the rightmost pair downwards. The representation of $\bar{X}$ and $\bar{Z}$ is unchanged from (c). Note that, as compared to (a), the representation of $\bar{X}$ is now along the diagonal, which is the same as the representation of $\bar{Y}$ prior to the $S$ gate. Hence, we have mapped $\bar{X}$ to $\bar{Y}$.}
\label{fig: S gate}
\end{figure*}

To measure an arbitrary product of logical Pauli operators, which may include Pauli $Y$ operators, we introduce a protocol for implementing a logical $S$ gate. We recall that conjugation by an $S$ gate maps the Pauli operators as:
\begin{align}
    X \to Y, \quad Y \to -X, \quad Z \to Z. 
\end{align}
Therefore, a logical Pauli $Y$ can be measured by applying an $S$ gate, measuring $\bar{X}$, then undoing the $S$ gate. A protocol for implementing an $S$ gate is shown in Fig.~\ref{fig: S gate}. Note that alternative schemes for implementing an $S$ gate can be found in Refs.~\cite{Bonderson2008Measurementonly, Knapp2016diabatic, Zheng2016Measurementonly, Tran2020optimizing}.

\section{$\ZZ_N$ Floquet codes with twist defects} \label{sec: ZN defects}

In this section, we generalize the construction of twist defects to $\ZZ_N$ Floquet codes, which exhibit ISGs with the same topological order as the $\ZZ_N$ toric code. The $\ZZ_N$ Floquet codes described below are parameterized by a pair of integers $(p,q) \in \ZZ_N \times \ZZ_N$, satisfying the condition $pq = 1 \text{ mod }N$. The pair $(p,q)$ specifies the automorphism undergone by the anyons after a period of the measurement schedule. More specifically, the automorphism is given by:
\begin{align} \label{eq: ZN automorphism}
    e \to m^p, \quad m \to e^q.
\end{align}
The condition that $pq = 1 \text{ mod }N$ ensures that the braiding relations are preserved by the automorphism. For the $\ZZ_2$ toric code, the only choice for $p$ and $q$ is $p=q=1$, which reproduces the automorphism that permutes $e$ and $m$ in Section~\ref{sec: review}. We point out that the $\ZZ_N$ Floquet codes parameterized by $(p,q)$ were also explored in the context of a parton construction in Ref.~\cite{sullivan2023floquet}.

Similar to Section~\ref{sec: Z2 defects}, we construct twist defects in the $\ZZ_N$ Floquet codes by condensing anyons along a path. To this end, we note that, given an automorphism specified by $(p,q)$, the anyon $me^q$ (and its powers) is invariant under the automorphism:
\begin{align}
    me^q \to e^q m^{pq} = me^q.
\end{align}
This is analogous to the fact that the emergent fermion $\psi$ is invariant under the dynamics of the $\ZZ_2$ Floquet code. Furthermore, this implies that there is a string operator with the property that it creates $me^q$ (and its inverse $m^{-1}e^{-q}$) for each of the ISGs. Therefore, defect lines can be constructed by condensing $me^q$ along a path. 

In what follows, we begin by defining $\ZZ_N$ Floquet codes without twist defects. We then introduce defect lines in Section~\ref{sec: ZN now with twists} by condensing $me^q$. Finally, in Section~\ref{sec: TQD Floquet codes}, we use the defect lines in the $\ZZ_N$ Floquet codes to construct Floquet codes corresponding to a subset of Abelian TQDs.

\subsection{Before adding twist defects} \label{sec: ZN measurement schedule}

The $\ZZ_N$ Floquet codes are defined on a hexagonal lattice with an $N$-dimensional qudit~at each vertex. To simplify the discussion, we assume that the hexagonal lattice has periodic boundary conditions -- although, we expect that the construction can be generalized to open boundary conditions, similar to Section~\ref{sec: Z2 defects boundaries}. We define generalized Pauli operators $X$ and $Z$ in the computational basis as:
\begin{align}
    X \equiv \sum_{\alpha \in \ZZ_N} |\alpha+1\rangle \langle \alpha |, \quad Z \equiv \sum_{\alpha \in \ZZ_N} \omega^\alpha |\alpha \rangle \langle \alpha |,
\end{align}
where $\omega$ is the $N$th root of unity $\omega = e^{2\pi i/ N}$. These operators satisfy the relations:
\begin{align}
    ZX = \omega XZ, \qquad X^N=Z^N=1.
\end{align}
For a choice of $(p,q) \in \ZZ_N \times \ZZ_N$ satisfying $pq = 1 \text{ mod }N$, we define the 2-body operator $K_{ij}$ for every edge $\langle ij \rangle$ as:
\begin{align} \label{eq: ZN checks}
        K_{ij} \equiv 
        \begin{cases}
            X_iX_j & \text{ if } \langle ij \rangle \in x\text{-edges}, \\
            (XZ^q)^\dagger_i(XZ^q)^\dagger_j & \text{ if } \langle ij \rangle \in y\text{-edges}, \\
            Z^q_iZ^q_j & \text{ if } \langle ij \rangle \in z\text{-edges}.
        \end{cases}
\end{align}
Here, the set of $x$-, $y$-, and $z$-edges are determined according to Fig.~\ref{fig: xyz edges}. 

In the absence of twist defects, the check operators of the $\ZZ_N$ Floquet code specified by the pair $(p,q)$ are precisely the operators $K_{ij}$. The $K_{ij}$ operators are measured according to the same schedule as the $\ZZ_2$ Floquet code [Eq.~\eqref{eq: Z2 measurement schedule}]. In particular, we define the $0$-, $1$-, and $2$-edges as in Fig.~\ref{fig: 012 edges} and refer to the $K_{ij}$ operators on $0$-, $1$-, and $2$-edges as the $0$-, $1$-, and $2$-checks, respectively. The schedule begins with four measurement rounds to initialize the code and proceeds by measuring the $0$-, $1$-, and $2$-checks in sequence:
\begin{align}
    [2,0,1,2](0,1,2)(0,1,2)(0,1,2) \ldots
\end{align}
We note that, since $q$ is coprime to $N$, one can measure $Z_iZ_j$ on $z$-edges, instead of $Z^q_iZ^q_j$.

The $r$-ISG of the $\ZZ_N$ Floquet code, i.e., the ISG after the $r$ round of measurements, is generated by the $r$-checks and the following plaquette stabilizer:
\begin{equation}
    S_p \equiv X_1 (X_2Z_2^q)^\dagger Z^q_3 X_4 (X_5Z_5^q)^\dagger Z^q_6,
\end{equation}
where the Pauli operators are labeled according to the vertices in Fig.~\ref{fig: 012 edges}. The $r$-ISG can be represented pictorially as:
\begin{align} \label{eq: ZN r-ISG}
    \mathcal{S}_{r} \equiv \Bigg \langle 
        \vcenter{\hbox{\includegraphics[scale=.5]{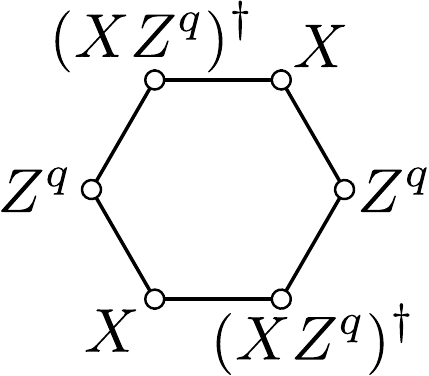}}}, \,\,\,
        \underbrace{\vcenter{\hbox{\includegraphics[scale=.6]{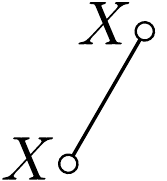}}}, \,\,\,
        \vcenter{\hbox{\includegraphics[scale=.6]{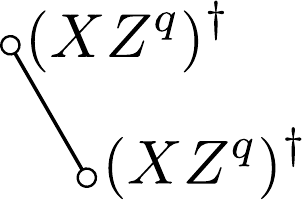}}}, \,\,\,
        \vcenter{\hbox{\includegraphics[scale=.6]{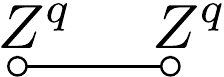}}}}_{\text{$r$-checks}}
    \Bigg \rangle.
\end{align}
Notice that, if we take $N=2$ and $q=1$, we recover the usual $r$-ISG of the $\ZZ_2$ Floquet code.

\begin{table}[tb]
    \centering
    \begin{tabular}{c|c|c|c}
         \hspace*{0mm}$\langle ij \rangle$ edge type \hspace*{0mm} & \hspace*{0mm}Subspace\hspace*{0mm} & \hspace*{3mm} $X_e$\hspace*{3mm} & \hspace*{3mm} $Z_e$ \hspace*{3mm}     
\\
         \hline \hline
        $x\text{-edge}$ & $X_i X_j = 1$  & $(X_iZ_i^q)^\dagger Z^q_j  $ & $X^p_iI_j$ \\
        $y\text{-edge}$ & $(X_iZ_i^q)^\dagger (X_jZ_j^q)^\dagger = 1$ & $Z^q_iX_j $ & $I_i(X_j^pZ_j)$\\
        $z\text{-edge}$ & $Z^q_iZ^q_j= 1$ & $(X_iZ_i^q)^\dagger X_j$ & $I_iZ_j$\\
    \end{tabular}
    \caption{After measuring the $r$-checks, we define an effective qudit at each $r$-edge $\langle ij \rangle$ by restricting to the $N$-dimensional subspace where $K_{ij}$ acts as the identity. The effective Pauli operators $X_e$ and $Z_e$ are defined up to multiplication by $K_{ij}$. On the $x\text{-edge}$ for example, $Z_e = X_i^pI_j \equiv X_i^{p+s}X_i^s$ for any $s \in \ZZ_N$ since $X_iX_j =1$. }
    \label{tab: effective qudits}
\end{table}

After each round of measurements, the ISG can be mapped to the stabilizer group of a $\ZZ_N$ toric code on an effective lattice. To see this, we consider the subspace stabilized by the $r$-checks and define an effective $N$-dimensional qudit at each $r$-edge. The effective qudits live on the edges of a hexagonal super-lattice, as shown in Fig.~\ref{fig: effective lattices}. The effective Pauli operators $X_e$ and $Z_e$ for the effective qudit are defined in Table \ref{tab: effective qudits} in terms of the Pauli operators at the vertices. 
The $S_p$ stabilizers at the $r$-plaquettes are mapped to the plaquette stabilizers\footnote{More precisely, the $S_p$ stabilizers are mapped to the $q$th power of the plaquette stabilizer. However, since $q$ is coprime to $N$, this generates the plaquette stabilizer of the $\ZZ_N$ toric code.} of the $\ZZ_N$ toric code on the super-lattice:
\begin{eqs}
\label{eq: ZN supperlattice 1}
    \vcenter{\hbox{\includegraphics[scale=.5]{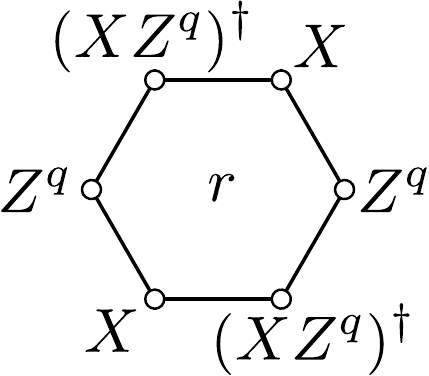}}} \, \longleftrightarrow \vcenter{\hbox{\includegraphics[scale=.5]{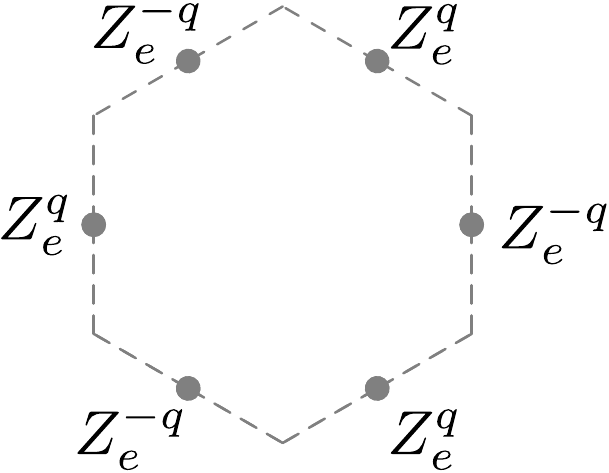}}},
\end{eqs}  
while the $S_p$ stabilizers at the $(r-2)$-, and $(r-1)$-plaquettes are mapped to the vertex stabilizers of the $\ZZ_N$ toric code:
\begin{eqs}    
\label{eq: ZN supperlattice 2}
    \vcenter{\hbox{\includegraphics[scale=.5]{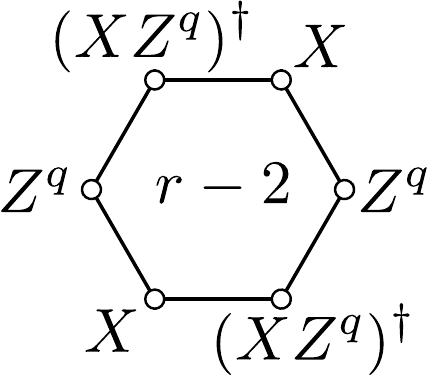}}} \, \longleftrightarrow \!\!\!  \vcenter{\hbox{\includegraphics[scale=.5]{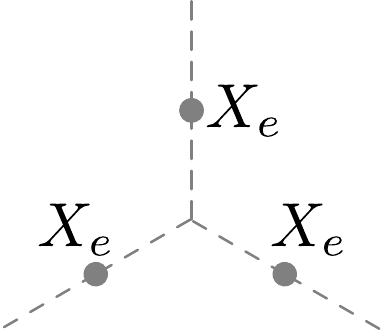}}}, \qquad
    \vcenter{\hbox{\includegraphics[scale=.5]{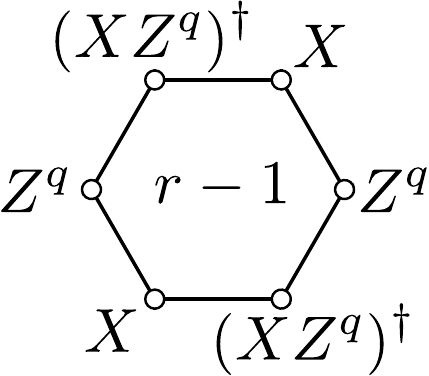}}} \, \longleftrightarrow \!\!\! \vcenter{\hbox{\includegraphics[scale=.5]{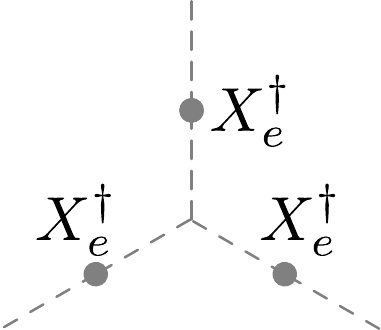}}}.
\end{eqs}
As a choice of convention, we take a change of $\omega^q$ in the measurement outcome of an $r$-plaquette stabilizer to be an $m$ anyon and a change of $\omega^q$ for an $(r-2)$- or $(r-1)$-plaquette stabilizer to be an $e^q$ anyon.  

\begin{figure}[t]
\centering
\includegraphics[width=.23\textwidth]{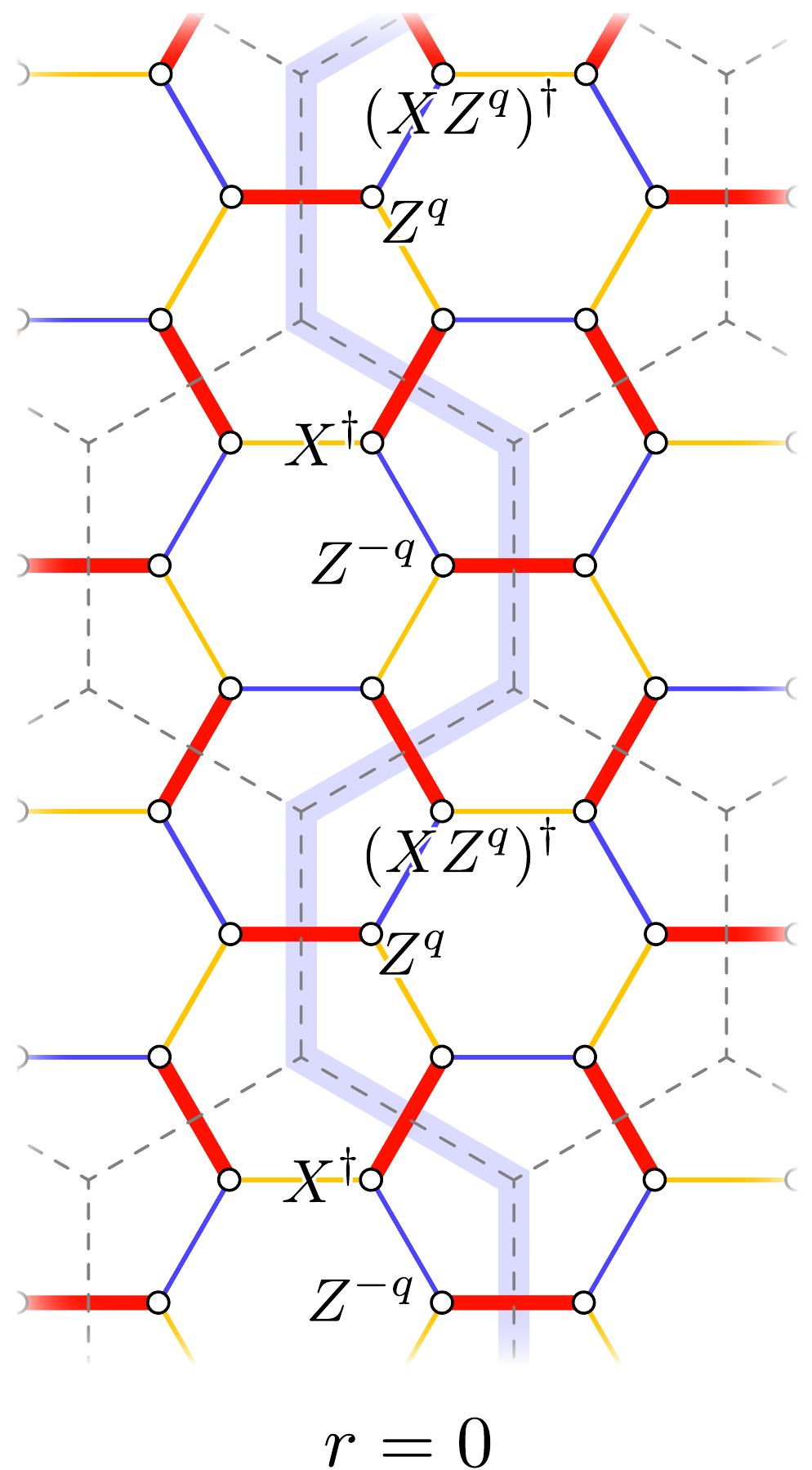}\,
\raisebox{.53cm}{\rule{.5pt}{6.1cm}}
\includegraphics[width=.23\textwidth]{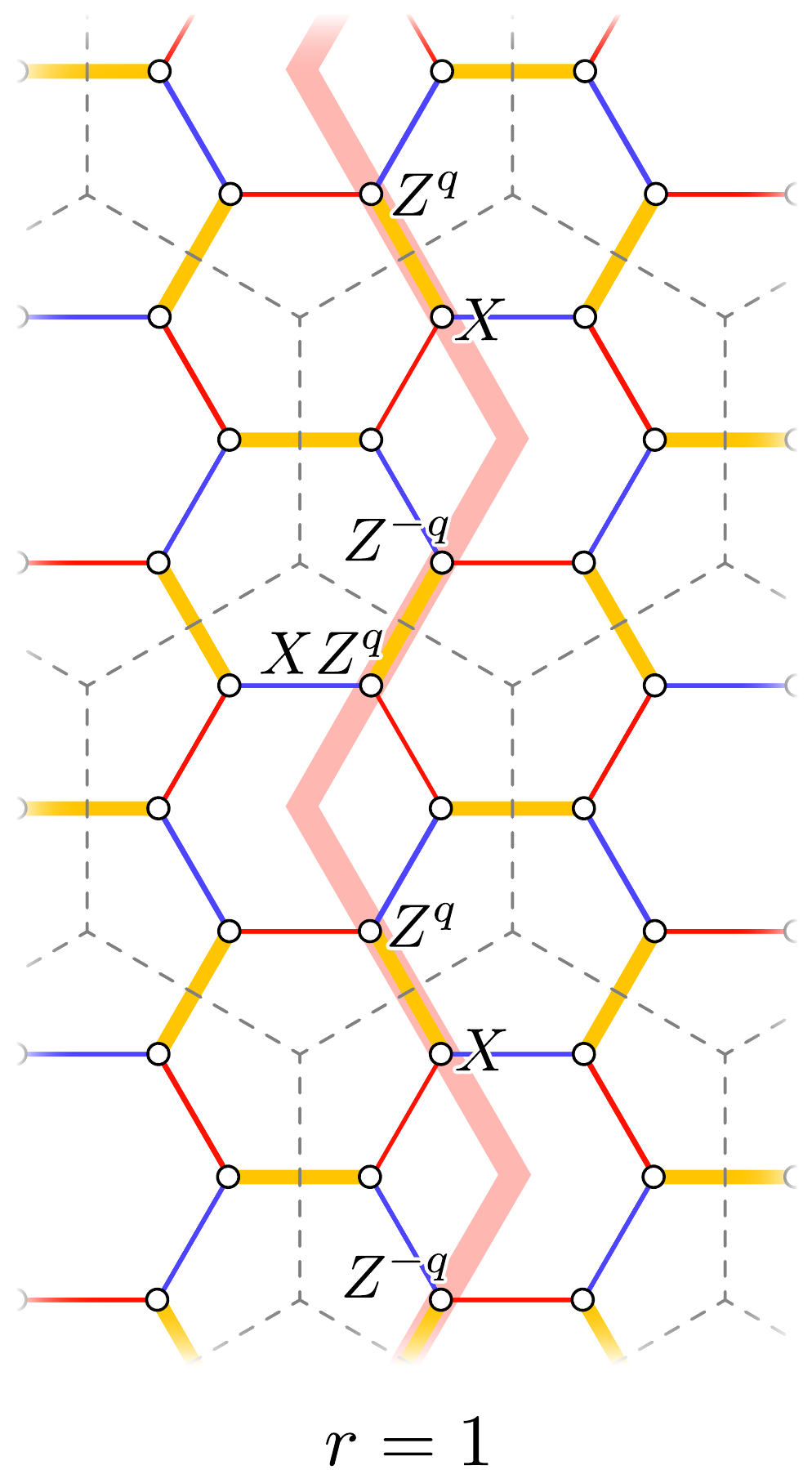}\,
\raisebox{.53cm}{\rule{.5pt}{6.1cm}}
\includegraphics[width=.23\textwidth]{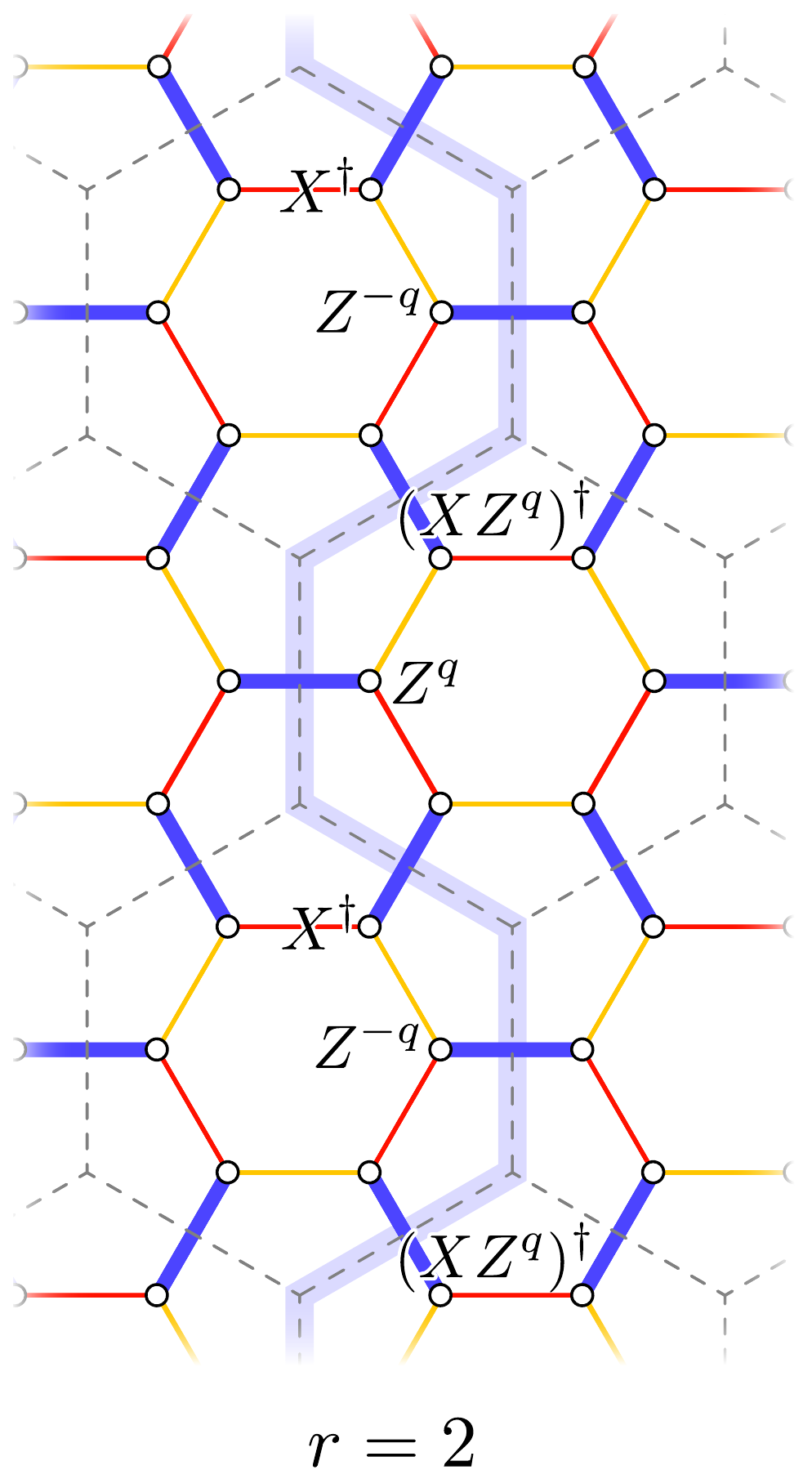}\,
\raisebox{.53cm}{\rule{.5pt}{6.1cm}}
\includegraphics[width=.23\textwidth]{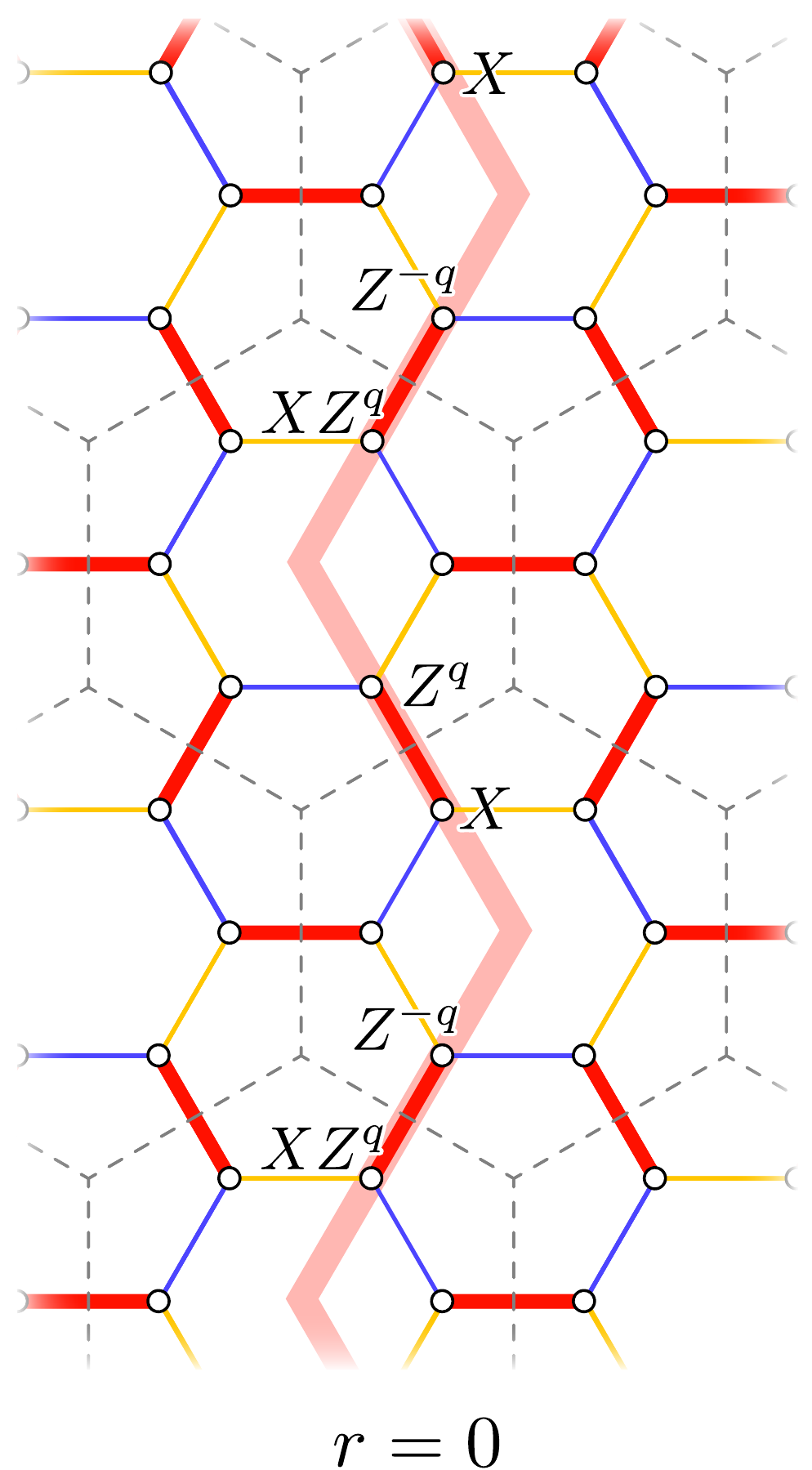}
\caption{Starting with an $e^q$ string operator (light blue) in the $r = 0$ round, we multiply by powers of the 0-checks
(red) to find a representation of the logical operator that commutes with the 1-checks (yellow). After the $r=1$ round, we next multiply the logical (light red) by powers of the 1-checks to find a representation of the logical operator that commutes with the 2-checks (blue). We repeat this process for the $r=2$ round, and the net result is that, after one full period of measurements, the initial $e^q$ string operator transforms into the $m$ string operator shown in the rightmost panel. }
\label{fig: ZN logical transformation}
\end{figure}

The logical operators of the $\ZZ_N$ Floquet code are given by the logical operators of the instantaneous $\ZZ_N$ toric codes. These are $e$ and $m$ string operators that are wrapped around non-contractible paths. The logical information is preserved throughout the measurement schedule, in the sense that, every logical operator of the $r$-ISG has a representation (up to instantaneous stabilizers) that is a logical operator of the $(r+1)$-ISG. We demonstrate this in Fig.~\ref{fig: ZN logical transformation} by starting with an $e^q$ string operator and tracking it through a full period. As claimed, the $e^q$ string operator transforms into an $m$ string operator after a period, implying that the string operators undergo the automorphism in Eq.~\eqref{eq: ZN automorphism}. Note that the product of instantaneous stabilizers that are used to transform the $e^q$ string operator into an $m$ string operator can be interpreted as an $me^{-q}$ string operator. This means that the $me^{-q}$ string operator is generated by check operators.

\subsection{Now with twist defects} \label{sec: ZN now with twists}

To insert twist defects into the $\ZZ_N$ Floquet code, we take the same approach as in Section~\ref{sec: Z2 defects} and condense anyons along an open path. This has the effect of creating a defect line along the path with twist defects hosted at the endpoints. More specifically, for the $\ZZ_N$ Floquet code with parameters $(p,q)$, we condense $me^q$ along an open path. These anyons (as well as their powers) have the property that they can be created by string operators that commute with all of the checks along their length. Importantly, this implies that the string operators create an $me^q$ and $m^{-1}e^{-q}$ in each of the ISGs and further implies that the twist defects persist throughout the dynamics. 

To define such a string operator for $me^q$ along a clockwise-oriented, contractible path $\gamma$, we take a product of all of the plaquette stabilizers $S_p$ enclosed by $\gamma$. The operators in the bulk cancel, and we are left with a clockwise-oriented string operator for $me^q$ along $\gamma$. This is indeed a string operator for $me^q$, as can be checked by considering it as a string operator in the instantaneous $\ZZ_N$ toric codes. Furthermore, it commutes with all of the checks, since the stabilizers $S_p$ belong to each of the ISGs. To create a string operator for $me^q$ along an oriented, open path $\gamma$, we first find an extension of $\gamma$ to a contractible loop with a clockwise orientation, then we define the string operator on the loop by taking the product of $S_p$ operators enclosed, and finally truncate the string operator to $\gamma$.

\begin{figure}[t]
\centering
\includegraphics[width=.23\textwidth]{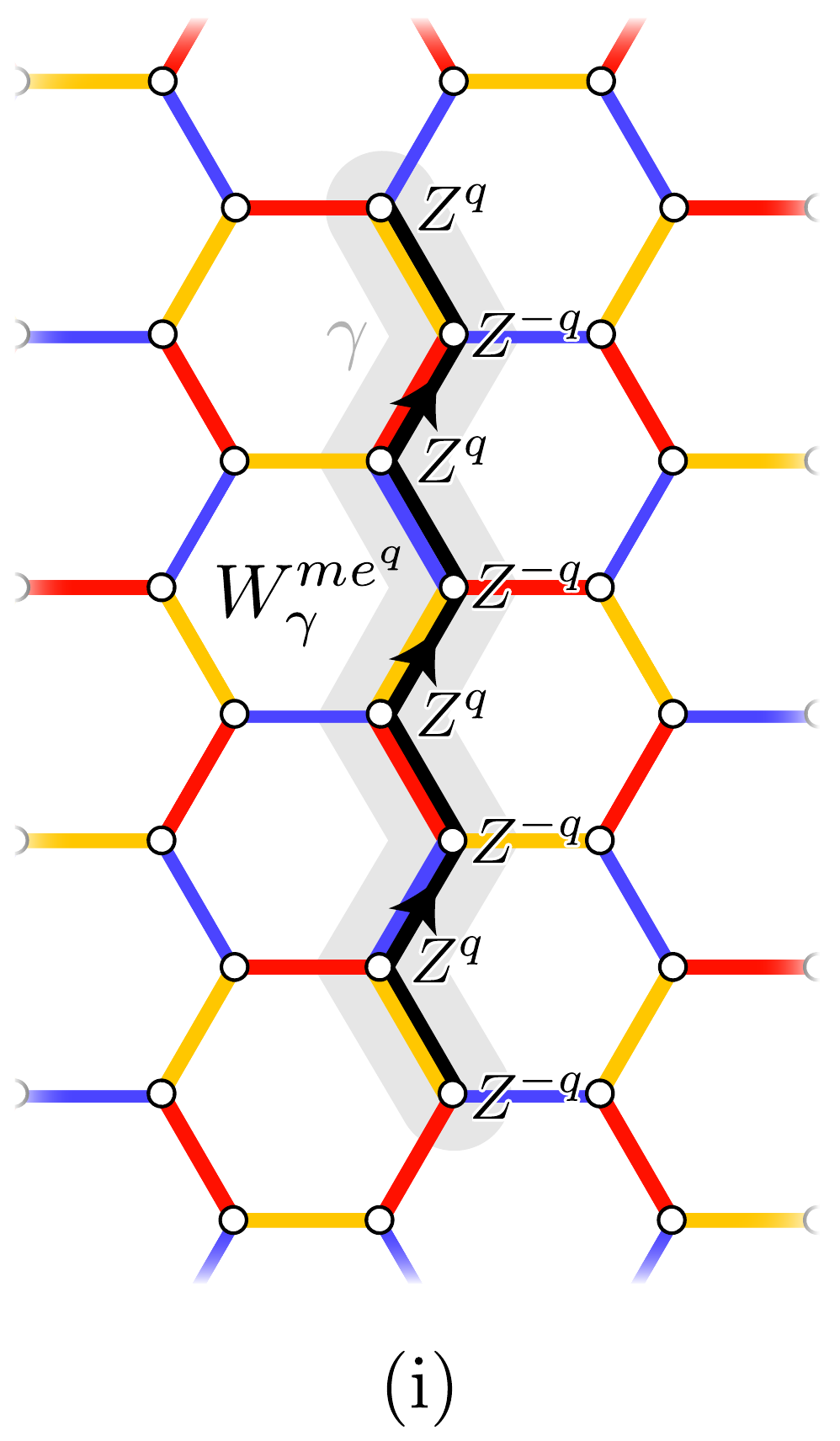}
\raisebox{.683cm}{\rule{.5pt}{5.545cm}}
\includegraphics[width=.23\textwidth]{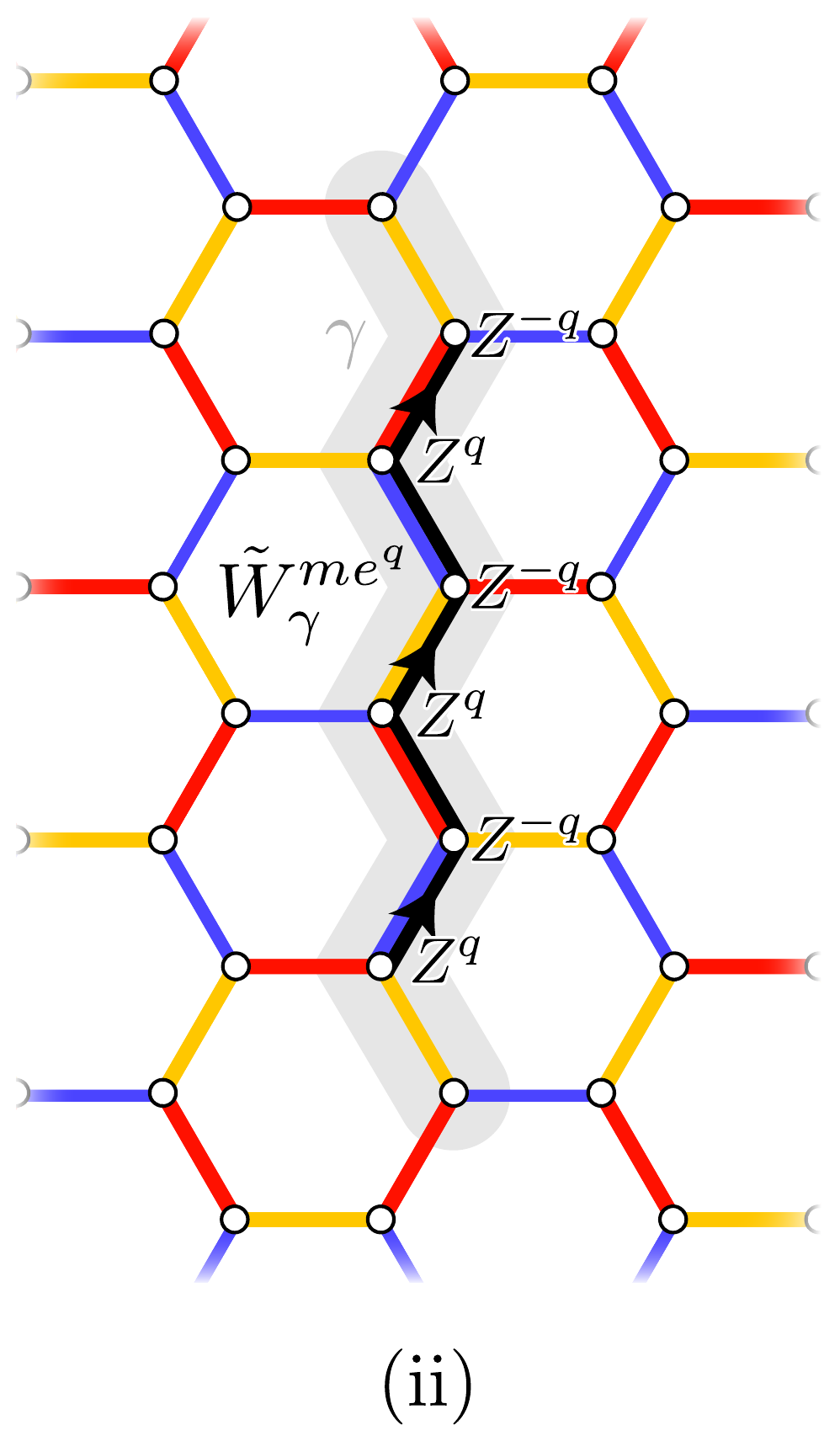}
\raisebox{.683cm}{\rule{.5pt}{5.545cm}}
\includegraphics[width=.23\textwidth]{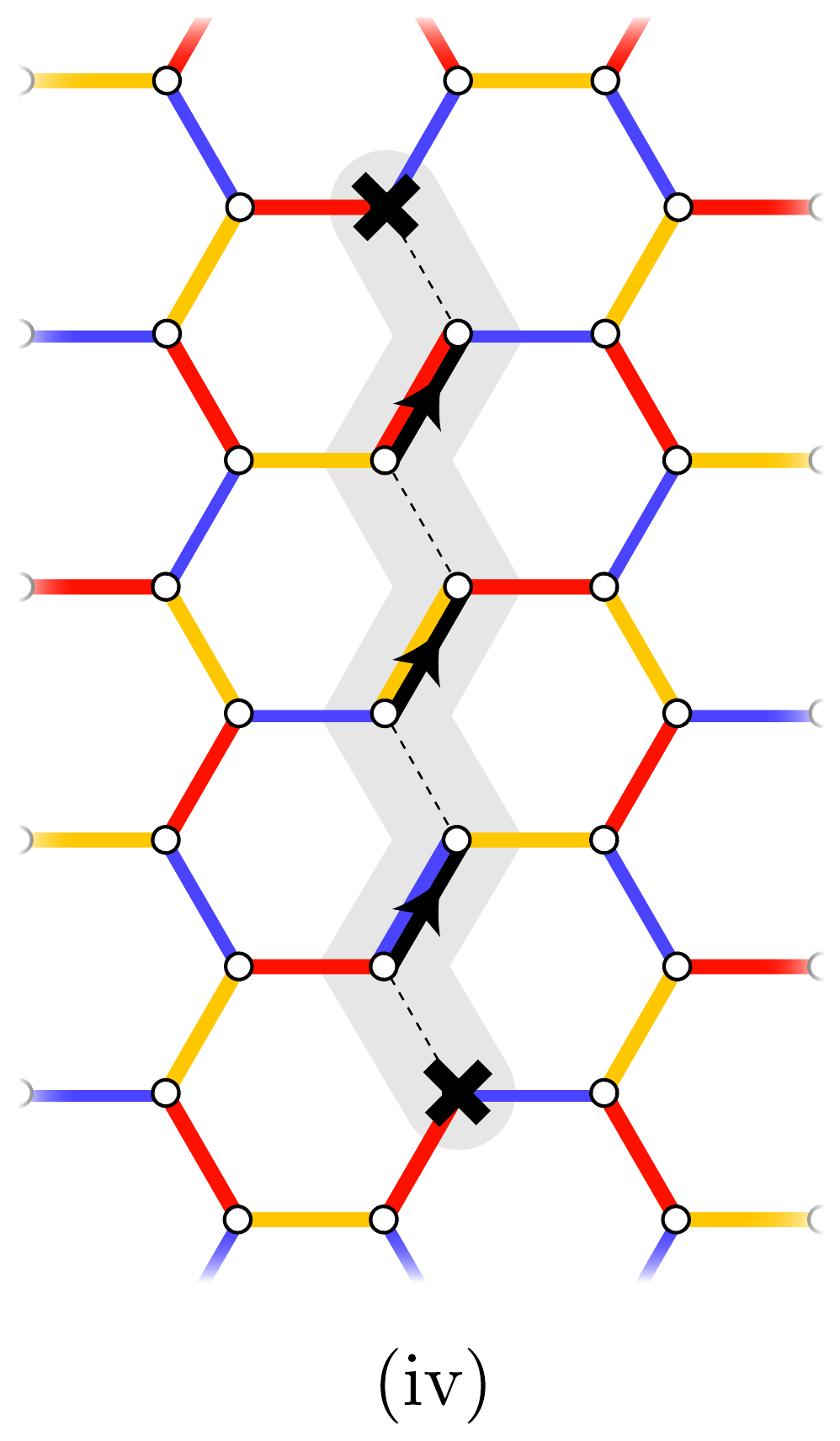}
\raisebox{.683cm}{\rule{.5pt}{5.545cm}}
\includegraphics[width=.23\textwidth]{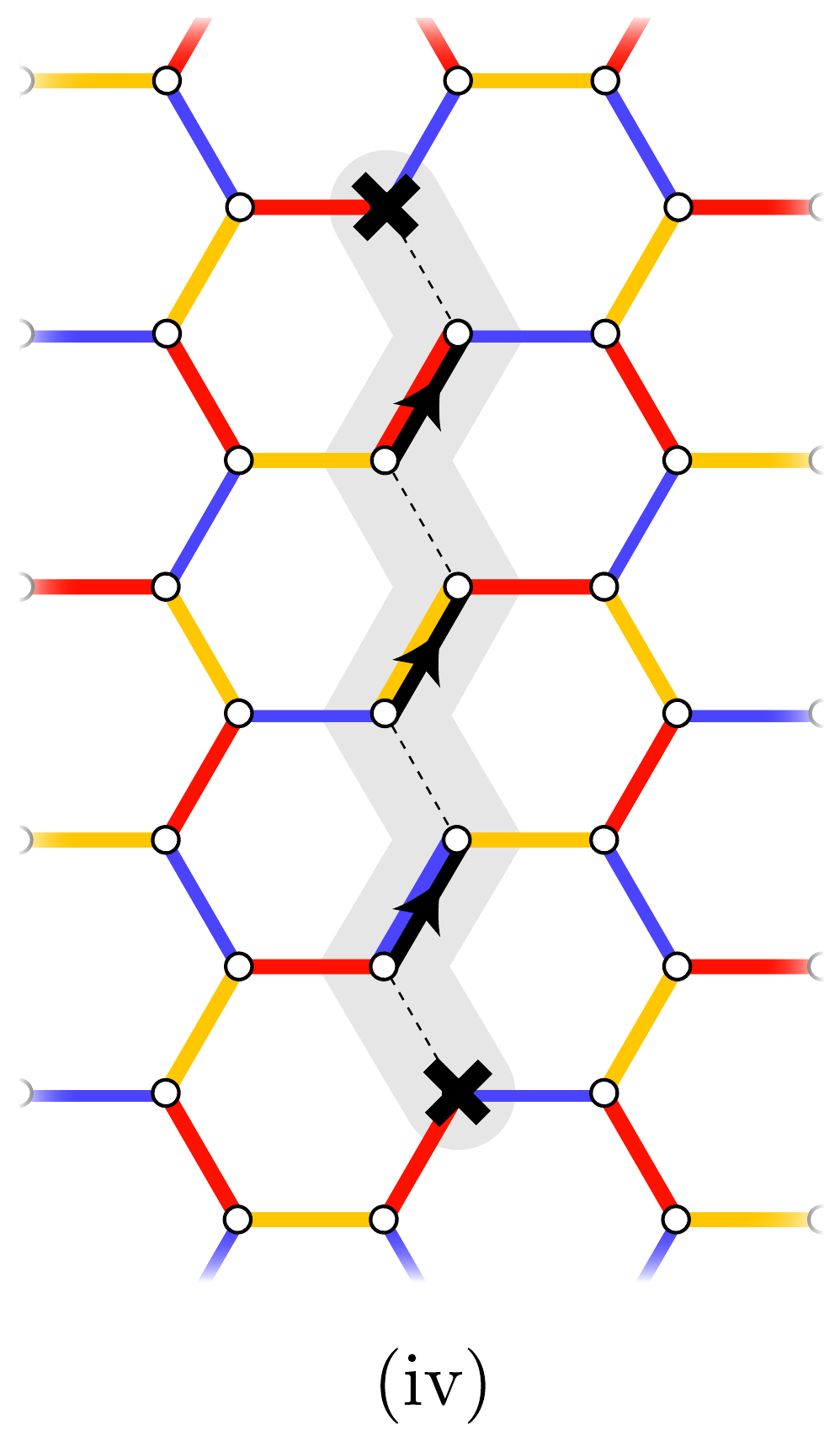}
\caption{The procedure for inserting a defect line along an oriented, open path $\gamma$ of odd length (light gray) consists of four steps. (i) We define an $me^q$ string operator $W^{me^q}_\gamma$ along $\gamma$. (ii) We remove the two Pauli operators at the endpoints to define the string operator $\tilde{W}^{me^q}_\gamma$. (iii) We decompose $\tilde{W}^{me^q}_\gamma$ into 2-body short string operators. (iv) We take the 2-body short string operators to be the defect checks (bold black) and remove all of the check operators that fail to commute with the defect checks (dashed lines). The twist defects (bold black crosses) live at the endpoints of $\gamma$.}
\label{fig: ZN defect construction}
\end{figure}

We are now prepared to create a defect line along an oriented path $\gamma$ of odd length by modifying the set of check operators along $\gamma$. The procedure for creating a defect line is detailed in Fig.~\ref{fig: ZN defect construction}. In summary, we define a new set of check operators, called the defect checks, which can be interpreted as short string operators that create an $me^q$ anyon and an $m^{-1}e^{-q}$ anyon. To ensure that the defect checks belong to all of the ISGs, we remove all check operators that fail to commute with the defect checks. The remaining $r$-checks are referred to as the $r^*$-checks.

With this new set of check operators along $\gamma$, the measurement schedule is modified to:
\begin{align} \label{eq: ZN defect measurement schedule}
    [2,0,1,2]\underbrace{(0,1,2)\ldots(0,1,2)}_{d-1}(\tilde{0}^*,1^*,2^*)(\tilde{0}^*,1^*,2^*)\ldots
\end{align}
Note that the first four rounds of measurements and the subsequent $d-1$ periods are needed to fault-tolerantly initialize the code. The notation $\tilde{0}^*$ denotes the simultaneous measurement of the defect checks and the $0^*$-checks, while $1^*$ and $2^*$ denote the measurements of the $1^*$- and $2^*$-checks. The ISG after the $r^*$-checks is given by:
\begin{eqs} 
    &\mathcal{S}_{r^*} \equiv  \\
    &\Bigg \langle 
        \vcenter{\hbox{\includegraphics[scale=.5]{Figures/ZN_Sp_no_p.pdf}}}, \,\,\,
        \underbrace{\vphantom{ \left(\frac{{{{{{{{{{{a^5}}^5}^5}^5}^5}^5}^5}^5}^5}^{.5}}{5}\right) }\vcenter{\hbox{\includegraphics[scale=.6]{Figures/ZN_x_check.pdf}}}, \,\,\,
        \vcenter{\hbox{\includegraphics[scale=.6]{Figures/ZN_y_check.pdf}}}, \,\,\,
        \vcenter{\hbox{\includegraphics[scale=.6]{Figures/ZN_z_check.pdf}}}}_{\text{$r^*$-checks}}, \,\,\,
        \underbrace{\vcenter{\hbox{\includegraphics[scale=.6]{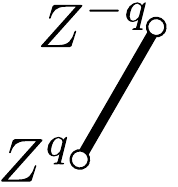}}}, \,\,\, 
        \vcenter{\hbox{\includegraphics[scale=.47]{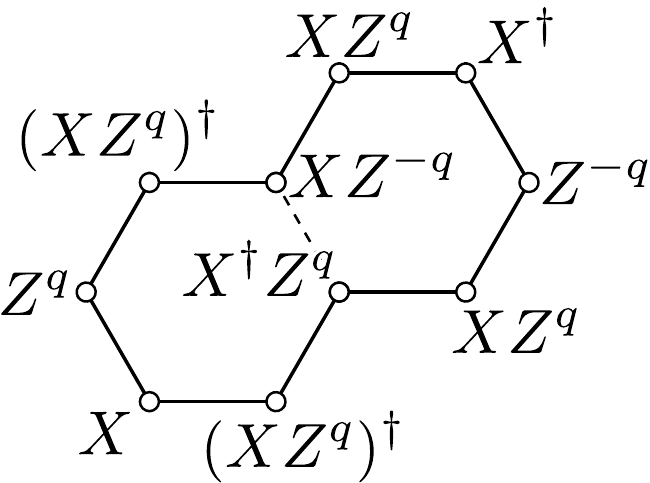}}}, \,\,\,
        \vcenter{\hbox{\includegraphics[scale=.5]{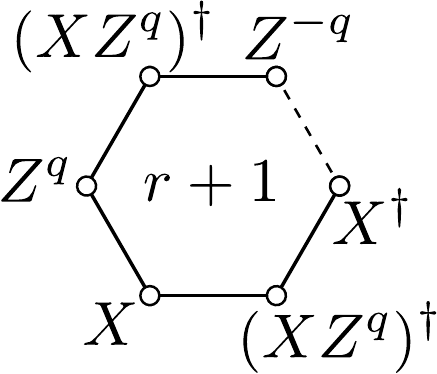}}}}_{\text{Along defect line}}
    \Bigg \rangle.
\end{eqs}

\begin{figure*}[t]
\centering
\subfloat[\label{fig: ZN em permute a}]{\includegraphics[width=.245\textwidth]{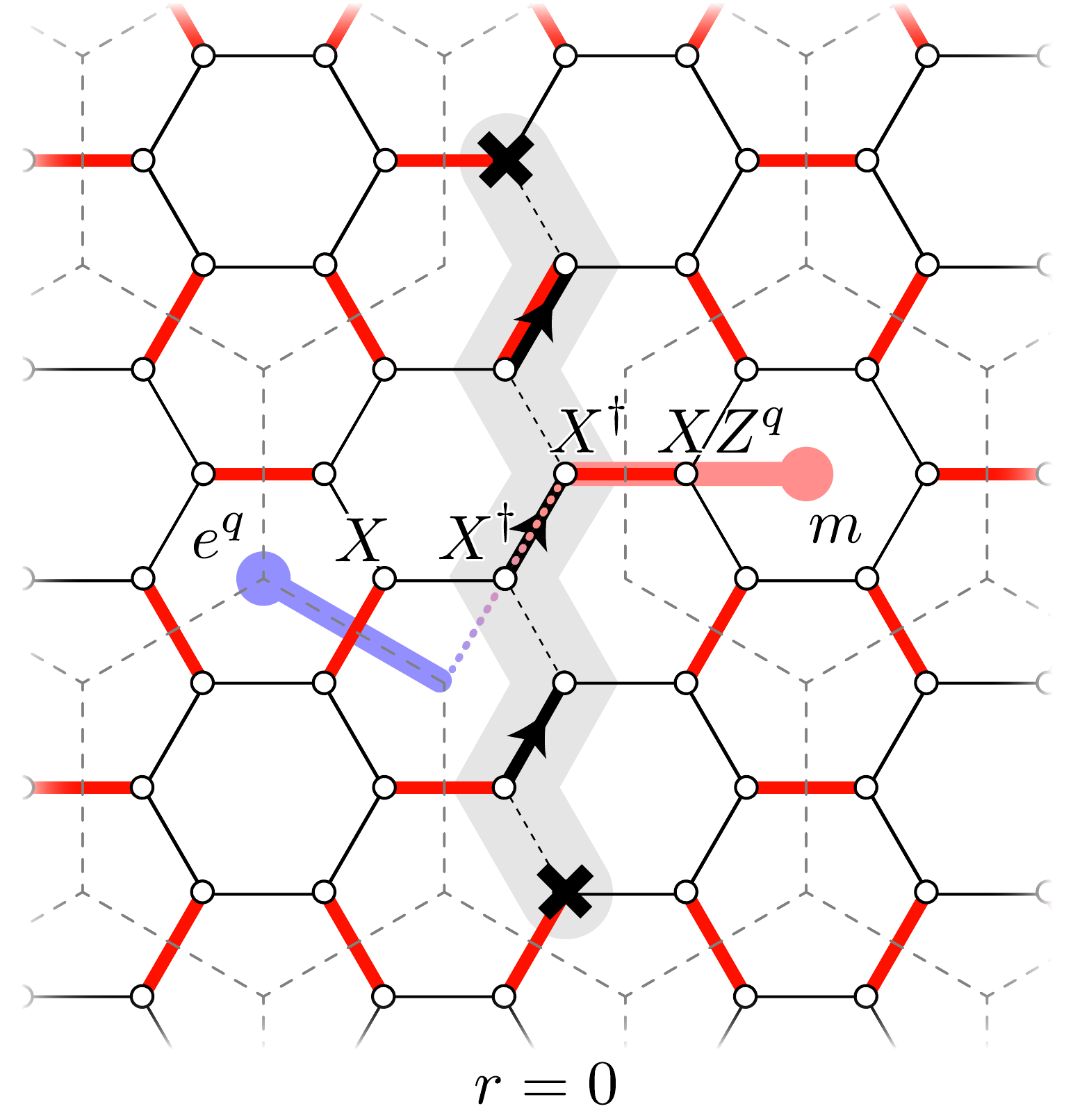}}\,
\raisebox{.25cm}{\rule{.5pt}{3.76cm}}
\subfloat[\label{fig: ZN em permute b}]{\includegraphics[width=.245\textwidth]{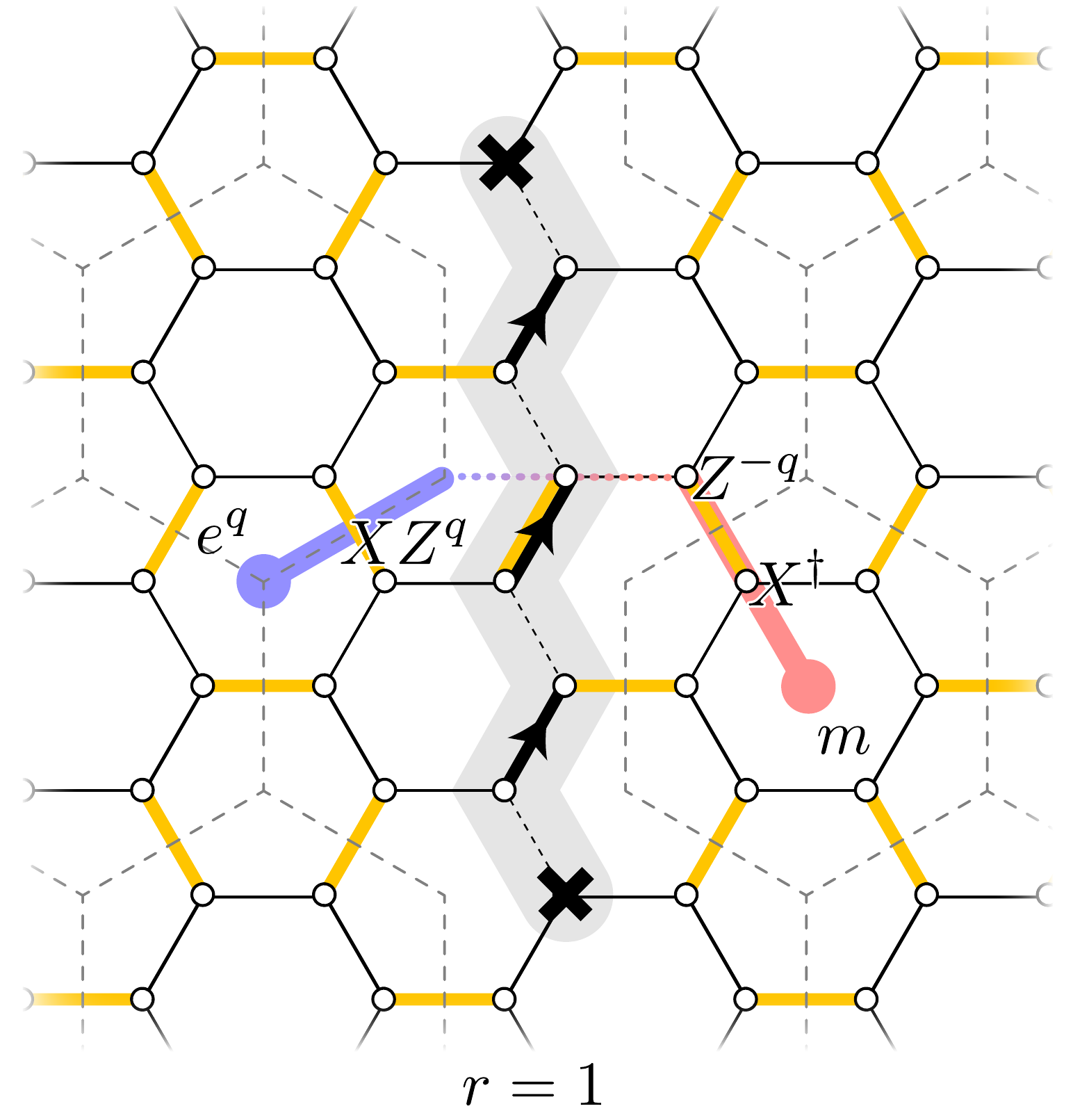}}\,
\raisebox{.25cm}{\rule{.5pt}{3.76cm}}
\subfloat[\label{fig: ZN em permute c}]{\includegraphics[width=.245\textwidth]{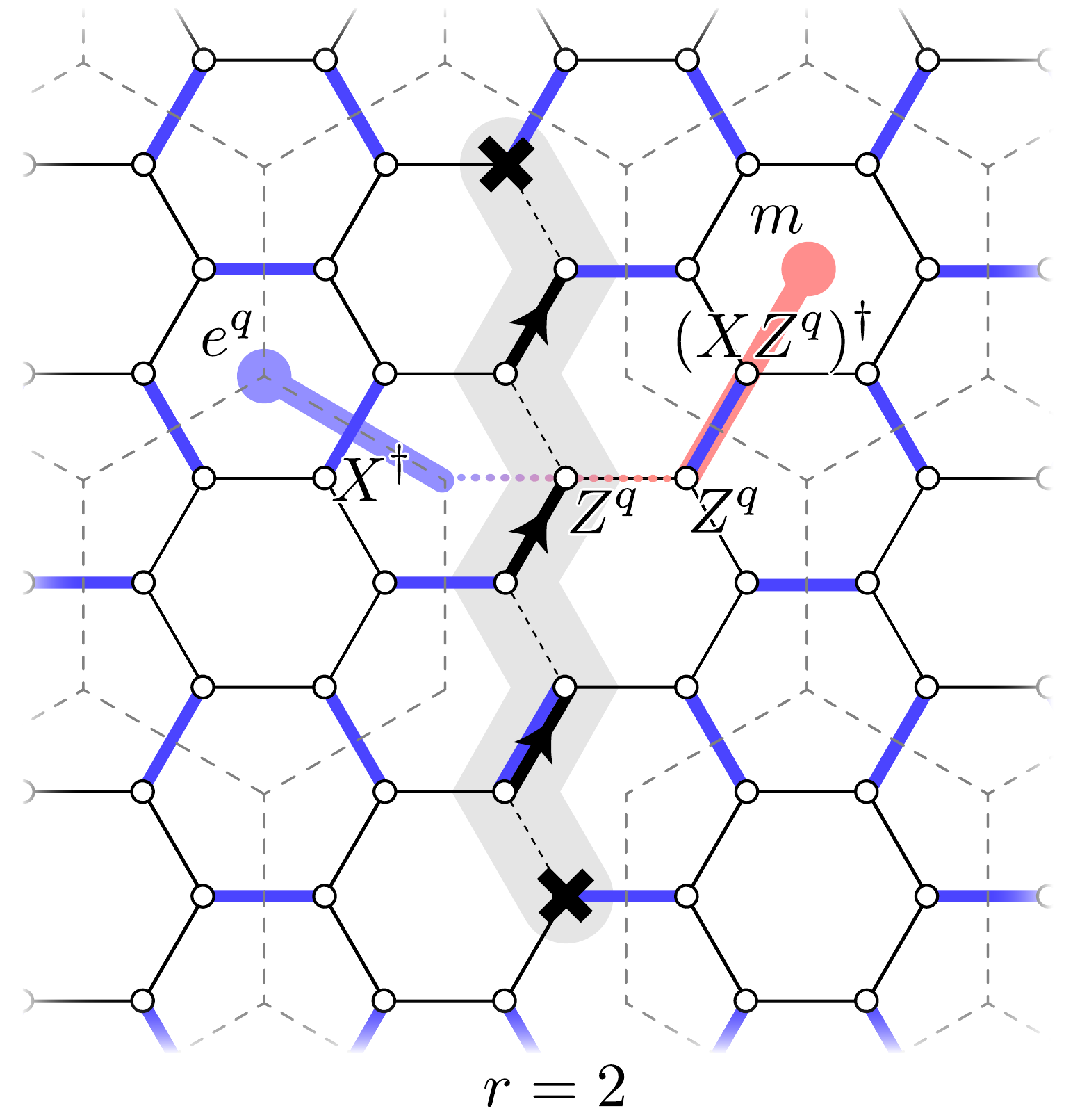}}\,\,
\subfloat[\label{fig: meq condense}]{\raisebox{.18cm}{\includegraphics[width=.2\textwidth]{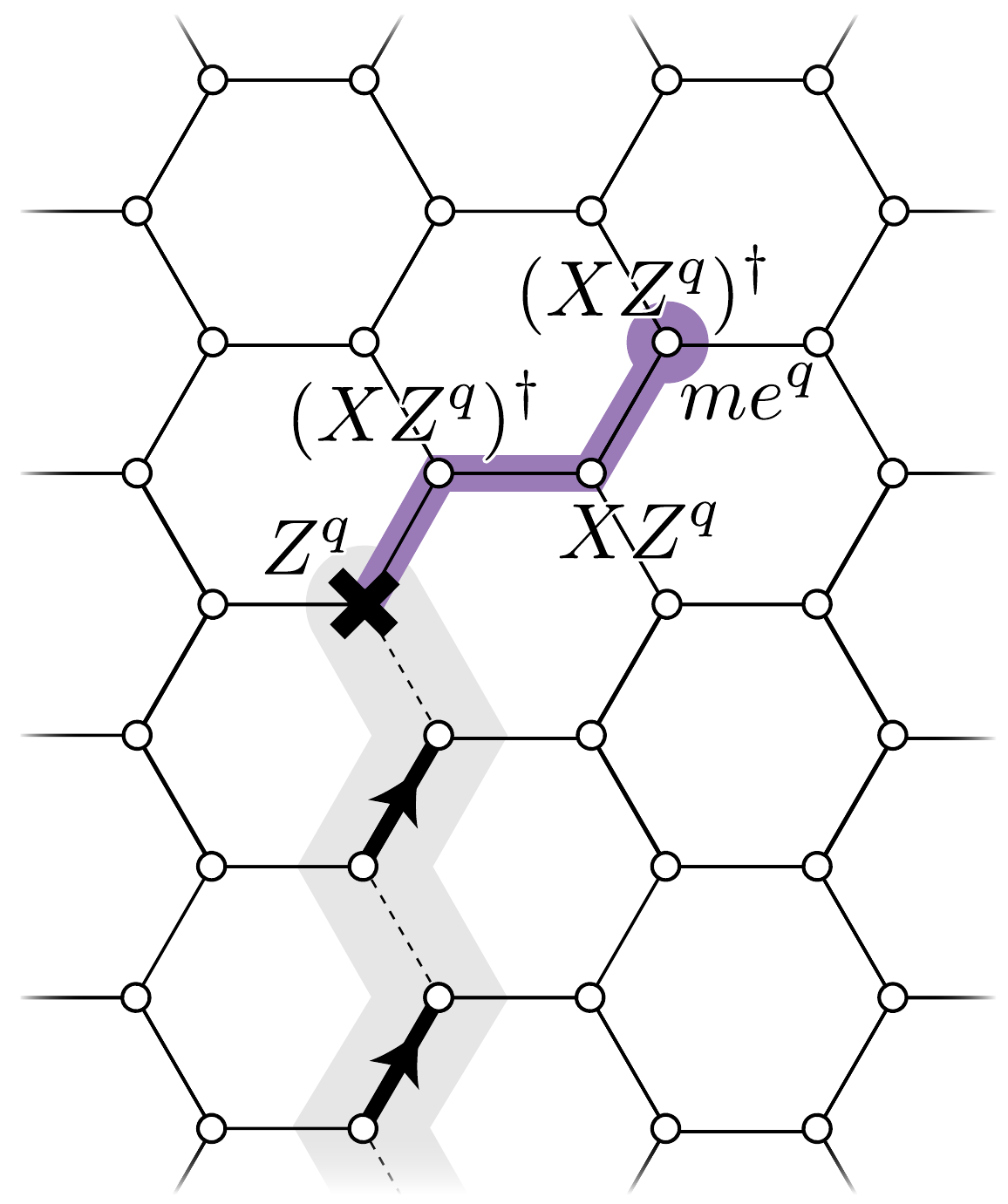}}}
\caption{(a-c) For each ISG, there is a defect line along the path $\gamma$ (light gray). In particular, an $e^{-q}$ anyon is transformed into an $m$ anyon when moved across the defect line. This is evidenced by the fact that there is a string operator (light blue and light red) that creates $e^q$ on one side of the defect line and $m$ on the other side, while commuting with the instantaneous stabilizers along its length.
(b) The endpoints of the defect lines host twist defects (bold black crosses). An $m^{-1}e^{-q}$ anyon can condense on the twist defect, since there is a string operator (purple) for $me^q$ that can terminate on the twist defect without violating any of the instantaneous stabilizers near the twist defect. 
}
\label{fig: ZN defect line properties}
\end{figure*}

In Figs.~\ref{fig: ZN em permute a}-\ref{fig: ZN em permute c}, we confirm that the ISGs have a defect line along $\gamma$, by identifying string operators that create an $e^q$ anyon to the left of the defect and an $m$ to the right. This implies that an $e^{-q}$ anyon is transformed into an $m$ anyon when it is moved across the defect line. It further implies that the anyon $me^q$ (or its powers) can be condensed at the endpoint of the defect line, as illustrated in Fig.~\ref{fig: meq condense}. 

To store logical information, we need to insert more than one defect line. In general, $k$ defect lines (with endpoints) encode $k-1$ logical qudits of dimension $N$. The logical operators can be represented by $me^q$ string operators that connect between twist defects, in analogy with with Fig.~\ref{fig: multiple logicals}. 

\subsection{Twisted quantum double Floquet codes} \label{sec: TQD Floquet codes}

We now use the $\ZZ_N$ Floquet codes to outline a construction of Floquet codes whose ISGs have the same topological order as certain Abelian TQDs. The Floquet codes constructed below correspond, in particular, to a subclass of type I TQDs. Type I TQDs are classified by two pieces of data: (i) a group $\ZZ_N$, and (ii) an integer $n$ in $\ZZ_N$. We consider the subclass for which $n$ is coprime to $N^2$, i.e., there exists an integer $p$ such that $pq = 1 \text{ mod } N^2$. The anyons of type I TQDs are generated by a gauge charge $c$ and an elementary flux $\varphi$ with the following fusion rules:
\begin{align} \label{eq: type I fusion}
    c^N =1, \qquad \varphi^N = c^{2n}.
\end{align}
Braiding a gauge charge around an elementary flux produces the Aharonov-Bohm phase:
\begin{align} \label{eq: type I braiding}
    B(c,\varphi) = e^{2\pi i/ N}.
\end{align}
The gauge charge $c$ is a boson, while the elementary flux $\varphi$ can have more exotic exchange statistics -- in general, exchanging two gauge charges or two elementary fluxes produces the phase:
\begin{align} \label{eq: type I statistics}
    \theta(c) = 1, \qquad \theta(\varphi) = e^{2\pi i/ N^2}.
\end{align}

In Ref.~\cite{Ellison2022stabilizer}, it was argued that a type I TQD characterized by $\ZZ_N$ and $n \in \ZZ_N$ can be created from a $\ZZ_{N^2}$ toric code by condensing the following boson:
\begin{align} \label{eq: type I boson}
    b \equiv m^{-N} e^{Nn}. 
\end{align}
This has the effect of confining certain anyons of the $\ZZ_{N^2}$ toric code, so that the remaining deconfined anyons are precisely those of the desired type I TQD. For example, the double semion (DS) topological order, characterized by $\ZZ_2$ and $n=1$, can be constructed from a $\ZZ_4$ toric code by condensing the boson $b=m^2e^2$. The deconfined anyons are generated by the $e^2$ and $em$ anyons of the $\ZZ_4$ toric code. Modulo the boson $b$, these anyons satisfy the relations in Eqs.~\eqref{eq: type I fusion}-\eqref{eq: type I statistics} for the DS topological order.

\begin{figure*}[t]
\centering
\subfloat[\label{fig: confined and deconfined}]{\includegraphics[width=.35\textwidth]{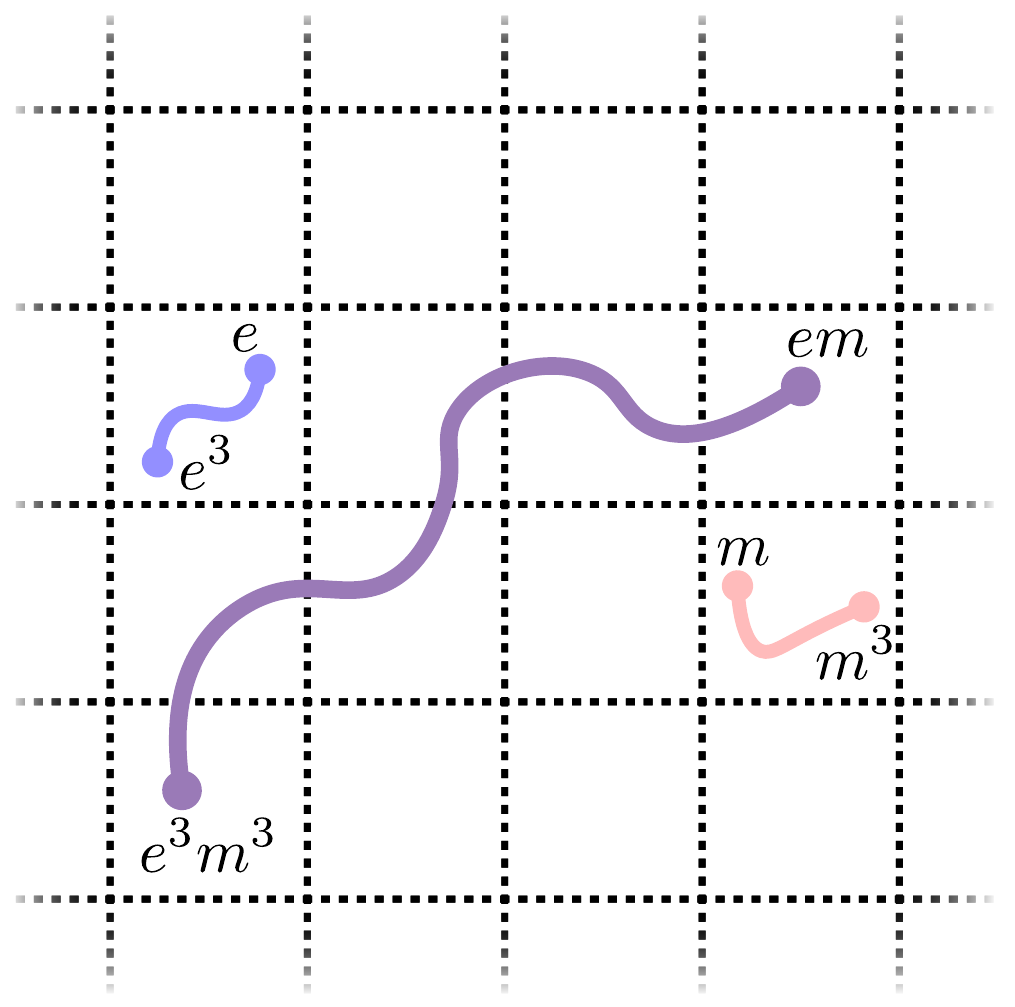}} \qquad
\subfloat[\label{fig: intersecting defect lines}]{\includegraphics[width=.45\textwidth]{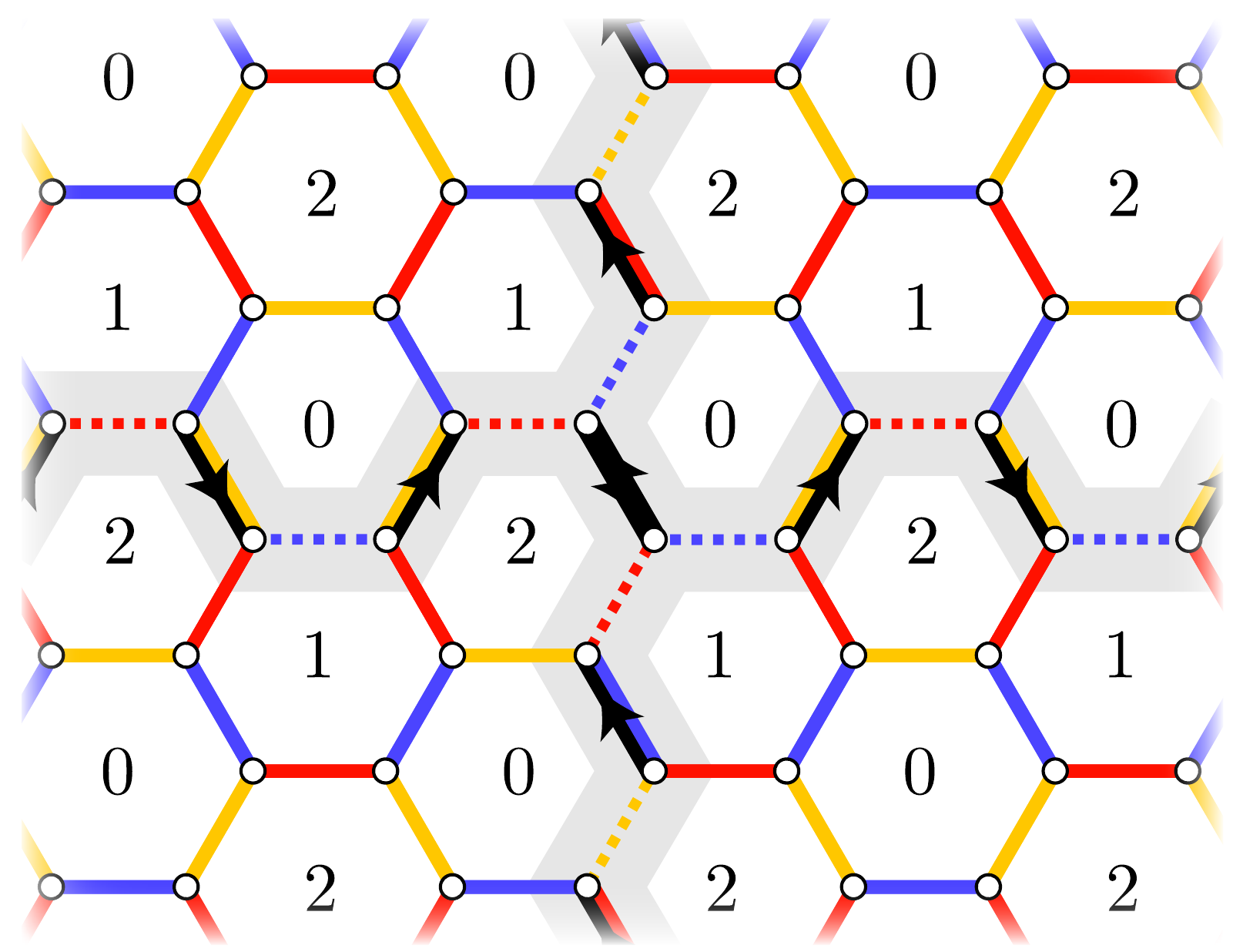}}
\caption{(a) We create a grid of defect lines (dashed lines) by condensing $b$. The spacing between the defect lines is assumed to be large compared to the lattice spacing of the underlying hexagonal lattice. Given our choice of $b$, the defect lines are semi-permeable, meaning that some anyons become confined, and others are able to pass through the defect lines. In the case of the construction of a Floquet code characterized by the DS topological order, the $e$ and $m$ anyons (light blue, and light red) of the $\ZZ_4$ toric code are confined, while the $em$ anyons (purple) are deconfined. (b) To ensure that the defect checks belong to each of the ISGs, and hence the boson $b$ has been condensed along the path, we remove the check operators that fail to commute with the defect checks. In the case of the $\ZZ_N$ Floquet codes in the previous section, this entails removing all of the checks on an edge. For the defect checks (bold black) that create $b$ anyons, there may be powers of checks that still commute with the defect checks (dashed colored lines). For the example of the DS Floquet code, the squares of the check operators are still measured along the defect lines.
The defect lines can be inserted so that, at their intersections, two defect checks share the same edge. The defect checks of the boson are mutually commuting. }
\label{fig: TQD construction}
\end{figure*}

Alternatively, instead of condensing $b$ over the entire system, the boson can be condensed in a grid-like pattern, as in Fig.~\ref{fig: confined and deconfined}. This creates a grid of semi-permeable (i.e., non-invertible) defect lines, where only the anyons that have trivial braiding relations with $b$ are able to cross the grid lines. The anyons of the $\ZZ_{N^2}$ toric code that have nontrivial braiding relations with $b$ become confined, since they are unable to pass through the defect lines. The deconfined anyons -- i.e., those that can pass through the semi-permeable defect lines -- then generate the type I TQD characterized by $\ZZ_N$ and $n\in \ZZ_N$. 

This gives a prescription for creating a Floquet code whose ISGs have the same topological order as a type I TQD characterized by $\ZZ_N$ and $n$ coprime to $N^2$. In particular, we start with the $\ZZ_{N^2}$ Floquet code with the property that $b$ in Eq.~\eqref{eq: type I boson} is invariant under the dynamics. This is the case for $q=-n$ and an arbitrary choice of $p$ satisfying $pq = 1 \text{ mod } N^2$. The automorphism maps $m \to e^q$ and $e \to m^p$, so that $b$ is invariant:
\begin{align}
    m^{-N}e^{Nn} \to m^{Nnp}e^{-Nq} = m^{-N}e^{Nn},
\end{align}
where on the right-hand side, we have used that $q=-n$ and $pq = 1  \text{ mod } N^2$. We then create a coarse-grained grid of defect lines (on a scale that is large compared to the lattice spacing of the hexagonal lattice) by condensing $b$, as shown in Fig.~\ref{fig: confined and deconfined}. The intersection of two grid lines is shown in Fig.~\ref{fig: intersecting defect lines}. 

We claim that each of the ISGs is equivalent to a $\ZZ_{N^2}$ toric code with a grid of defect lines. Since the deconfined anyons are those of the desired TQD (modulo $b$), this suggests that we have constructed Floquet codes with ISGs characterized by the topological order of type I TQDs with $n$ coprime to $N^2$. As an example, we can construct a Floquet code with ISGs characterized by the DS topological order by starting with a $\ZZ_4$ Floquet code and creating a grid of defect lines by condensing the boson $b = m^2e^2$. We note that an alternative construction of a Floquet code characterized by the DS topological order was recently presented in Ref.~\cite{bauer2023topological}.

\section{Discussion} \label{sec: discussion}

In this work, we have constructed twist defects in the $\ZZ_2$ Floquet code of Ref.~\cite{HH_dynamic_2021}~by~condensing emergent fermions along paths. We have argued that the twist defects yield~dynamically generated logical qubits, which can be used to fault-tolerantly store and process quantum information. We further showed that, by constructing twist defects on a system with boundary, we obtain a planar variant of the $\ZZ_2$ Floquet code. This construction~of a planar $\ZZ_2$ Floquet code has advantages over the construction in Ref.~\cite{Haah2022boundarieshoneycomb}, since the twist defects do not require any modifications to the connectivity of the qubits, and the measurement schedule consists of only three rounds of 2-body measurements. Finally, we described $\ZZ_N$ Floquet codes, whose ISGs have the same topological order as $\ZZ_N$ toric codes, and generalized the construction of twist defects to these codes. Using defect lines in the $\ZZ_N$ Floquet codes, we built Floquet codes with ISGs characterized by the topological orders of certain type I TQDs. We now conclude by commenting on potential avenues for future work. 

Firstly, given that the $\ZZ_2$ Floquet code consists of making 2-body measurements on a hexagonal lattice, it is well suited for IBM's quantum processors, where the qubits are connected according to a heavy hexagonal lattice.~Notably, the heavy hexagonal lattice features qubits on the edges, which can be used as ancilla for the 2-body measurements. In Ref.~\cite{Wootton2022Floquet}, a first demonstration of the $\ZZ_2$ Floquet code on the IBM Quantum hardware was given by implementing 7 rounds of the measurement schedule. Our construction of twist defects requires only mild changes to the measurement schedule, suggesting that it too can be implemented on one of IBM's quantum processors. We look forward to discussing an experimental realization of our twist defects. 

Furthermore, it would also be valuable to have a more in-depth analysis of the spacetime overheads of the $\ZZ_2$ Floquet code with twist defects.~An important next step in this direction is to assess the error thresholds and sub-threshold scaling for the code with twist defects. Error thresholds have already been computed for the $\ZZ_2$ Floquet code in the presence of circuit-level noise~\cite{Gidney2021faulttolerant, Paetznick2023Performance, Gidney2022benchmarkingplanar}. We expect that the error thresholds with twist defects will be comparable to the high error thresholds of the $\ZZ_2$ Floquet code without defects. We also expect that there is room to optimize the computational scheme described in Section~\ref{sec: Z2 defect gate set}. We leave this analysis to future work.

In our construction of the twist defects, we relied on the fact that the checks of the $\ZZ_2$ Floquet code have an underlying 1-form symmetry~\cite{Qi2021higherform,Ellison2022subsystem} -- that is, they commute with the emergent fermion string operator supported on closed paths. This guaranteed that, for each instantaneous stabilizer code, the emergent fermions could be condensed with a single choice of defect checks.
Recently, it has been discovered that there are Floquet codes without any underlying 1-form symmetries (corresponding to anyons)~\mbox{\cite{kesselring2022anyon, davydova2022floquet}}. This suggests that new methods are needed to insert twist defects in these Floquet codes. One possibility is to construct a non-invertible defect with a rough boundary on one side and a smooth boundary on the other. This type of defect has many of the same computational properties as the (invertible) twist defects described in this work. Another possibility is to consider the~spacetime perspective on Floquet codes introduced in Refs.~\cite{Williamson2022spacetime, bombin2023unifying, bauer2023topological} and develop methods for inserting defect membranes in spacetime, potentially by proliferating loops of fermion worldlines on a membrane. Yet another possibility comes from stacking multiple copies of the Floquet codes on top of one another. This introduces a (0-form) symmetry corresponding to swapping layers. Layer exchange defects can then be created by redefining the checks along a branch cut so that they connect between the different layers. This approach, however, would require increasing the qubit overhead and changing the connectivity of the qubits. 

As noted in Ref.~\cite{HH_dynamic_2021}, the check operators of the $\ZZ_2$ Floquet code can be interpreted as the gauge generators of a subsystem code. Likewise, the checks of the $\ZZ_N$ Floquet codes can be interpreted as gauge generators for the subsystem codes described in Refs.~\cite{Barkeshli_2015,Ellison2022subsystem}. Similar to the procedures above, it is possible to insert twist defects into the subsystem codes. In general, a pair of twist defects in the subsystem code encodes an $N$-dimensional qudit if $N$ is odd, and an $(N/2)$-dimensional qudit if $N$ is even. This implies that, the twist defects of the $\ZZ_N$ Floquet codes yield dynamically generated logical qudits only for even $N$, and hence, the dynamics are not essential to storing quantum information in twist defects for odd $N$. We hope to discuss the twist defects of the $\ZZ_N$ subsystem codes for odd $N$ in more detail in future work.

The underlying subsystem code of the $\ZZ_2$ Floquet code is based on Kitaev's honeycomb model in Ref.~\cite{Kitaev_2006}. This model has been generalized to three spatial dimensions, such as in Ref.~\cite{Mandal_2009}. The 3D generalization in Ref.~\cite{Mandal_2009} exhibits a $2$-form symmetry that is generated by loops of emergent fermion string operators. This suggests that, with an appropriate measurement schedule, the ISGs are described by a 3D topological order with an emergent fermion~\cite{Levin2003Fermions, Yuan2019bosonization3D, Yuan2021disentangling}. Similar to the construction presented in this work, twist defects can be inserted, at least in principle, by condensing emergent fermions. The potential advantage over the 2D Floquet code is that a stack of four copies\footnote{Assuming that the 3D topological order is that of a 3D toric code with an emergent fermion, we can redefine the point-like excitations by binding the emergent fermion from one copy to the emergent fermions of the other three copies. This converts the three copies into three copies of the 3D toric codes with an emergent boson, which admits a transversal $CCZ$ gate~\cite{yoshida2017boundaries, Vasmer2019transversal}.} of this 3D topological order with an emergent fermion potentially admits a fault-tolerant non-Clifford gate~\cite{yoshida2017boundaries,yuan2022highergroup,yuan2023codimension2},
thus warranting future investigation.

\vspace{0.2in}
\noindent{\it Acknowledgements -- } We would like to acknowledge Nathanan Tantivasadakarn for useful discussions about a related upcoming work on Floquet codes.
T.D.E. thanks Yu-An Chen and Meng Cheng for the conversations about condensation defects that inspired this work. T.D.E. is also grateful to Julio Carlos Magdalena de la Fuente and Dominic J. Williamson for valuable discussions about Floquet codes and the defect network construction of TQDs. J.S. would like to thank Andrew Potter and Rui Wen for helpful conversations related to this and overlapping work.
J.S. is supported by DOE DE-SC0022102.
A.D. is supported by the Simons Foundation through the collaboration
on Ultra-Quantum Matter (651438, AD) and by the Institute for Quantum
Information and Matter, an NSF Physics Frontiers Center (PHY-1733907).

\appendix

\section{$\ZZ_2$ toric code with condensation defects} \label{app: TC condensation defects}

Here, we demonstrate that the twist defects of the $\ZZ_2$ toric code can be constructed~by condensing emergent fermions along open paths. A similar approach was taken in Ref.~\cite{pandey2022topological} to construct defect lines along non-contractible paths. We note that this construction is derived from the concept of higher gauging in Ref.~\cite{Roumpedakis2022Higher}, where the defect lines are referred to as condensation defects.
We also note that this construction mirrors the construction of twist defects in Ref.~\cite{yuan2023codimension2}, where twist defects are constructed by first creating a 1D topological superconductor in a system of physical fermions and then mapping the physical fermions to emergent fermions via a bosonization duality~\cite{Chen2018bosonization}.

\begin{figure*}[t]
\centering
\subfloat[\label{fig: Wpsi string}]{\includegraphics[width=.35\textwidth]{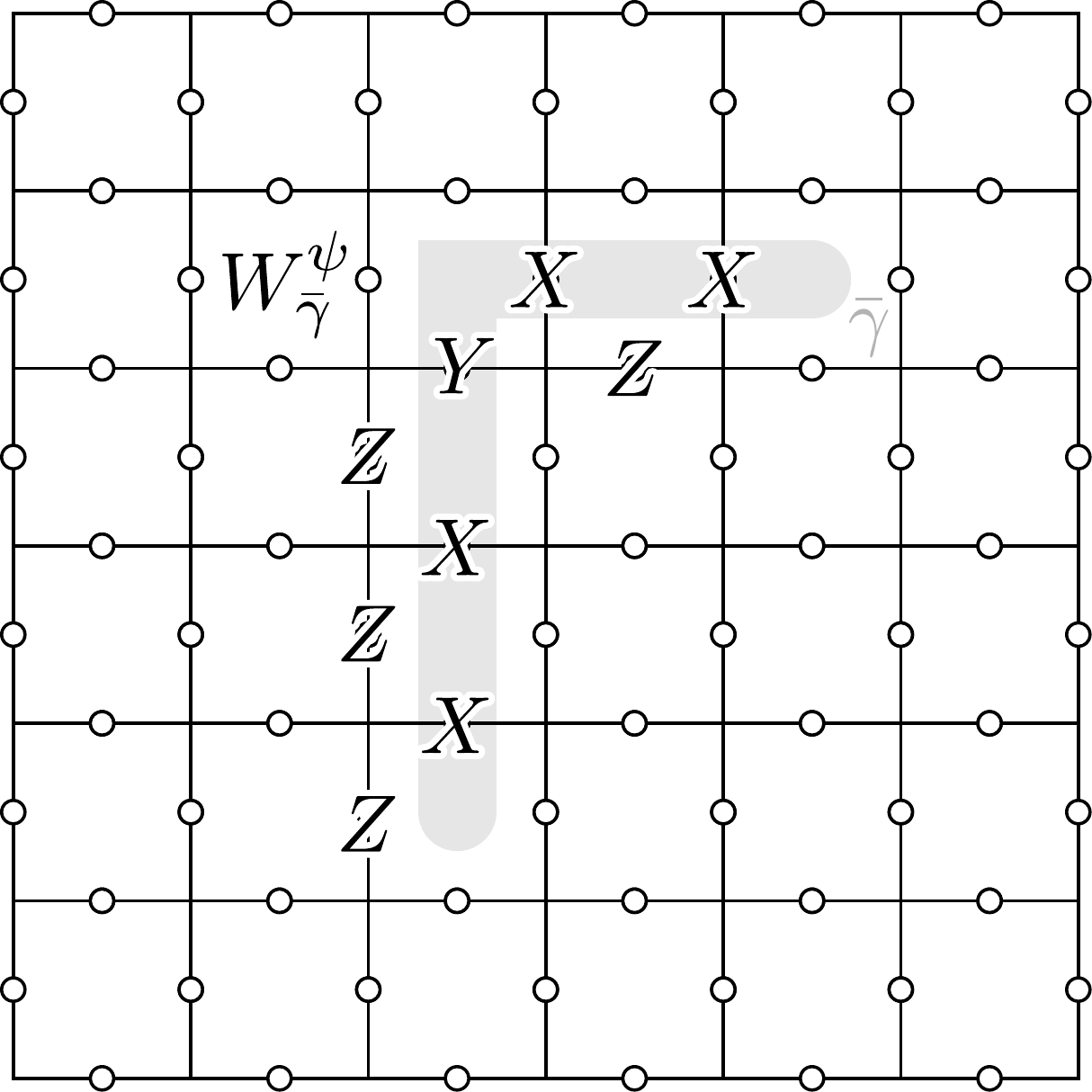}} \qquad
\subfloat[\label{fig: Wpsi string decomposition}]{\includegraphics[width=.35\textwidth]{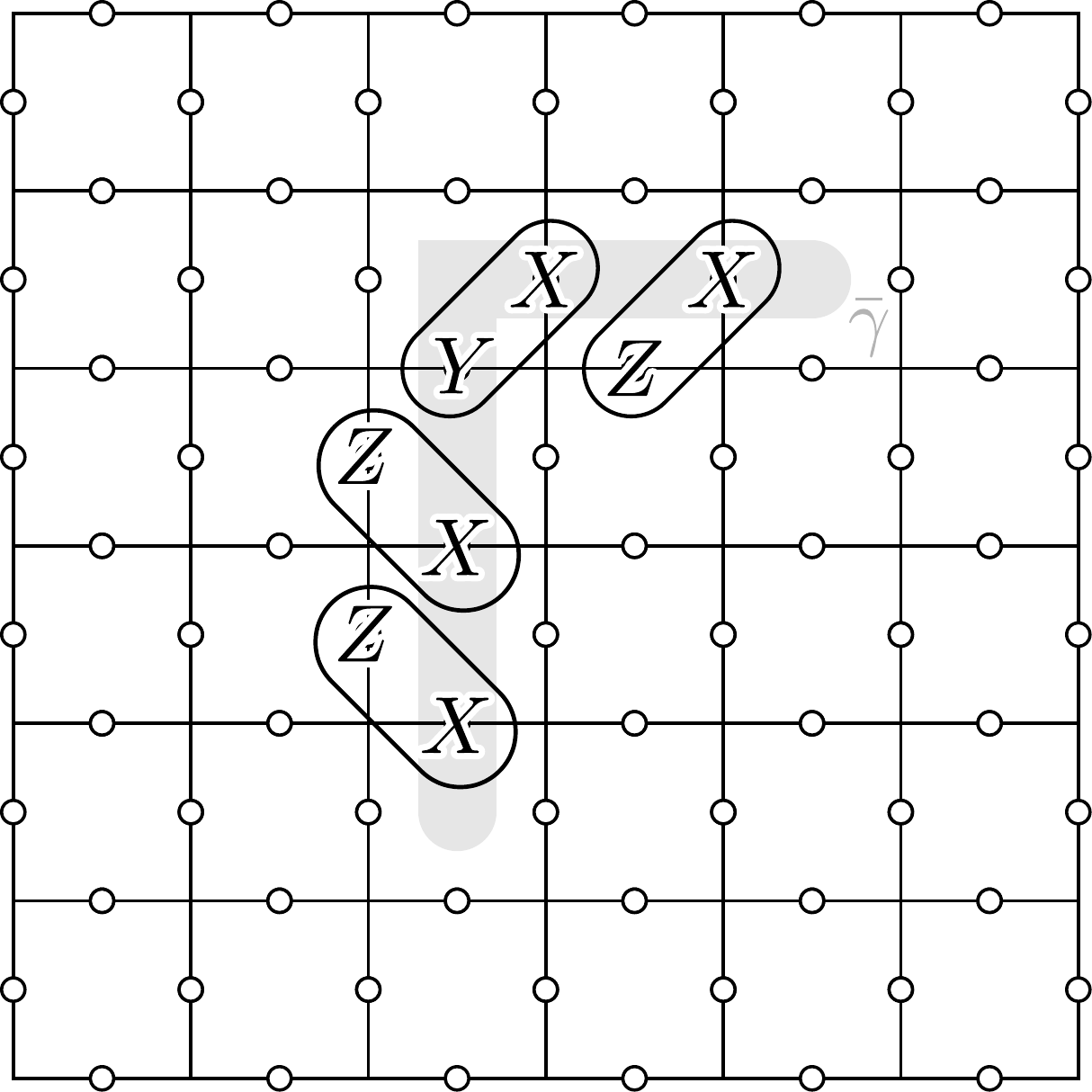}}
\caption{(a) The string operator $W_{\bar{\gamma}}^\psi$ is a product of short string operators $W_\ell^\psi$ in Eq.~\eqref{eq: Z2 TC short strings} along the path $\bar{\gamma}$ (light gray). 
(b) Up to Pauli operators at the endpoints, the string operator $W_{\bar{\gamma}}^\psi$ can be decomposed into mutually commuting 2-body short string operators that create $\psi$ anyons. Note that the short string operators $W_\ell^\psi$ are not mutually commuting in this case. To decompose the string operator into mutually commuting 2-body short string operators, we have paired $X$ with $Y$ at the corner, instead of $X$ and $Z$.}
\label{fig: TC Wpsi examples}
\end{figure*}

We consider a $\ZZ_2$ toric code defined on a square lattice with a qubit on each edge, as can be seen in Fig.~\ref{fig: TC Wpsi examples}. Although the construction generalizes straightforwardly to arbitrary 2D lattices, we consider a square lattice for concreteness. Without loss of generality, we also assume that the lattice is embedded in an infinite plane. We then take the stabilizer group $\mathcal{S}_\text{TC}$ to be generated by the vertex term $A_v$ and plaquette term $B_p$ given below:
\begin{align}
     A_v \equiv \vcenter{\hbox{\includegraphics[scale=.6]{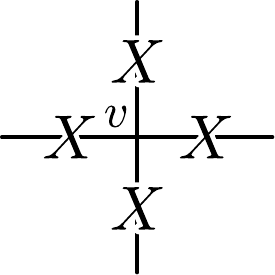}}}, \quad 
     B_p \equiv \vcenter{\hbox{\includegraphics[scale=.6]{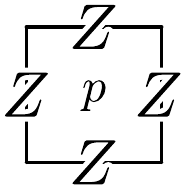}}}.
\end{align}

We label the anyons of the toric code by the elements of $\{1, e, m, \psi\}$. We take the violation of a vertex stabilizer to be an $e$ anyon, and the violation of a plaquette stabilizer $B_p$ to be an $m$ anyon. The violation of both a vertex stabilizer and a nearby plaquette stabilizer is then a $\psi$ anyon, corresponding to a composite of an $e$ anyon and an $m$ anyon. 
The string operators that move the $e$, $m$, and $\psi$ anyons are built from short string operators, which can be chosen to be:
\begin{align} \label{eq: Z2 TC short strings}
    W^e_\ell \equiv \vcenter{\hbox{\includegraphics[scale=.6]{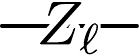}}}, \,\,
    \vcenter{\hbox{\includegraphics[scale=.6]{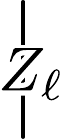}}}, \qquad 
    W^m_\ell \equiv \vcenter{\hbox{\includegraphics[scale=.6]{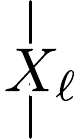}}}, \,\,
    \vcenter{\hbox{\includegraphics[scale=.6]{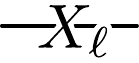}}}, \qquad 
    W^\psi_\ell \equiv \vcenter{\hbox{\includegraphics[scale=.6]{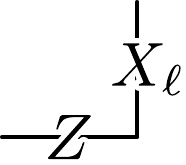}}}, \,\,
    \vcenter{\hbox{\includegraphics[scale=.6]{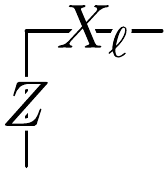}}},
\end{align}
where $\ell$ is an arbitrary edge. Note that, our only requirement of the short string operators is that their product along a path yields a string operator that commutes with the stabilizer generators along its length and creates the associated stabilizer violations at the endpoints. It can be checked from the commutation relations of the string operators that $e$ and $m$ have bosonic statistics, while $\psi$ has fermionic statistics (see Refs.~\cite{Kawagoe2020microscopic, Ellison2022subsystem}). 

To create twist defects at the endpoints of a path $\bar{\gamma}$ in the dual lattice,\footnote{Here, for simplicity, we assume that the path $\bar{\gamma}$ is not self-intersecting.} we condense $\psi$ anyons along $\bar{\gamma}$. This amounts to adding short string operators to the stabilizer group, which proliferate the $\psi$ anyons along the path $\bar{\gamma}$. In general, we say that fermions have been condensed along a path $\bar{\gamma}$ if there exist fermion string operators $W_{pq}^\psi$ connecting plaquettes $p$ and $q$ along $\bar{\gamma}$ such that 
the expectation value of $W_{pq}^\psi$ in a code state goes to a constant $C$ in the separation between $p$ and $q$:
\begin{align} \label{eq: fermion string expectation}
   \langle W_{pq}^\psi \rangle \to C \text{ in the limit of large }|p-q|.
\end{align}
This expression mimics the correlator used to detect boson condensation in Refs.~{\mbox{\cite{Fredenhagen1983charged,Fredenhagen1988dual,Gregor2011diagnosing,Verresen2021prediction}}}. However, here, we have restricted the correlator to string operators along the path $\bar{\gamma}$.
The expression in Eq.~\eqref{eq: fermion string expectation} is certainly satisfied if the string operators $W_{pq}^\psi$ belong to the stabilizer group, as is the case in what follows.

The first step to condensing the fermion $\psi$ is to identify a set of mutually commuting short string operators along $\bar{\gamma}$. To this end, we use the choice of short string operators $W_{\ell}^\psi$ in Eq.~\eqref{eq: Z2 TC short strings} to build a string operator $W_{\bar{\gamma}}^\psi$ supported along $\bar{\gamma}$, as illustrated in Fig.~\ref{fig: Wpsi string} 
We then decompose $W_{\bar{\gamma}}^\psi$ into a (possibly different) product of short string operators, such that the new short string operators (i) are mutually commuting, (ii) have disjoint supports, and (iii) each creates a pair of $\psi$ anyons. Note that the short string operators $W_\ell^\psi$ might not satisfy these conditions, since they may fail to be mutually commuting, such as in Fig.~\ref{fig: Wpsi string}. We claim that it is always possible to find a decomposition $W_{\bar{\gamma}}^\psi$ into short string operators satisfying (i)-(iii), up to redefining the string operator $W_{\bar{\gamma}}^\psi$ by Pauli operators at the endpoints (Fig.~\ref{fig: Wpsi string decomposition}). We let $\mathcal{S}^\psi_{\bar{\gamma}}$ denote the group generated by a set of short string operators satisfying (i)-(iii).

\begin{figure*}[t]
\centering
\subfloat[\label{fig: TC em permute}]{\includegraphics[width=.35\textwidth]{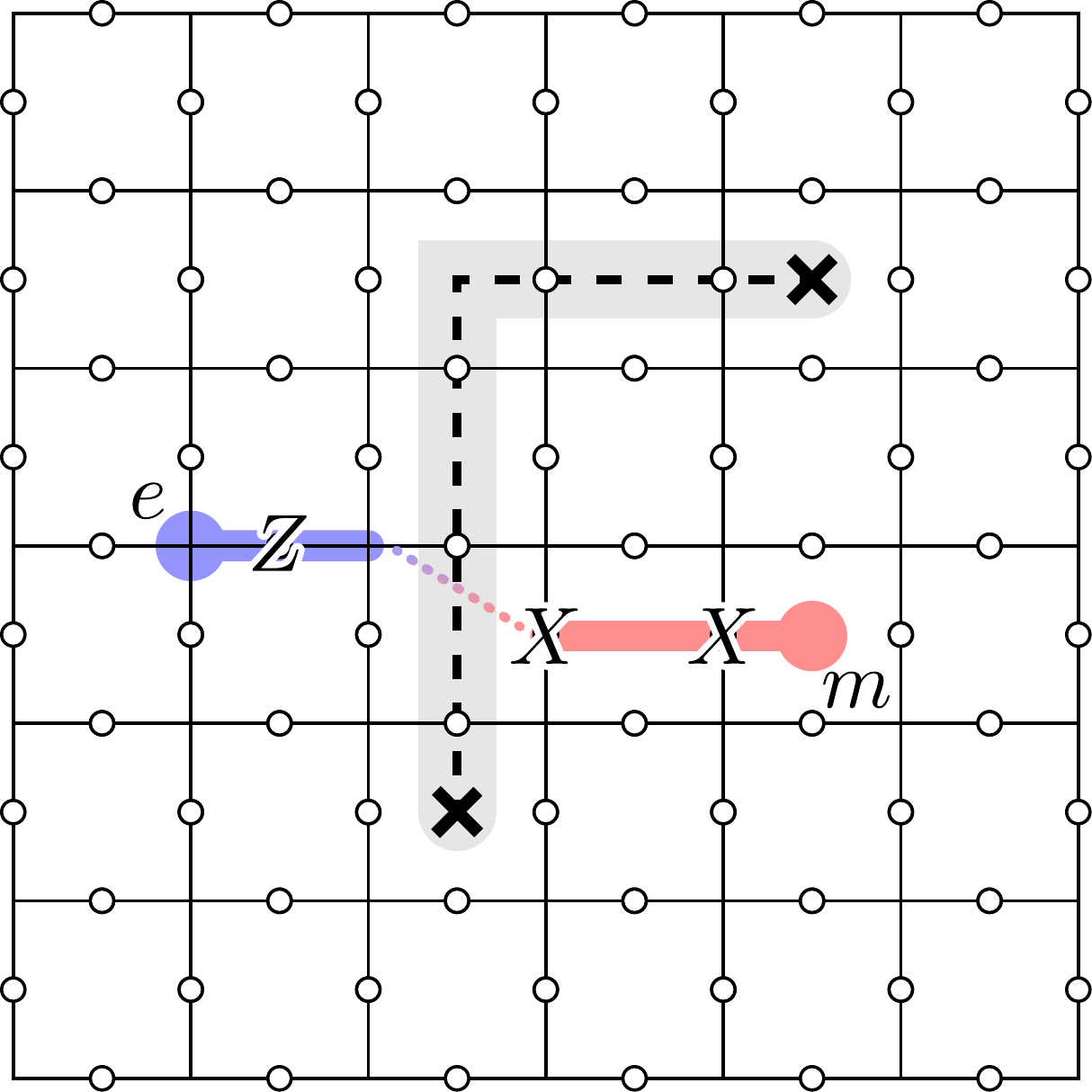}} \qquad 
\subfloat[\label{fig: TC em condense}]{\includegraphics[width=.35\textwidth]{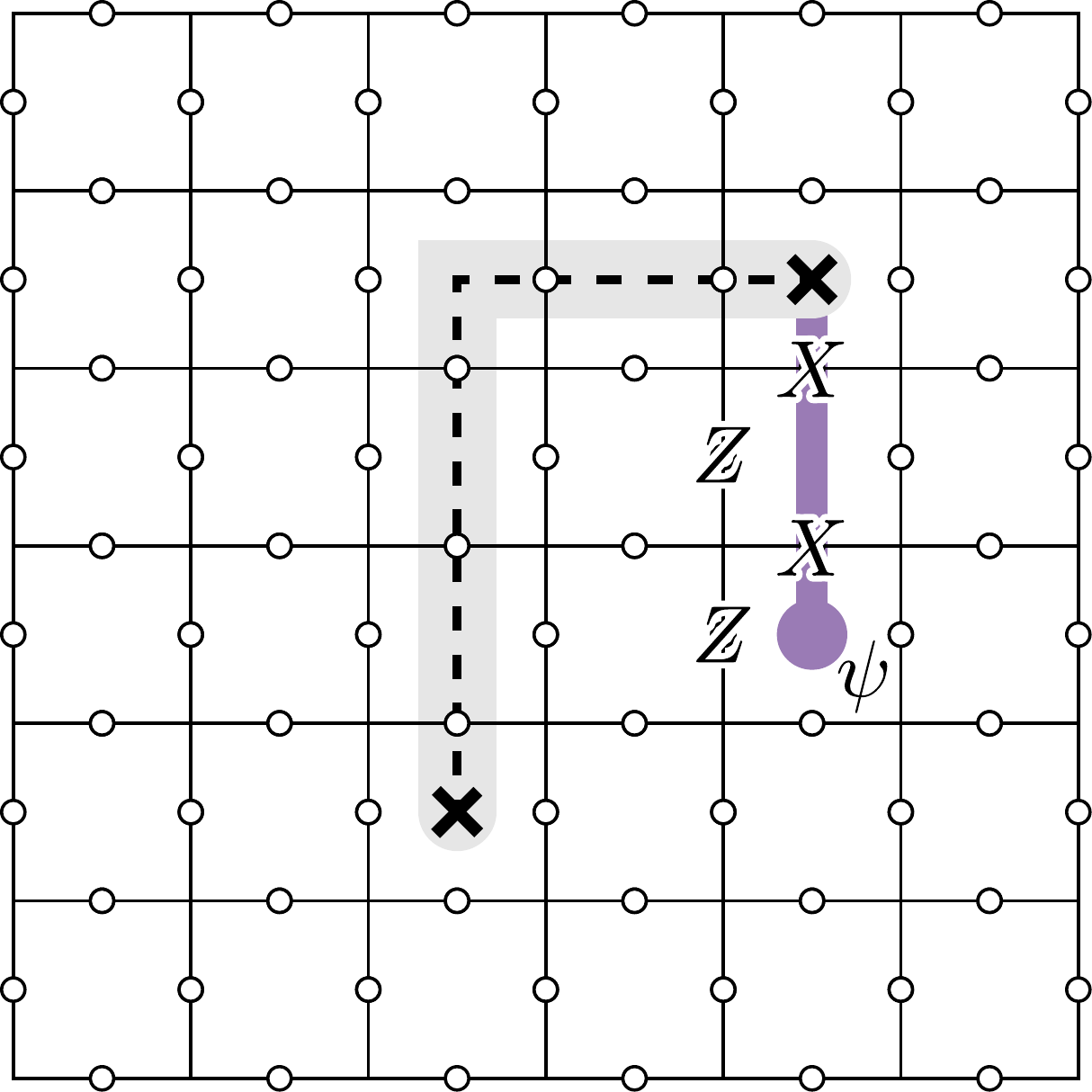}}
\caption{(a) The $e$ and $m$ anyons are permuted across the defect line. This is exemplified by the fact that there is a string operator (light blue and light red) that creates an $e$ anyon on one side, passes through the defect line, and creates an $m$ anyon on the other side, all while commuting with the stabilizers along its length. (b) The $\psi$ anyons can be condensed on the twist defects (bold black crosses). This is because there is a $\psi$ string operator (purple) that commutes with all of the stabilizers in the vicinity of the twist defect and creates a $\psi$ anyon at its endpoint. In other words, the $\psi$ string operator can be terminated on the twist defect without violating any stabilizers near the twist defect. }
\label{fig: TC defect signatures}
\end{figure*}

Next, we add the group $\mathcal{S}^\psi_{\bar{\gamma}}$ to the stabilizer group $\mathcal{S}_\text{TC}$ and remove all of the stabilizers that fail to commute with the operators in $\mathcal{S}^\psi_{\bar{\gamma}}$. We let $\mathcal{S}'_\text{TC}$ be the resulting stabilizer group, which is defined explicitly as:
\begin{align}
    \mathcal{S}'_\text{TC} \equiv \{ S \in \langle \mathcal{S}_\text{TC}, \mathcal{S}_{\bar{\gamma}}^\psi \rangle : ST = TS, \,\, \forall T \in \mathcal{S}_\text{TC}\},
\end{align}
where $\langle \mathcal{S}_\text{TC}, \mathcal{S}_{\bar{\gamma}}^\psi \rangle$ is the group generated by $\mathcal{S}_\text{TC}$ and $\mathcal{S}_{\bar{\gamma}}^\psi$.
Intuitively, the elements of $\mathcal{S}^\psi_{\bar{\gamma}}$ create, annihilate, and move the $\psi$ anyons along $\bar{\gamma}$. This means that for states in the $+1$ eigenspace of the stabilizers $\mathcal{S}'_\text{TC}$, we can freely add and remove pairs of $\psi$ anyons along $\bar{\gamma}$. In this sense, $\psi$ anyons have been proliferated along $\bar{\gamma}$. More importantly, the group $\mathcal{S}^\psi_{\bar{\gamma}}$ contains string operators that satisfy the criterion for fermion condensation in Eq.~\eqref{eq: fermion string expectation}.

We now give two arguments that this indeed produces a defect line along the path $\bar{\gamma}$. First, we argue that the $e$ anyons are permuted with $m$ anyons across the path $\bar{\gamma}$. In particular, there are string operators, intersecting the path $\bar{\gamma}$, that commute with the local stabilizer generators along the length of the string and create a vertex violation at one endpoint (an $e$ excitation) and a plaquette violation (an $m$ excitation) at the other, as depicted in Fig.~\ref{fig: TC em permute}. This tells us that $e$ anyons become $m$ anyons when moved across $\bar{\gamma}$ and vice-versa.
Second, we notice that the $\psi$ string operators can be terminated at the endpoints of $\bar{\gamma}$, as shown in Fig.~\ref{fig: TC em condense}. This is because, by construction, the short string operators in $\mathcal{S}^\psi_{\bar{\gamma}}$ fail to commute with the stabilizers of $\mathcal{S}_\text{TC}$ that detect $\psi$ anyons -- hence, those are removed in $\mathcal{S}'_\text{TC}$.
We say that the $\psi$ anyons can be condensed at the endpoints of $\bar{\gamma}$.

\section{Inferring products of plaquette stabilizers} \label{app: inferring stabilizers}

We argue here that the measurement outcome of a product of plaquette stabilizers along the defect line can be inferred from the measurements of the checks, following the schedule in Eq.~\ref{eq: Z2 defect measurement schedule}. We also show that the six-round measurement schedule in Eq.~\eqref{eq: six round schedule} is sufficient for inferring the measurement outcomes of the products of plaquette stabilizers, as long as the removed checks are 0- and 1-checks. 

The first thing to notice is that, along a defect line, every other check has been removed. This implies that, if a plaquette has more than one removed check, the removed checks must be of the same type, i.e., 0-, 1-, or 2-checks. More generally, this means that the product of plaquette stabilizers sharing edges with removed checks must have the same type of removed checks. As an example, in Fig.~\ref{fig: 0edge defect}, the (extensive) product of plaquette stabilizers that commutes with the defect checks have 0-edges in common. 

Let us assume that the removed checks shared by a given product of plaquette stabilizers $P$ are $r$-checks. We then go through the schedule to determine when the measurement outcome of $P$ can be inferred. For pedagogical reasons, suppose we measure the $r^*$-checks first. The measurement outcomes of the $r^*$-checks along the defect line are randomized by the measurements of the $(r+1)^*$-checks. This is because the $r^*$-checks along the defect line only commute with the $(r+1)^*$-checks in triplets, and by construction, one of the three checks is missing. The measurement outcomes of the $(r+1)^*$-checks are subsequently randomized after the $(r-1)^*$-checks. This follows from the fact that the $(r+1)^*$-checks and $(r-1)^*$-checks belong to different plaquettes (for plaquettes sharing $r$-edges). The measurement outcomes of the $r^*$-checks then commute with the product of three $(r-1)^*$-checks, because the $(r-1)^*$-checks come in triplets. Finally, the product of $(r-1)^*$-checks and $r^*$-checks commutes with the $(r+1)^*$-checks, since the commutation relations of the $(r-1)^*$-checks make up for the missing $r$-checks. Hence, the product of plaquette stabilizers sharing $r$-edges can be inferred from the following sequence of measurements, starting with the $(r-1)^*$-checks:
\begin{align}
    (r-1)^*, r^*, (r+1)^*.
\end{align}

For the six-round measurement schedule, this sequence only appears for $r=0,1$. More specifically, the sequences $\tilde{0}^*, 1^*, 2^*$, and $2^*, \tilde{0}^*, 1^*$ appear in the six-round sequence $\tilde{0}^*, 1^*, 2^*, 1^*, \tilde{0}^*, 2^*$, while the sequence $1^*,2^*,\tilde{0}^*$ does not. This suggests that the products of plaquettes sharing 2-edges cannot be inferred from the schedule. Moreover, this implies that the removed checks along the defect lines need to be 0- and 1-checks, as claimed.

\bibliographystyle{quantum} 
\bibliography{bib}

\end{document}